\documentclass[rmp,aps,twocolumn,epsfig,eqsecnum,floatfix,amsmath,amssymb]{revtex4}
\usepackage{graphicx}
\usepackage{dcolumn}
\usepackage{bm}
\usepackage{epsfig}

\def\be{\begin{equation}}
\def\ee{\end{equation}}
\def\ba{\begin{eqnarray}}
\def\ea{\end{eqnarray}}

\def\LSCO{La$_{2-x}$Sr$_x$CuO$_4$}
\def\LCO{La$_2$CuO$_4$}
\def\LCOplus{La$_2$CuO$_{4+\delta}$}
\def\LSNiO{La$_{2-x}$Sr$_x$NiO$_{4+\delta}$}
\def\LBCO{La$_{2-x}$Ba$_x$CuO$_4$}
\def\YBCO{YBa$_2$Cu$_3$O$_{6+y}$}

\def\BSCCO{Bi$_2$Sr$_2$CaCu$_2$O$_{8+\delta}$}
\def\C60{A$_x$C$_{60}$}
\def\LNSCO{La$_{1.6-x}$Nd$_{0.4}$Sr$_x$CuO$_{4}$}

\def\HgCu3{HgCa$_2$Cu$_3$O$_{8+y}$}
\def\HgCu4{HgBa$_2$Ca$_3$Cu$_4$O$_{10+y}$}
\def\TlCu{Tl$_2$Ba$_2$CuO$_{6+\delta}$}
\def\TlCu3{Tl$_2$Ba$_2$Ca$_2$Cu$_3$O$_{10+y}$}
\def\TlCu4{Tl$_2$Ba$_2$Ca$_3$Cu$_4$O$_{12+y}$}

\def\BiCu3{Bi$_2$Sr$_2$Ca$_{2}$Cu$_3$O$_y$}

\def\BiCaMnO{Bi$_{1-x}$Ca$_x$MnO$_3$}
\def\NCCO{Ne$_{2-x}$Ce$_x$CuO$_{4\pm\delta}$}
\def\8LSCO{La$_{1.88}$Sr$_{.12}$CuO$_4$}
\def\110LNSCO{La$_{1.5}$Nd$_{0.4}$Sr$_{0.1}$CuO$_{4}$}
\def\stage4LCO{La$_{2}$CuO$_{4+\delta}$}
\def\Y248{YBa$_2$Cu$_4$O$_8$}
\def\PCCO{Pr$_{2-x}$Ce$_x$CuO$_{4\pm\delta}$}
\def\Tn{T_{\rm n}}
\def\F{{\rm F}}

\newcounter{savefoot}
\newcounter{savefootb}

\begin{document}

\title{How to detect fluctuating stripes in the high-temperature superconductors}
\author{S. A. Kivelson}
\affiliation{Department of Physics,
University of California at Los Angeles,
405 Hilgard Ave., Los Angeles, CA 90095}
\author{E. Fradkin}
\affiliation{Department of Physics, University of Illinois,
1110 W.\ Green St., Urbana IL 61801-3080}
\author{V. Oganesyan}
\affiliation{Department of Physics, Princeton University,
Princeton NJ 08544}
\author{I. P. Bindloss}
\affiliation{Department of Physics,
University of California at Los Angeles,
405 Hilgard Ave., Los Angeles, CA 90095}
\author{J. M. Tranquada}
\affiliation{Physics Department, Brookhaven National Laboratory, Upton, NY
11973-5000}
\author{A. Kapitulnik}
\affiliation
{Department of Physics, Stanford University, Stanford, CA 94305-4045}
\author{C. Howald}
\affiliation
{Department of Physics, Stanford University, Stanford, CA 94305-4045}

\date{\today}
\begin{abstract}
We discuss  fluctuating order in a quantum disordered phase proximate
to a quantum critical point, with particular emphasis on fluctuating
stripe order. Optimal strategies for extracting information concerning such
local order from experiments are derived with emphasis on
neutron  scattering and scanning tunneling microscopy.
These ideas are
tested by application to two model systems - the exactly solvable one
dimensional electron gas  with an impurity, and a weakly-interacting 2D
electron gas. We extensively review  experiments on
the cuprate high-temperature superconductors which can be analyzed
using these strategies.  We
adduce evidence that stripe correlations are widespread in the cuprates.
Finally, we compare and contrast the advantages of two limiting
perspectives on the high-temperature superconductor: weak coupling, in
which correlation effects are treated as a perturbation on an underlying
metallic (although renormalized) Fermi liquid state, and strong coupling,
in which the magnetism is associated with well defined localized spins,
and stripes are viewed as a form of micro-phase separation.
We present quantitative indicators that the latter view better accounts
for the observed stripe phenomena in the cuprates.
\end{abstract}
\maketitle

\tableofcontents

\section{Introduction}

Ordered states of matter are characterized by 
broken symmetry.  Depending on various real-world details, this may be
relatively easier or harder to detect experimentally, but once detected
it is unambiguous.  The notion that order parameter fluctuations are
important in the disordered phase proximate to an ordered state is a
rather obvious extension of the idea of broken symmetry;  however,
this notion is more difficult to define precisely.
To develop this important concept, we
will discuss  strategies for detecting quantum fluctuating order in
the particular context of high-temperature superconductor. We will
focus on the detection of stripe order in the putative stripe liquid
phase of these systems.\footnote{For
brief reviews of the current evidence on
various types of stripe order in the cuprates, see
\protect\cite{emer99,zaan00,sach02a,carl03,oren00,neto03}} More generally, 
we are interested
in electronic liquid crystalline states and their associated
fluctuations \cite{kive98}, but the results are also easily generalized to
other forms of order.\footnote{Fluctuations associated with off-diagonal 
({\it i.e.\/}, superconducting) order are best studied using
different strategies;  see, for example,
\protect\cite{xu00,wang01,ussi02,carl00,iguc01,jank99}}

Since the whole notion of fluctuating order is based on proximity of
an ordered state, it is essential first to establish  the existence of
the ``nearby'' ordered state by directly detecting the relevant broken
symmetry.   The ordered phase may be  induced by making small
changes to the chemical composition of the material, applying
pressure or magnetic fields, etc.
Unless an actual ordered state can be reached, it is dangerous to
speculate about the effects of related fluctuating order.

Typically, the best way to detect both the broken-symmetry state and
the relevant
fluctuations is by measuring the appropriate dynamical structure
factor, $S(\bf q,\omega)$.
Indeed,  X-ray and neutron scattering studies have provided the best
evidence\footnote{For reviews of neutron and X-ray evidence of
stripes, see \protect\cite{emer99,tran98b,tran98a}.} of ordered and
fluctuating stripe phases, as we shall discuss.  However, in many
interesting materials,
appropriate crystals are not available and so such experiments are
not possible.  Here, probes of
local order, such as nuclear magnetic resonance (NMR), nuclear quadrupole
resonance (NQR), muon spin rotation ($\mu$SR), and scanning
tunneling  microscopy (STM) techniques, may be the best available.  All of
these are quasi-static ({\it i.e.\/}, nearly zero frequency) probes.  In
a pure quantum system, in its disordered  phase, the order-parameter
fluctuations are
not static, but rather fluctuate with a characteristic  frequency
that grows  with the distance ${\ell}$ to the quantum critical
point.  Thus, unless something is done to ``pin'' these
fluctuations, they are invisible to
local probes.  Such pinning is induced by
boundaries, vortices, crystal-field effects, weak quenched disorder,
etc.

{\bf In this article}, we first discuss results obtained for solvable
model systems, which we analyze in various ways to illustrate the optimal
strategies for extracting information about local order.  In particular,
much of this discussion addresses the character of the local order in a
quantum disordered phase close to a quantum critical point;  we believe
that the  intuitions gleaned from this study are more generally valid,
but without the nearness to the critical point as a small parameter, it
is difficult to treat the
problem in a controlled fashion.
We also review in some depth, although by no means exhaustively, the
experimental evidence of various forms of stripe order in the cuprate
superconductors.   There are several related topics that we do not
cover in this article;  instead, we direct the interested reader to
recent reviews.  Specifically, the mechanism of stripe
formation\footnote{Recent reviews of the mechanisms of stripe formation
in the cuprates and, more generally in doped antiferromagnets, often with
rather different perspectives on the issue, can be found in Refs.
\protect\cite{emer99,zaan98b,vojt99,whit00,whit02,carl03,zaan94,ichi99b,
hass02}.}
is
given short shrift, and the possible relevance\footnote{Various
perspectives concerning the relevance of local stripe order to other
properties of the high-temperature superconductor are reviewed in
\protect\cite{carl03,zaan01,lore01a,lore01b,neto01}.} of local stripe
order to the mechanism of high temperature superconductivity and to the
various remarkable normal state properties observed in the cuprates is
only touched on briefly.
More broadly, the context for this review is the role of ``competing orders'' in high
temperature superconductors.\footnote{Several recent discussions of the
role of competing orders in determining the phase diagram of the high
temperature superconductors are contained in
\cite{emer99,zaan00,sach00,oren00,chak01,varm97,kive01}.} 

In Section \ref{general}
we discuss basic scaling considerations
that govern the relevant length,
frequency, and energy scales of fluctuating order in the neighborhood
of a quantum critical
point, and the consequences of the pinning induced by disorder and
other perturbations.
We show that it is the low frequency part of
the dynamical structure factor, rather than its integral over
all frequencies, that contains the clearest information concerning
local order.  We also define a new
response function which determines the local density-of-states
modulations induced by a weak-impurity potential.  Finally, we sketch
various possible versions of the phase diagram of a system with competing
stripe and superconducting phases.

In Section \ref{1DEG} (and Appendix \ref{sec:luttinger}),
we consider a theoretically well understood system, the one
dimensional electron gas (1DEG)
in the presence of an impurity.  We
compute the local density of states (LDOS), as would be measured in STM, and the
dynamical structure factor, and show by this explicit example how the
general considerations articulated in the present paper are manifest in
this solvable model.  (The 1DEG can also be viewed as a quantum critical
state associated with charge density wave (CDW) order.)  In addition to its 
pedagogical value,
this section contains explicit results that should be useful for
analyzing STM experiments on quasi 1D systems with dilute impurities, such
as the chain-layers of YBCO \cite{derr02}, or STM experiments on
carbon nanotubes \cite{horn02,odom02}. 

In Section \ref{2DEG} we calculate the 
LDOS in the context of a  simple model-system
with quasiparticles and incipient order - 
the weakly-interacting electron gas in two dimensions.

In Section
\ref{detection} we discuss applications of these ideas to experiments
in the cuprates. In Section
\ref{diffraction} we discuss diffraction studies, and in Section
\ref{STM} STM studies.  This latter section contains new insights
concerning the optimal way to
analyze STM data to extract information about fluctuating order;
applying these ideas to the recent  experimental results of
\textcite{hoff02,hoff02b}, and \textcite{howa02,howa02b},  reveals
the existence of a nearby stripe ordered phase in optimally doped BSCCO.
~\footnote{This conclusion is in agreement with the inferences drawn from the data
by \textcite{howa02,howa02b}, but in disagreement with those
drawn by \textcite{hoff02b};  we will discuss the
origins of the differing conclusions.}    Section
\ref{nematic} contains a discussion of experiments to detect local
nematic order, and a discussion of recent STM evidence of such order on
the same BSCCO surfaces. In Section
\ref{18}, we discuss indirect evidence of stripe order that comes from the
``1/8 anomaly.'' In Section \ref{other} we briefly discuss other probes of 
stripe order, including NMR, NQR and $\mu$SR.

In Section\ref{weak}, we address an important issue of
perspective: In the weak coupling limit, the electronic properties
of a solid are essentially determined by the quasiparticle band
structure, while collective modes and various forms of order reflect
slight rearrangements
of the states near the Fermi surface.  Often these effects can be
plausibly treated in the context of a Hartree-Fock or RPA
treatment of the residual interactions. Conversely, in the strong
coupling limit, the physics is more simply understood in terms of
interacting  collective modes, such as spin waves, superconducting
phase fluctuations, and the ``phonons'' of a charge-ordered state.
Here, we discuss the interrelation between these opposite
perspectives. In particular we present evidence that the
stripe order observed in the cuprates and related compounds is
best thought of as arising from the strong-coupling
antiferromagnetism of the undoped Mott insulating parent compounds
than from the more conventionally metallic physics of the strongly
overdoped materials.

Our most important conclusions
are embodied in five  numbered ``Lessons,'' which are stated in 
Sections \ref{general}, \ref{1DEG} and \ref{2DEG}, and summarized in
Section \ref{conclusion}.

\section{General considerations}
\label{general}

Since the discovery\cite{bedn86} of high temperature
superconductivity in the cuprate superconductors, there
has been intense interest in the question of how a Mott insulating
antiferromagnet is converted, upon doping, into a high temperature
superconductor.  Generally, in electronic systems, there is a
competition between the kinetic energy (Fermi pressure) which
favors a uniform Fermi liquid phase with sharply defined,
itinerant quasiparticles, and the Coulomb repulsion between
electrons, which favors various forms of
insulating magnetic and/or charge ordered states. Thus, it should
not be at all surprising to find various forms of charge ordered
states appearing in doped antiferromagnets.  In particular,
``stripe'' states have consistently turned up in theoretical
studies of doped antiferromagnets, including  early Hartree-Fock
solutions of the Hubbard model\footnote{\cite{zaan89,schu89,mach89}},
studies of Coulomb frustrated phase separation\footnote{
\cite{emer93,low94,seul95,cast95,hass99}},
slave-boson mean-field theories of the t-J model\footnote{
\cite{han01,lore02,seib98}},
and various Monte-Carlo and DMRG studies of t-J and Hubbard
models\footnote{ For varying perspectives, see
\cite{whit00,whit98a,hell97,hell00}}. 
The discovery~\footnote{\cite{hayd92,chen93,tran94a}} of stripe
order in {\LSNiO} 
and soon after in {\LNSCO} \cite{tran95a} added considerable credibility
to the suggestion that stripe states form an important bridge between the
Mott insulator, and the more metallic state at heavy doping.

``Stripes" is a term that is used to describe unidirectional density wave
states, which can involve unidirectional charge modulations (``charge
stripes") or coexisting charge and spin density order (``spin
stripes").\footnote{One can also imagine ``orbital stripes'' which
involve a unidirectional modulation of a staggered flux or d-density wave
order \protect\cite{scho02}; in large part, the analysis of the present
paper would apply equally well to this form of order, as well.}
CDW states can occur in the
weak coupling limit if there are sufficiently well nested portions
of the Fermi surface.  From the strong coupling perspective, stripes are a
real-space pattern of micro-phase separation (hence the name), driven
largely by a lowering of the doped-hole kinetic energy\footnote{As
discussed, for instance, in \cite{poil89,schu89,neto03,cher00,
zach00,emer99,whit98a}.}, in which the doped holes are itinerant in
metallic rivers, and the antiferromagnetic correlations of the parent
insulator are preserved in between.  Although the characteristic dynamics
of the stripes may be quite different in these two limits, as are the
possible implications for other physical properties, from a broken
symmetry viewpoint there is no difference between a unidirectional CDW and
a stripe-ordered state.  

Since the principal purpose of the present
paper is to address questions concerning the existence of local stripe
order we will for the most part consider
unidirectional CDW and stripe order as two limits of the same physics.
Indeed, it is important to
stress that the microscopic mechanism of stripe formation in particular
materials is still not clear.  Questions such as whether or not the
long-range Coulomb interactions are important have been widely debated,
but remain unresolved.
There is always a  linear coupling between any form of charge order and
equal period  lattice distortions, so it is obvious that lattice
distortions play a significant role in enhancing all the observed stripe
phenomena, whether or not they are fundamental to the mechanism.  This
observation is further corroborated by the observed stabilization
 of stripes by particular orientational orders of the
apical oxygens in some of the cuprates~\cite{buch94a,axe94}, and by a series of thermal
transport ~\cite{babe98,sun03} and isotope effect~\cite{craw90}
anomalies associated with the onset of stripe order~\cite{tran95a}.

\subsection{Stripe ordered phases}
\label{stripes}

In 2D, an ordered stripe phase
with the stripes running in the
$y$ direction gives rise to new Bragg peaks in the electronic charge
scattering (and corresponding peaks in the nuclear
scattering) at
${\bf k}=\pm {\bf Q}_{\rm ch} = (2\pi/a)(\pm\delta_{\rm ch},0)$ and
harmonics, where
$\delta_{\rm ch}=1/\lambda_{\rm ch}$ and $\lambda_{\rm ch}a$ is the
charge-stripe  period.  Where
spin order coexists with charge order, new magnetic Bragg peaks occur
at harmonics of ${\bf k}={\bf Q}_{\rm s}={\bf Q}_{\rm AF}\pm \frac12
{\bf Q}_{\rm ch}$ where ${\bf Q}_{\rm AF} = (2\pi/a)(\frac12,\frac12)$ is
the N{\'e}el ordering vector.\footnote{Note that, in a crystal, wave
vector equalities are always to be interpreted as meaning equal up to  a
reciprocal lattice vector.}

Charge stripes break rotational symmetry and translation symmetry
perpendicular to
the stripes-in a crystal, these are to be interpreted as breaking of
the crystal symmetry group, rather than the continuous symmetries of
free space.  The relevant
order parameter, $\langle\rho_{{\bf Q}_{\rm ch}}\rangle$, is the Fourier
component of
the electron charge density at the ordering wave vector.  If the state is, in
addition, a conducting or superconducting electron fluid, it is a charge stripe
smectic.  Spin stripes, in addition, break spin-rotational and time reversal
invariance (although a particular combination of time reversal and translation
is preserved).  Wherever there is spin-stripe order, there is necessarily
\cite{zach98} charge-stripe order, as well.  The new order parameter which
distinguishes the spin stripe phase,
$\langle{\bf s}_{{\bf Q}_{\rm s}}\rangle$, is
the Fourier component of the spin-density.

There is more than one possible stripe-liquid phase.  In
particular, there is the possibility of a ``stripe nematic''
phase, in which thermal or quantum fluctuations have caused the
stripe-ordered state to melt ({\it i.e.\/}, translational
symmetry is restored) but orientational symmetry remains broken
({\it i.e.\/}, a snapshot of the system is more likely to see
stripe segments oriented in the $y$ than $x$ direction).  The
nematic shares with a charge-ordered
state a precise definition in terms of broken symmetry.  The
order parameter can be taken to be any physical  quantity
transforming like a traceless symmetric tensor. For instance, the
traceless part of the dielectric tensor is a measure of nematic
order of the charged fluid.  In two dimensions, a useful nematic
order parameter is also~\cite{chai95}
\begin{equation}
{\cal Q}_{\bf k}\equiv \frac{S({\bf k})-S({\cal R}[{\bf k}])}{S({\bf
k})+S({\cal R}[{\bf k}])},
\label{N}
\end{equation}
where ${\cal R}$ is a rotation by
$\pi/2$, and
\begin{equation}
S({\bf k})=\int_{-\infty}^{\infty} \frac{d\omega}{2\pi} \; S({\bf k},\omega)
\label{S}
\end{equation}
is the thermodynamic (equal-time) structure factor.
More exotic states can also occur, such as a ``nematic spin-nematic,"
which breaks rotational, spin rotational, and time reversal symmetry, but
not  the  product of
${\cal R}$ and time reversal.\footnote{A nematic spin nematic is most
easily pictured as a state in which the spin up and spin down electrons
each form a nematic state, with their principle axes at 90$^\circ$ to each
other.}

\subsection{Fluctuating order near a quantum critical point}
\label{critical}

We consider a system in a quantum disordered phase near to a quantum
critical point beyond which
the ground state would be ordered.\footnote{For extremely clear reviews,
see  \protect\cite{sach00} and \protect\cite{sach99}, as well as
\cite{sond97}.}  To be concrete, we will discuss charge-stripe order.
By the quantum
disordered phase,
we mean one in which there is no long-range stripe order as $T\to 0$,
but there may be other
forms of order, for instance superconducting order.

The charge-density dynamical structure factor,
$S_{\rm ch}({\bf k},\omega)$,  for ${\bf k}$ in the
neighborhood of ${\bf Q}_{\rm ch}$, measures the collective fluctuations
which are most sensitive to
the proximity of the quantum critical point.  The scaling theory of
quantum critical phenomena
tells us that, on the quantum disordered side of the quantum critical
point, there are diverging
length and time scales,
$\xi \sim \ell^{-\nu}$ and $\tau \sim \ell^{-\nu z}$  where
$\nu$ is a critical exponent, $z$ is the dynamical critical exponent,
and $\ell$ is the dimensionless distance to the quantum
critical point.\footnote{
The values of the critical exponents are ``universal" in the sense that
they are a small discrete set of numbers depending on some 
general features
of the critical point.  Under
many circumstances, $z$ is either 1 or 2.  $\nu$ depends on the
effective dimension, $D+z$ and, to a lesser extent,
on the symmetry of the order parameter;  typically, for $D\ge 2$,
$\nu$ is in the range $2/3 \ge \nu \ge 1/2$.}

In the quantum disordered phase, and in the absence of quenched
disorder, the fluctuation spectrum
has a characteristic frequency scale, $E_{\rm G}/\hbar$.
In the
absence  of dissipation\footnote{
In the presence of zero temperature dissipation, such as
one expects in a normal metal, the
collective-mode spectrum may be gapless, in which case the interpretation of
$\tau$ is more subtle.  Most of the scaling arguments we make in the present
paper are readily generalized to this case, as well. For a recent 
perspective on the effect of
dissipation on quantum critical phenomena, see
\cite{kapi01}.  See also \textcite{mill93} and \textcite{sach99b} on
quantum critical points in metals.},
which is the case we will treat for concreteness,
$E_{\rm G}$
typically is a gap in the collective
mode spectrum.
$E_{\rm G}$ is related to the correlation time by the scaling law
$E_{\rm G} \sim \hbar/\tau\sim \ell^{\nu z}$. For $\hbar\omega$ slightly
larger than $E_{\rm G}$, $S({\bf k},\omega)$ has a pole
     corresponding to a sharply defined elementary excitation;  this
``soft-mode" is the quantity which condenses across the quantum
critical point, so its quantum
numbers and characteristic wave vector directly encode the nature of the
nearby ordered state. At somewhat higher energy (typically,  for
$\hbar\omega >3E_{\rm G}$) $S({\bf
k},\omega)$ exhibits a multi-particle continuum. Deep in the
continuum the system effectively exhibits scale invariance and it
behaves  in much the same way
as if it were precisely at the quantum critical point. Both in this
high frequency regime and at the quantum critical point the notion of a
quasiparticle is, truly speaking, ill-defined, although
for systems with small anomalous dimension $\eta$ (the typical case
for $D\ge 2$) the continuum will often exhibit
features (branch-points, not poles) whose dispersion resembles that
of the Goldstone modes of
the ordered phase. \footnote{The naive argument, which is often made, that
at frequencies larger than $E_G$ the quantum disordered phase looks like
the ordered phase is thus not completely correct.  There is, however, a
more nearly correct version of this statement:  at high frequencies, on
both the ordered and disordered sides of the quantum critical point, the
system looks quantum critical, and so looks the same whichever state
is being probed.}\setcounter{savefootb}{\value{footnote}}

A classical critical point is described by thermodynamics alone,
so none of these dynamical
considerations affect the critical phenomena.  In particular, it
follows from the classical limit of the
fluctuation-dissipation theorem that $S({\bf k})=T \chi({\bf k},\omega=0)$, so
$S$ has the same critical behavior as the
(static) susceptibility, $\chi$, even though $S$ involves an integral over the
dynamical structure factor at all frequencies.
Consequently, a growing peak in $S({\bf k})$ at the ordering vector,
${\bf k}={\bf Q}_{\rm ch}$ with width
$|{\bf k}-{\bf Q}_{\rm ch}|\sim 1/\xi$,
and  amplitude $S({\bf Q}_{\rm ch})\sim |T-T_c|^{-\gamma}$
reflects the presence of fluctuating stripe order near a thermal
transition.  Here $\gamma=\nu(2-\eta)$ is
typically about 1 or greater, so the amplitude is strongly
divergent.

The situation is quite different at a quantum critical point, where
the dynamics and the thermodynamics are
inextricably linked.  Here, the fact that
the largest contribution
to $S({\bf k})$ comes from the high frequency  multi-particle
continuum which does not directly probe the
fluctuating order means that the relevant structure in $S({\bf k})$
is relatively small, and can be difficult to
detect in practice. To see this, we compare the expressions in terms
of the dissipative response function,
$\chi^{''}({\bf k},\omega)$, for the susceptibility
(obtained from the Kramers-Kr\"onig relation) and $S({\bf k})$
(obtained from the fluctuation-dissipation theorem):
\ba
\chi({\bf k},\omega=0)=&&\int \frac{d\omega}{2\pi} \; \chi^{\prime\prime}(
{\bf k},\omega)/\omega  \\
\label{chi}
S({\bf k})=&&\int \frac{d\omega}{2\pi} \;
\coth\left(\frac{\beta\hbar\omega} 2\right) \;
\chi^{\prime\prime}({\bf k},\omega)
\nonumber
\ea
In the limit $T\to 0$,  $\coth(\beta\hbar\omega) \to 1$;  then
clearly, the factor $1/\omega$ weights the important
low frequency part of the integral expression for  $\chi$ much more
heavily than in the expression for $S$.
Consequently, it follows under fairly general circumstances\cite{chubasubir}
   from the
scaling form of $\chi^{\prime\prime}$ that $S({\bf Q}_{\rm ch})
\sim \tau^{-1} \chi({\bf Q}_{\rm ch})$, or in other words that
$\chi({\bf Q}_{\rm ch}) \sim \ell^{-\gamma}$ is
strongly divergent in the neighborhood of the quantum critical point,
while $S({\bf Q}_{\rm ch}) \sim \ell^{\nu z-\gamma}
=\ell^{-\nu(2-z-\eta)}$ is much more weakly divergent, or, if perchance
$z\ge 2-\eta$, not divergent at all.
At criticality ($\ell=0$), similar
considerations imply that $S({\bf k},\omega)$ is a less singular
function at small momentum/frequency than the corresponding
susceptibility.

These considerations apply even on the ordered side of the quantum
critical point, where the
Bragg peak makes an exceedingly small (although sharp)  contribution
to the total structure factor which vanishes in
proportion to $\ell^{2\beta}$.   By contrast, the Bragg peak
constitutes the entire contribution to $S(
{\bf k},\omega=0)$, and is the dominant piece of the structure
factor integrated over any small frequency
window which includes $\omega=0$.

{\bf Lesson \#1:}  There is an important lesson to
be gleaned from this general discussion concerning the best
way to analyze experiments . It is the low frequency part of
$S({\bf k},\omega)$ which is most directly  sensitive to the stripe
order.  Rather than analyze $S({\bf k})$,
information concerning local stripe order is best obtained by
studying $\chi({\bf k},\omega=0)$
or by analyzing the partially-frequency-integrated spectral
function,
\begin{equation}
\tilde S({\bf k},\Omega)\equiv (2 \Omega)^{-1}\int_{-\Omega}^{\Omega}
\frac{d\omega}{2\pi}\;
S_{\rm ch}({\bf k},\omega)
\label{eq:Sch-int}
\end{equation}
where on the ordered side of the critical point, the smaller $\Omega$
the better, while on the
quantum disordered side, $\Omega$ must be taken larger than
$E_{\rm G}/\hbar$, but not more than a
few times $E_{\rm G}/\hbar$.  This quantity is less severely contaminated
by a ``background" arising
from the incoherent high energy excitations.  As we will discuss in
Sec. \ref{diffraction}, this
is precisely the way local stripe order is best detected in clean,
high temperature
superconductors.

\subsection{How weak disorder can make life simpler}

In the absence of quenched disorder, there is no useful
information available from the structure factor at frequencies
less than $E_{\rm G}$, which means  in particular that static
experiments ($\omega=0$) are blind to  fluctuating order. However,
a small amount of quenched disorder, with characteristic energy
scale $V_{\rm dis}\sim E_{\rm G}$  but much smaller than the
``band width" of the continuum of the spectral function, has
important effects on the low-energy states. It is intuitively
clear that induced low-frequency structure of $S({\bf k},\omega)$
will be largest for values of ${\bf k}$ where, in the absence of
disorder, $S$ has spectral weight at the lowest frequencies, {\it i.e.\/}, 
for ${\bf k}\sim{\bf Q}_{\rm ch}$. 
Although the effects
of quenched randomness on the pure critical theory can be subtle
\cite{grif69,mcco68,fish92}, and are rarely well understood (unless
disorder happens to be irrelevant), low-energy
states are frequently produced.

  {\bf Lesson \# 2:} The upshot is that the effect of
weak quenched randomness in a quantum disordered phase is to produce
a low frequency ``quasi-elastic" part of the spectral function
$S({\bf k},\omega)$.   In other words, disorder will eliminate the
spectral gap, but will only weakly affect the partially integrated
spectral function, Eq.~(\ref{eq:Sch-int}), with the integration
scale set by $V_{\rm dis}$. In particular, in the presence of weak
disorder, the static structure factor, $S({\bf k},\omega=0)$,
should exhibit similar ${\bf k}$ dependence as $\tilde  S({\bf k},\Omega)$
of the pure system, and so  can be used to reveal the nature of the
nearby ordered phase.

\subsubsection{Response functions}
\label{sec:response}

To formalize some of these notions, we consider the somewhat simpler
problem of the response of
the system to a weak, applied external field which couples to the
order parameter.
In a quantum disordered phase the existence of an order parameter can
only be made apparent
directly by coupling the system to a suitable symmetry-breaking field.
For a charge-stripe smectic, a non-uniform potential
couples to the order parameter and thus serves as a suitable symmetry
breaking field. Thus, the
Fourier component $V_{\bf k}$ of a weak potential induces a
non-vanishing expectation value of
the order parameter which, in linear response, is
\begin{equation}
\langle \rho_{\bf k} \rangle =\chi_{\rm ch}({\bf k}) V_{\bf k} + \ldots
\label{eq:nQ}
\end{equation}
     The linear response law of Eq.~(\ref{eq:nQ}) is valid provided $V_{\bf
k}$ is sufficiently small. However, if the typical magnitude of $V$ with
Fourier components in the range $\left\vert \bf k-\bf Q_{\rm ch}
\right\vert \xi \lesssim 1$
is not small,
$\bar V_{\bf k_{\rm ch}}
\gg E_{\rm G}$,
then the  critical region is accessed and Eq.~(\ref{eq:nQ}) is
replaced by the law
\begin{equation}
\langle \rho_{\bf k_{\rm ch}}\rangle \sim  [\bar V_{\bf
k_{\rm ch}}]^{1/\delta}
\label{eq:nQ-critical}
\end{equation}
where $\delta=2/(D-2+\eta)$ is another critical exponent.

Many local probes, including STM and NMR, are more sensitive to the
local density of electronic states, ${\cal N}({\bf r},E)$.
Again, in the quantum disordered state, translation invariance
implies that, in the absence of an external
perturbation, ${\cal N}({\bf r},E)={\cal N}_0(E)$ is independent of
$\bf r$.   In the presence of a weak potential,
it is possible to define a  relation for the local density of states
\begin{equation}
N({\bf k},E)= \chi_{\rm DOS}({\bf k},E) V_{{\bf k}} + \ldots
\label{linear}
\end{equation}
where $N({\bf k},E)$ is the Fourier transform of ${\cal N}({\bf
r},E)$. {From} linear response theory it follows that
\ba
\chi_{\rm DOS}({\bf k},E)&&  =  (2 \pi)^{-1}\int d{\bf r}\, dt\, d\tau\,
e^{iEt-i{\bf k}\cdot {\bf r}}\theta(\tau)
\nonumber\\
\label{chiEk}
\times
&& \langle[\{\Psi^{\dagger}_{\sigma}({\bf r},t+\tau),
      \Psi_{\sigma}({\bf r},\tau)\},\hat n({\bf 0})]\rangle
\label{eq:chi-dos}
\ea
where $\theta$ is the Heaviside function, $\Psi_{\sigma}^{\dagger}$
is the electron creation operator, and $\hat
n=\sum_{\sigma}
\Psi_{\sigma}^{\dagger}\Psi_{\sigma}$ is the electron density
operator.  Note that, despite appearances, $E$ is not a
frequency variable, but is rather the energy at which the time
independent density of states is measured.  A simple
sum-rule relates $\chi_{\rm ch}$ to $\chi_{\rm DOS}$:
\begin{equation}
\chi_{\rm ch}({\bf k})=\int dE\, f(E) \chi_{\rm DOS}({\bf k}, E)
\label{sum-rule}
\end{equation}
where $f(E)$ is the Fermi function.

It is also interesting to consider stripe orientational
(nematic)
order.  Since $\chi_{\rm ch}(\bf k)$ is a property of the
uniform fluid phase, it respects all the symmetries of the underlying
crystal; in particular, if rotation by $\pi/2$ is a symmetry of the
crystal, then $\chi_{\rm ch}(\bf k)=\chi_{\rm ch}({\cal R}[\bf k])$, so
no information about incipient nematic order can be obtained to linear
order in the applied  field.  However, the leading non-linear response
yields a susceptibility for the nematic order parameter, defined in
Eq.~(\ref{N}), as
\ba
&&\!\!\!\!\!\!\!\!\!\!\! {\cal Q}_{{\bf k}}
=\int  {d{\bf p}}\
\chi_{\rm nem}({\bf k};{\bf p}) [V_{{\bf p}}-V_{{\cal R}[{\bf p}]}]
[V_{-{\bf p}}+V_{-{\cal R}[{\bf p}]}]
\nonumber \\
&&
\ea
This non-linear susceptibility ($\chi_{\rm nem}$ contains a four-density
correlator) diverges at the nematic to isotropic  quantum critical point,
and we would generally expect it to be largest for ${\bf k}\approx{\bf
Q}_{\rm ch}$.  It is  important to note that the density
modulations, reflecting the proximity of stripe order, are a first
order effect, while the nematic response, which
differentiates ${\bf Q}_{\rm ch}$ from ${\cal R}[{\bf Q}_{\rm ch}]$, is
second order, and so will tend to be weaker, even if the
nematic quantum critical point is nearer at hand than that involving
the stripe-ordered state.

\subsection{Phase diagrams}

\begin{figure}[h!]
\begin{center}
\leavevmode
\vspace{.2cm}
\noindent
\hspace{-0.5 in}
\includegraphics[width=0.4\textwidth]{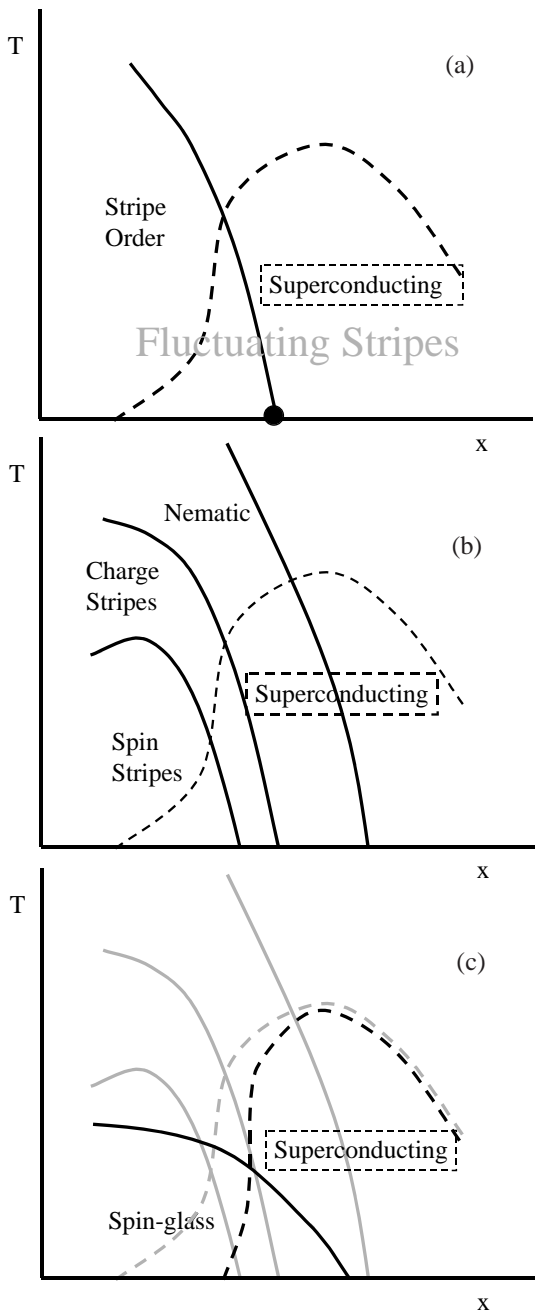}
\vspace{.2cm}
\end{center}
\caption
{Schematic phase diagrams showing various forms of stripe order, and its
interactions with high temperature superconductivity;  see text for
discussion.}
\label{phase}
\end{figure}

When there is more than one competing ordered phase, the phase diagram
can be very complicated, and moreover can be quite different
depending
on various microscopic details.  Nonetheless, it is often
useful to have a concrete realization of the phase diagram in mind,
especially when thinking about issues related to order, and order
parameter fluctuations.  In this subsection, we  sketch
schematic phase diagrams which characterize some of the ordered phases
and fluctuation effects discussed above;  the reader is cautioned,
however, that the shape, topology, and even the number of ordered phases
may vary
for material-specific reasons.

Figure~\ref{phase}(a) represents the essential features we have in mind,
in
the absence of any quenched disorder.  Here, there is a stripe ordered
phase which occurs at low doping and low temperature, with a phase
boundary which ends at a quantum critical point deep in the
superconducting phase.  In the region marked ``fluctuating stripes,''
there is significant local stripe order whose character is governed by the
proximity to the quantum critical point.  Notice that this fluctuation
region extends into the stripe ordered phase itself; although there is
true long-range stripe order in this region of the phase diagram,
close to the quantum critical point the
degree of striping at the local level is much greater than the small
ordered component.  There is no universal
statement possible concerning how far the fluctuating stripe region
extends beyond the ordered phase.  Clearly, if the region in which
there is {\it local} stripe order does not at least include the
entire region of
the phase diagram in which high temperature superconductivity occurs, it
cannot be essential for the mechanism of superconductivity.

Under many
circumstances, the stripe ordered phase will not end at a quantum
critical point, but rather will terminate at a first order
line \cite{kive01}.  If the transition is only weakly first order,
this will not affect the general considerations presented here.  If it
is strongly first order, fluctuation effects are much weaker, and the
effects of even very weak quenched disorder are much more severe than
in the case of a
continuous
transition.
This limit warrants further study \cite{card99}, both for application to
the cuprates and more generally; it is likely relevant for the organic
superconductors.\footnote{A first order transition of this sort has been
reported in the BEDT system
\protect\cite{lefe00} and in (TMTSF)$_2$PF$_6$ \protect\cite{brow02}.}

Figure~\ref{phase}(b) is a more ornate version
\footnote{In sketching the global shape of the
phase diagram, we have included the prejudice that static stripe order,
especially static spin-stripe order, competes strongly with
superconductivity, but that a degree of local stripe order is necessary
for high temperature  pairing-the latter prejudice is reflected in the
vanishing of the superconducting
$T_c$ as the local stripe order is suppressed upon overdoping.  However,
these prejudices affect only the interplay between stripe order and
superconducting order in the phase diagram, and not the central features
on which we focus in the present article, concerning the character of
fluctuating stripe order.}
of the phase diagram, in
which all the different broken-symmetry phases discussed in the present
paper are exhibited.  Here, we
illustrate the case in which the nematic, charge, and spin ordering occur at
distinct transitions, under which circumstances the nematic ordering
must precede the
charge ordering, which must in turn precede
the spin ordering \cite{zach98}.    In general, there are also
distinctions between commensurate and incommensurate, diagonal and
vertical, bond-centered and site centered stripe order, so
the same general considerations
could lead to significantly more complicated phase diagrams.

Since the superconducting order is
suppressed in the vortex state, and eliminated above $H_{c2}$, studies of
the magnetic-field-induced changes in the local stripe  order have recently
emerged as one of the  best ways of determining
the nature of the interplay between stripe and superconducting order.
There has been a flurry of recent papers, both experimental\footnote{Recent
experimental studies of induced stripe order in the vortex state of high
temperature superconductors have been carried out in
\protect\cite{lake01,lake02,khay02,hoff02,kata00}.} and
theoretical\footnote{Theoretical studies of the effect of magnetic 
fields on the
competition between stripe and superconducting order have recently been
extensively studied; see, for instance,
\protect\cite{deml01,zhan01,kive02} and references therein.  For an
earlier related discussion, see
\protect\cite{zhan97,arov97}.} on this subject.  Since this
subject is fairly involved, and is well reviewed in the literature, we
will not elaborate here on the phase diagram in a field, although we
will briefly mention some of the salient findings in Sec.
\ref{detection}.

Finally, Fig.~\ref{phase}(c) shows the effect of weak
disorder on the phase diagram illustrated in Fig.~\ref{phase}(b).
Phase transitions involving breaking of spatial symmetries are generally 
rounded by
quenched disorder. The resulting crossovers should have a glassy character 
{\it i.e.\/}, 
the apparent transition temperature is frequency-dependent.
Thus, the only true phase transitions in the presence of disorder are a
superconducting and a spin-glass transition, although the spin-glass
should have a local stripe character - it is \cite{emer93} a ``cluster
spin glass."  Note that quenched disorder generally introduces some
frustration \cite{zach00}, so where in the absence of quenched
disorder the spin-freezing temperature is large, quenched disorder tends
to suppress it.  Conversely, where quantum fluctuations
have destroyed the ordered state, the additional ``pinning'' effect of
quenched disorder can lead to a spin-glass phase extending beyond the
border that the magnetic phase would have in the zero disorder limit.

\section{One-dimensional Luttinger liquid}
\label{1DEG}

In this section, we  show how the general features described above
play out in the case of the 1DEG.  For
simplicity, we present results for a spin-rotation-invariant
Tomanaga-Luttinger liquid (TLL); in a forthcoming
publication \cite{bind02} we will provide details of the derivations, and
will also discuss the  Luther-Emery
liquid (LEL), {\it i.e.\/}, the
spin-gap case.  In this section we will describe the most salient
aspects of the ways fluctuating order manifests itself in these
critical systems. In Appendix \ref{sec:luttinger} we give more details of
this theory.  (We use units in which $\hbar=k_B=1.$)

Both the TLL and the LEL can be regarded as quantum
critical systems\footnote{For reviews of the the theory of the 1DEG, see
\protect\cite{emer79,frad91,stone94,gogo98}.}
\setcounter{savefoot}{\value{footnote}}
with dynamic critical exponent $z=1$. So long as
the charge Luttinger parameter is in the range $0< K_{\rm c} <1$ (which is
typically the case for repulsive interactions),  the  charge
susceptibility of the TLL  diverges as $k\to 2k_\F$ as
\begin{equation}
\chi(2k_\F+q)\sim |q|^{-g}.
\label{eq:static}
\end{equation}
where $k_\F$ is the Fermi wave vector and $g=1-K_{\rm c}$.
This system can rightly be viewed as a quantum critical CDW state.
(The CDW fluctuations are still stronger in the LEL case, where the
susceptibility exponent is replaced by $K_{\rm c}-2$.)
Thus, it is the perfect laboratory for testing the validity of the
general scaling considerations of Section
\ref{general}.

The charge-density structure factor of the TLL is the
Fourier transform of the more readily evaluated space
time structure factor
\begin{equation}
S( k, \omega)=\int_{-\infty}^{\infty} dt \int_{-\infty}^{\infty} dx\;
{\cal S}(x,t) e^{ikx-i\omega t}
\end{equation}
where the charge correlation function ${\cal S}(r,t)$ is given by
\begin{eqnarray}
{\cal S}(r,t) =&& {\cal S}_0(r,t)+[e^{i2k_\F r}{\cal S}_{2k_\F}(r,t)+
{\rm c.c.}]\nonumber \\
&&+[e^{i4k_\F r}
{\cal S}_{4k_\F}(r,t) + {\rm c.c.}] +\ldots
\end{eqnarray}
Explicit
expressions  for ${\cal S}_0$,
${\cal S}_{2k_\F}$, etc. are given in the
literature.$^{\arabic{savefoot}}$ These expressions can be
Fourier transformed (although in general, this must be done numerically)
to yield expressions for the dynamical structure factor.  Since the LL is
quantum critical, this (and other) correlation functions have a scaling
form; for example, near $2k_\F$ ({\it i.e.\/}, for small $q$)
\begin{equation}
S(2k_\F+q,\omega) = \frac 1 {v_c}\left(\frac D { v_{\rm c}q}\right)^{g}
\Phi_{2k_\F}\left(\frac{\omega}{v_cq},\frac {\omega}{T}\right)
\label{eq:S2kF}
\end{equation}
where
$D$ is an ultraviolet cutoff, {\it i.e.\/}, the
band width of the TLL. The scaling function
$\Phi_{2k_\F}$ of Eq. (\ref{eq:S2kF}) depends implicitly on the
dimensionless parameters
$K_c$ and $\frac{v_{\rm c}}{v_{\rm s}}$ with $v_{\rm c}$ and $v_{\rm s}$
respectively, the charge and spin velocities. (Spin rotation invariance
constrains the spin Luttinger exponent
$K_{\rm s}$ to be equal to 1.)   To exhibit the general features we are
interested in, while making the Fourier transform as simple as possible,
we can consider the limit $T=0$ and set $v_{\rm s}=v_{\rm c}=v$.
Then, for $q$ small
\begin{equation}
S(2k_\F+q,\omega) \sim
     \frac 1 v\left(\frac D { vq}\right)^{g}
\theta\left(\frac {\omega^2}{v^2q^2}-1\right)
\left[\frac {\omega^2}{v^2q^2}-1\right]^{-g/2}
\end{equation}
where $\theta(x)$ is the step function.

\begin{figure}[h!]
\begin{center}
\leavevmode
\vspace{.2cm}
\noindent
\includegraphics[width=0.48\textwidth]{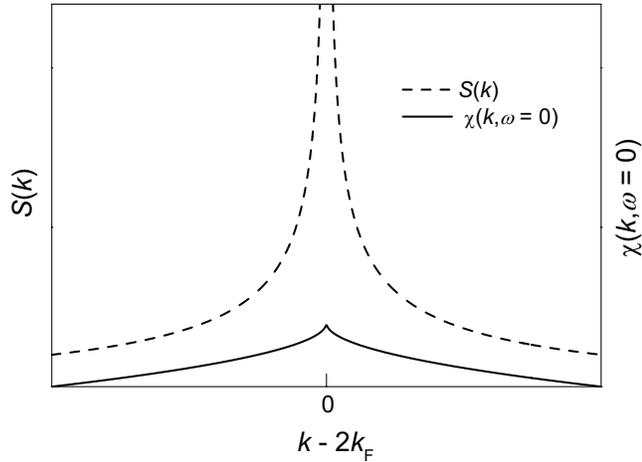}
\vspace{.2cm}
\end{center}
\caption
{The $T=0$ charge structure factor, $S(k)$ (solid line), and the static
susceptibility, $\chi(k,\omega=0)$ (dashed line), of
the TLL with $v_{\rm s}=v_{\rm c}$ and $K_{\rm c}=0.5$.}
\label{Sofq}
\end{figure}

For fixed $\omega$, as a function of $q$, the dynamic structure factor
exhibits a multi-particle continuum for $-\omega/v \le q \le
\omega/v$, but it does
have singular structure, which can
be thought as an image of the Goldstone (phason) modes one would find
were there true
CDW order:
\begin{equation}
S(2k_\F+q,\omega) \sim   [\ \omega- v|q|\ ]^{-g/2}
\ \ {\rm as}
\ |q|\to \omega/v
\end{equation}
The equal-time structure factor can also be readily computed by integrating the
dynamical structure factor;  as $q \to 0$,
\begin{equation}
S(2k_\F+q)\sim A -A'\left({|q| \alpha}/{2}\right)^{K_{\rm c}}
\end{equation}
where $A$ and $A'$ are numbers of order 1
and $\alpha$ is a short distance cutoff which we take to be
$\alpha\sim  v_F/D$.
As promised, this singularity is much weaker than that exhibited by
$\chi$, and indeed
the $2k_\F$ component of the
structure factor remains finite (but not differentiable)
as $q \to 0$. (See Fig. \ref{Sofq}.)

We now turn to the spatial structure in the
1DEG induced by a single impurity
at the origin.
For $K_{\rm c}$ in this range,
the $2k_\F$ component of an impurity potential, whose amplitude we
will denote by
$\Gamma$, is a {\sl relevant} perturbation \cite{kane92,kane94} with
(boundary) scaling
dimension  $d=\frac{1}{2}(K_{\rm c}+1)$.
Thus, there is a
crossover  {\sl energy scale} $T_K \propto \Gamma^{2/(1-K_{\rm c})}$ such
that excitations with energy
large compared to $T_K$ see a weak backscattering potential, and are
thus only weakly perturbed by the impurity. Conversely, for
energies low compared to
$T_K$ the system is  controlled \cite{kane92,kane94} by the fixed point
at $\Gamma \to \infty$.
However, at this fixed point, the high energy
cutoff is replaced by a renormalized cutoff, $D\to T_K$.

\begin{figure}[b]
\begin{center}
\vspace{.2cm}
\noindent
\includegraphics[width=0.5\textwidth]{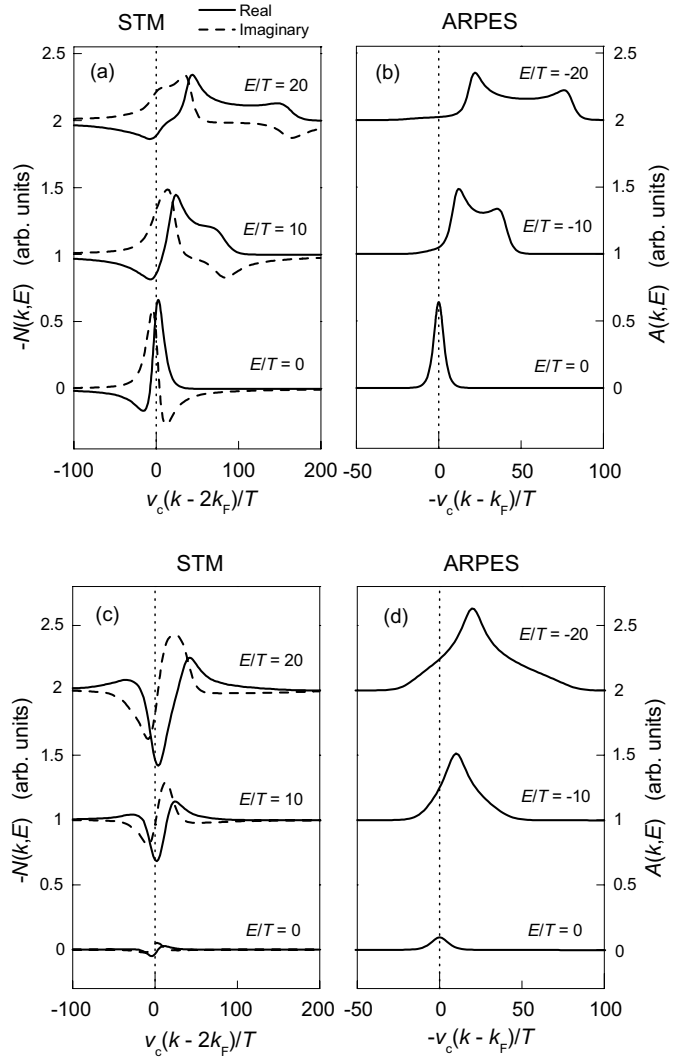}
\vspace{.2cm}
\end{center}
\caption
{Thermally scaled STM spectra $N(k,E)$ in the neighborhood of $k=2k_\F$ - (a)
and (c) - and ARPES spectra - (b) and (d) - near $k=k_\F$.  In all cases,
$v_{\rm c}/v_{\rm s}=4$ and $K_{\rm s}=1$;  for (a) and (b), $K_{\rm
c}=0.5$ ($b=1/16$) while for
(c) and (d) $K_{\rm c}=0.17$ ($b=1/2$).  In (a) and (c), the density
of states oscillations are
induced by an impurity scatterer at the origin with $T_K\gg E$.  In
the STM spectra, the real and
imaginary parts are represented by the solid and dashed lines,
respectively.  The curves for $E/T= 10$ and $E/T = 20$ are offset by 1 and 2 
respectively, in arbitrary
 units.}
\label{Nofk}
\end{figure}

We begin, therefore, by considering the limit $\Gamma = \infty$,
{\it i.e.\/}, a semi-infinite system with $x\geq 0$ and the
boundary  condition that no current can flow past $x=0$.
For finite $\Gamma$, this solution is applicable
 for all energies $|E |\ll T_K$. The Fourier
transform of the impurity induced LDOS
\ba
N( k, E)=&& N^>(k,E)+N^<(k,E) \\
N^<(k,E)=&& \frac{1}{2\pi}\int_{-\infty}^{\infty} dt \int_0^{\infty} dx
\; g^<(x,x;t)\;
e^{i(Et-kx)}
\nonumber
\ea
can be computed using exact expressions for the appropriate single
hole Green function \cite{egge00,egge96,matt97}:
\ba
&&g^<(x,x;t)=\sum_{\sigma}\langle \Psi^{\dagger}_{\sigma}(x,t)
\Psi_{\sigma}(x,0) \rangle\\
&&\equiv  g_0(x,t) + [e^{i2k_\F x}
g_{2k_\F}(x,t) + e^{-i2k_\F x} g_{-2k_\F}(x,t)] +\ldots
\nonumber
\ea
where $g_0(x,t)$ is the long wavelength part
and $g_{2k_\F}(x,t) $ is the $2k_\F$ part;  the single electron
piece can be computed similarly from $g^>$, or more simply from the relation between their respective spectral densities,
$N^<(k,E)=e^{-E/T}N^>(k,E)$.
We will be interested in the $2k_\F$ component
which clearly contains information  about CDW correlations in this
semi-infinite 1DEG, and for which a general
expression in space and time has been given by Eggert and
others \cite{egge00,egge96,matt97}.
As with the structure factor, the
$2k_\F$ part of $N(k,E)$ can also be expressed in terms of
a scaling function $\Phi$ as (see Appendix \ref{sec:luttinger})
\begin{eqnarray}
N(&&\!\!\!\!\!\!q+2k_\F,E)= \displaystyle{
\frac{ B}{2E}\left(\frac{\alpha E}{v_{\rm c}}\right)^{2b}}\!\! 
\Phi\left(\frac{2E}{ v_{\rm c}q},\frac {E}{T}\right)
\label{eq:R2kF}
\end{eqnarray}
where $b=(1-K_{\rm c})^2/8K_{\rm c}$  and $B$ is a dimensionless constant;
   we have left implicit the dependence on the dimensionless parameters
$K_c$ and
$v_{\rm c}/v_{\rm s}$.

The scaling form of $N(k,E)$, given in Eq. (\ref{eq:R2kF}), expresses the
fact that the Luttinger liquid is a quantum critical system. The spectrum
of the Luttinger liquid and the existence of a charge-ordered  state
induced by the impurity dictate entirely the structure of the scaling
function
$\Phi\left(x,y\right)$:
it has a multi-soliton continuum for both right and
left moving excitations, each
with leading thresholds associated with one-soliton states
carrying separate spin and charge quantum numbers and moving at their
respective velocities, as well as a non-propagating feature associated
with the
pinning of the CDW order by the impurity. Thus,
the STM spectrum
exhibits all the striking features of the Luttinger liquid:
spin-charge separation and quantum criticality, {\it i.e.\/},
fluctuating order. The beauty of the 1DEG is that much of this can be
worked out explicitly.

In Figs. \ref{Nofk}a and \ref{Nofk}c, we show the scaling function
computed numerically for various representative values of the
parameters.  The plots
show the real and imaginary parts of $N(2k_\F+q,E)$
at fixed
$E/T$ as a function of of the scaled momentum, $ v_cq/T$, for
$|q|\ll k_\F$.   For comparison, we also show \cite{orga01} in
Figs.~\ref{Nofk}b and \ref{Nofk}d, the single hole spectral function
$A(k_\F+q,\omega)$ that would be
measured in an angle resolved photoemission experiment (ARPES) on the
same system in the absence of an impurity.

\begin{figure}[h!]
\begin{center}
\leavevmode
\vspace{.2cm}
\noindent
\includegraphics[width=0.45\textwidth]{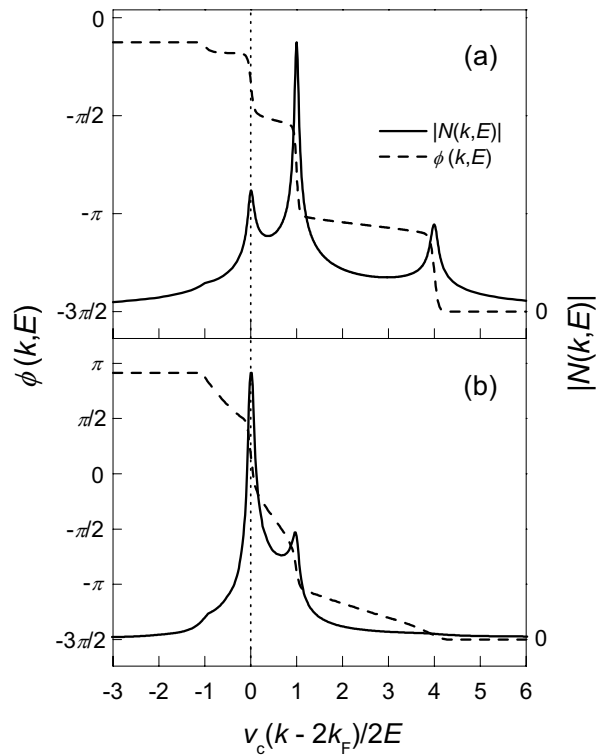}
\vspace{.2cm}
\end{center}
\caption
{Low-temperature ($E/T=100$) form of the STM spectra with $k$ near
$2k_\F$ for the
same parameters as in Fig. \ref{Nofk};  a) is for $K_{\rm c}=0.5$ and b)
$K_{\rm c}=0.17$.  Here, we have
expressed
$N(k,E)=|N(k,E)|\exp[i\phi(k,E)]$ where the amplitude is shown as a solid
line and the phase as a dashed line.}
\label{lowT}
\end{figure}

There are several things to
note about these plots\footnote{Details of the asymptotic analytic
behaviors of the singularities exhibited in these plots can be found
in Appendix \ref{sec:luttinger}.}: 
\begin{enumerate}
\item  
Because of the quantum
critical scaling
form, high energy and low temperature are equivalent.  Note, however,
that in interpreting the large $E/T$ spectra as representative of the
low-temperature behavior of the system, it is important to remember that
the thermal scaling of the {\it k}-axis of the figure hides the fact that
all features of the spectrum are becoming sharper as $T\to 0$;  this is
made apparent in Fig. \ref{lowT}. 
\item 
It is clear that
there are right-dispersing features of the scaling functions
characterized by the spin and charge velocities. The interference between
dispersing features of the ARPES spectrum near
$k_\F$ and $-k_\F$ can be loosely thought of as giving rise to the
dispersing features in the STM spectrum;  indeed,  as $K_{\rm c}\to 1$
(the non-interacting limit),
${\rm Re}\{N(2k_\F+q,E)\}\propto A(k_\F-q/2,-E)$ at fixed $E>0$.  However, 
it is also
clear from the figure that the stronger the interactions, the less direct is
the resemblance between $N(2k_\F+q,E)$ and $A(k_\F-q/2,-E)$. 
\item
There is also a very weak feature in the spectrum, visible only at quite
large
$E/T$ (compare Figs. \ref{Nofk} and \ref{lowT}), which disperses in the
opposite direction (left) to the main features of the spectra with
velocity $-v_c$.  Because these spectra are shown only for $k$'s in the
neighborhood of
$+2k_\F$ (or $+k_\F$ for $A(k,E)$), there is no symmetry between right and
left moving excitations.  If we showed the spectra on a larger scale,
there would of course be the mirror symmetric spectra near $-2k_\F$ (or
$-k_\F$) as required by Kramer's theorem. 
\item
 In the STM
spectrum, but not in the ARPES spectrum, there is a feature near
$q=0$ which does not disperse with increasing energy;  this is
directly related to the pinned CDW order.  Note that for $K_{\rm c}=0.5$,
this feature is weak in Fig.~\ref{Nofk}, and only becomes prominent
at very large $E/T$, as shown in Fig. \ref{lowT}; but this is the
most important feature of the
data if one is interested in evidence of pinned CDW order.
\end{enumerate}

{\bf Lesson \#3:}  From this explicit example we learn
that dispersing features in an STM measurement that resemble the
interference effects that arise from non-interacting
quasiparticles do not necessarily imply the existence of well defined
quasiparticles!

\begin{figure}[h!]
\begin{center}
\leavevmode
\vspace{.2cm}
\noindent
\includegraphics[width=0.45\textwidth]{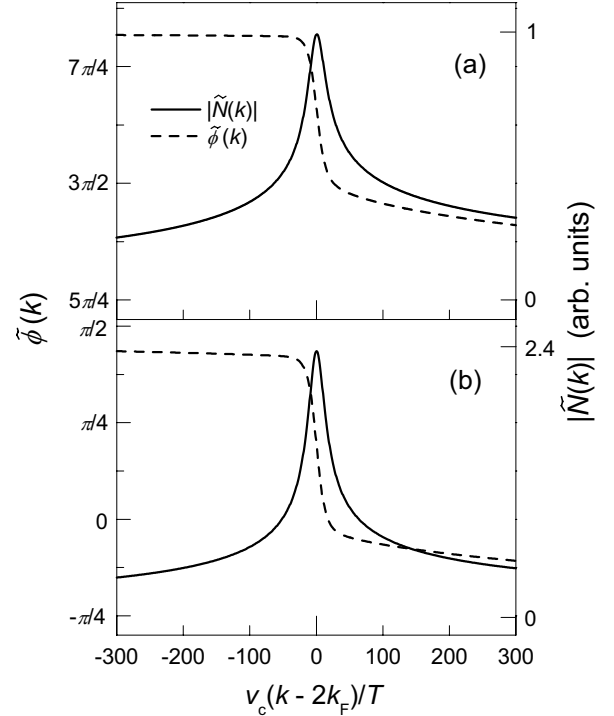}
\vspace{.2cm}
\end{center}
\caption
{
Thermally scaled integrated STM spectra $\tilde N(k)\equiv 
|\tilde N(k)|e^{i\tilde \phi(k)}\sim
\tilde N(k,T_K)$ in the neighborhood of
$k=2k_\F$ for
$v_{\rm c}/v_{\rm s}=4$ and $K_{\rm s}=1$ with a)  $K_{\rm c}=0.5$
and  b) $K_{\rm c}=0.17$.  The
solid line is the amplitude and the dashed line is the phase.$^{26}$}
\label{tildeN}
\end{figure}

It is interesting to compare $N(k,E)$
with the
LDOS
averaged over some energy scale,
\begin{equation}
\tilde N(2k_\F+q,E)\equiv E^{-1} \int_{-E}^0 d\epsilon N(2k_\F+q,\epsilon).
\end{equation}
Note that at $T=0$, $E\tilde N(k,E)\to
\langle\rho_k\rangle$ as $E\to \infty$.
However, the expression we have used for $N(k,E)$ was derived
for an infinite strength scattering potential, and so is only valid
up to energies of the order of $T_K \ll D$.
This integrated quantity is shown in Fig.
\ref{tildeN} for
representative
parameters\footnote{For technical reasons, what is actually plotted in
Fig.\
\ref{tildeN} is
$\tilde N(k) \equiv \lim_{T_K\to \infty} \int_{-\infty}^{\infty}
dE [1-f(E)] N(k,E)$, where $f$ is the Fermi function,
but the distinction between this and $\tilde N(k,T_K)$
is not important here.}
at $T=0$, and in the limit as $q \to 0$,
$\tilde N(2k_\F+q,T_K)$ has the asymptotic behavior
\begin{equation}
\tilde N(2k_\F+q,T_K) \propto
\left(\frac {1} { T_K}\right)
\left(\frac{\alpha T_K}{v_{\rm c}}\right)^{2b}
\left(\frac{T_K}{v_{\rm c} q}\right)^{g/2}.
\label{eq:integrated}
\end{equation}
Both
the integrated induced tunneling
density of states and the induced tunneling density of states at
fixed voltage $E$ have a singular behavior of the form
$q^{-g/2}$ (See Appendix \ref{sec:luttinger}).
However, the big difference is that $N(k,E)$
has dispersing singularities
in addition to the
non-dispersing singularity at
$k\to 2k_\F$,
while the singularity at $k\to 2k_\F$ is the only singular feature of
$\tilde N(k,T_K)$. Thus,  $\tilde N(k,{T_K})$ is more
easily analyzed for evidence of an almost ordered CDW state.

At energies $E > T_K$, the response of the system to the presence of the
impurity is weak ({\it i.e.\/}, proportional to $\Gamma$),
and can be computed in
perturbation theory.  For instance, in the weak-impurity limit
$T_K\to 0$ the integrated response of the system over all
energies is dominated by the high energy, perturbative regime where at
$T=0$ we find, in agreement with the sum rule of Eq. (\ref{sum-rule}), 
(see also Appendix \ref{sec:luttinger})
\begin{equation}
\tilde N(2k_\F+q,D)=  
\chi_{\rm ch}(2k_\F+q)\;
\Gamma+ O(\Gamma^2)
\label{eq:int-high}
\end{equation}
By comparing the weak $\Gamma$ behavior of $\tilde N(2k_\F+q,D)$ 
(Eq.\  (\ref{eq:int-high})) with its behavior in the 
opposite $\Gamma \to \infty$ limit, Eq.\  (\ref{eq:integrated}), we see that 
since  $\chi_{\rm ch} \sim |q|^{K_c-1}$ 
(as $q \to 0$), the integrated LDOS for $\Gamma$ small is much more singular 
as $q \to 0$  than for large 
$\Gamma$. This is a necessary consequence of the fact that the backscattering 
impurity potential is a 
relevant perturbation ~\footnote{The meaning of $\tilde N(2k_\F+q)$, 
the Fourier transform LDOS integrated up to some high 
energy scale, depends on this scale. 
For $\Gamma \to 0$ the high-energy cutoff is the band width $D$,  
and  $\tilde N(2k_\F+q)$ 
is the Fourier  transform of the induced charge density. Similarly, 
for $\Gamma \to \infty$,  where the high-energy cutoff $T_K$ 
is large, $\tilde N(2k_\F+q)$ is the Fourier  transform of the charge 
density of the semi-infinite system. 
However, for intermediate values of the backscattering amplitude $\Gamma$ 
the partially integrated LDOS $\tilde N(2k_\F+q)$ cannot be interpreted as 
a modulation of the charge density.} . 

It is also worth noting that the impurity induced 
LDOS at a fixed finite distance $x$ from the  impurity
\begin{equation}
\delta {\cal N} (x,E) \sim E^{(1-K_{\rm c})/{2K_{\rm c}}}.
\end{equation}
is always large at
low energies $|E|\ll v/x$ compared to the background DOS
of the clean TLL
\begin{equation}
{\cal N}(E) \sim E^{\frac{(1-K_{\rm c})^2}{4K_{\rm c}}}.
\end{equation}
This is yet another illustration of the way impurities enhance the low
energy effects of fluctuating order.

\section{Two-dimensional Fermi liquid}
\label{2DEG}

In this section, we consider the application of these ideas to the
case of weakly-interacting electrons in 2D.
To obtain explicit results,
we will consider electrons with a quadratic dispersion
$\epsilon_{\bf k}=\hbar^2{\bf k}^2/2m$.

Although, as seen in Eq. \ref{eq:chi-dos}, $\chi_{DOS}$ is a two particle
correlator, for non-interacting quasiparticles, it can be expressed as a
convolution~\cite{polk02b,byer93} of two single-particle Green functions:
\begin{equation}
\chi_{DOS}({\bf k},E) =- \frac{1}{\pi}
{\textrm Im}
\left\{
\int 
\frac
{d {\bf q}}
{ (2\pi)^d}
G({\bf k}+{\bf q},E) G({\bf q},E)
\right\}
\label{convolution}
\end{equation}
Therefore, in weakly-interacting systems, it can be analyzed to
obtain information about the
single-particle spectrum.
Explicitly, it is easy to see that for free electrons in 2D
$\chi_{DOS}({\bf k},E)$ is given by
\begin{eqnarray}
\chi_0({\bf k},E)
&=&
\frac{m}{\pi\hbar^2}
\frac{\theta(\epsilon_{{\bf k}}-4E)}{
\sqrt{\epsilon_{{\bf k}}\left(\epsilon_{{\bf k}}-4E\right)}}
\label{chi2DEG}
\end{eqnarray}
Here, the
subscript $0$ is introduced for later convenience to signify $\chi$
in the non-interacting
limit, and once again $\theta(x)$ is the Heaviside (step) function. For fixed
$E$ as a function of
${\bf k}$, this quantity diverges along curves in ${\bf k}$ space where
$\epsilon_{\bf k}=4E$ ($|{\bf k}|=2k_\F$).
Note that the fact that $\chi_0$ vanishes for $\epsilon_{\bf k}<4E$
is a peculiarity of the 2D case with infinite quasiparticle lifetime.
In 3D,
\begin{equation}
\chi_{0}({\bf k},E)=\frac{1}{8\pi^2}\left(\frac{2m}{\hbar^2}\right)^{3/2}
\frac{1}{\sqrt{\epsilon_{\bf k}}} \ln \left|\frac{2\sqrt{E}+\sqrt{\epsilon_{\bf
k}}}{2\sqrt{E}-\sqrt{\epsilon_{\bf k}}}\right|
\label{chi3DEG}
\end{equation}
which also has a singularity as $\epsilon_{\bf k} \to 4E$,
but
is non-zero (and positive) for all
$E>0$. In contrast, as shown in Section \ref{1DEG},  in a 1D
Luttinger liquid
the induced density of
states has a phase jump as $k$ crosses any of the propagating or
non-propagating singularities. It is straightforward to see that both in
2D and 3D a finite lifetime of the quasiparticles leads to a
rounding of the singularities and that in 2D it also leads to a
positive induced density of states for all $E>0$. It is also
simple to verify that  to non-linear order in the external
potential,  the induced density of states in 2D may or may  not
have a phase jump across the singularity depending on the details of the
perturbing potential.  For instance, the induced LDOS produced by a
single impurity can be computed (from the impurity $t$-matrix);  it
exhibits sign  reversal  ($\pi$ phase shift)
across the singularity at $\epsilon_{\bf k}=4E$ for a repulsive
potential, but not for an attractive one.
In contrast, as we will show below,
that
any ${\bf k}$ space structure that arises from proximity
to a quantum critical point is derived from the susceptibility, $\chi({\bf k})$, 
which is real and positive, so it always produces a signal
whose  phase is constant.

\noindent{\bf Lesson \# 4:}  The density of states modulations
induced by weak disorder in a
non-interacting metal are quite different in character from those
expected from the proximity to a
CDW quantum critical point, both in that they disperse strongly as a
function of energy, and the
peak intensities lie along curves (surfaces in 3D) in ${\bf k}$ space,
as opposed to the
structure associated with isolated points in $\bf k$ space expected
from near critical
fluctuations.
\footnote{In practice, the distinction between an $N(\bf k,E)$ which is peaked 
along curves, 
indicative of quasiparticle interference effects, and an $N(\bf k,E)$
peaked at isolated ``ordering'' vectors, ${\bf k}={\bf Q}$, may not
always be straightforward to establish in experiment. Consider the case in
which there is an anisotropy of strength $\alpha$ in the effective mass of the
2DEG, {\it i.e.\/}, $\epsilon_{\bf k}=\hbar^2 (k_x^2+\alpha k_y^2)/2m$.
In the limit of large anisotropy,
$\alpha\gg 1$, when the effects of the finite $\bf k$ resolution of
actual experiments are taken into account, apparent peak-like
structures can emerge.  If we represent the effect of finite
resolution by integrating the
expression in Eq.~\ref{chi2DEG} over a range of momenta
around different points along the ellipse
$\epsilon_{\bf k}=4E$, the integrated expression is peaked near
${\bf k}=\pm 2\sqrt{2mE}\ \hat e_x$, where it is a factor of
$\sqrt{\alpha}$ larger than where it is minimal near
$\sqrt{\alpha}\ {\bf k}=\pm 2\sqrt{2mE}\ \hat e_y$.}

The charge susceptibility itself, $\chi_0({\bf k})$, is well known from many
studies of the 2DEG.  It has an extremely weak non-analyticity,
whenever
$|{\bf{q}}|=2k_\F$
\ba
\chi_0({\bf q})&=&\frac{m}{2\pi \hbar^2}\
\left(1-\theta(q-2 k_\F)\frac{\sqrt{q^2-{4k_\F^2}}} q\right)
\ea
The inverse Fourier transform of this meager non-analyticity is what
gives rise to the famous Friedel
oscillations  in the neighborhood of an isolated impurity,
\begin{equation}
\langle \rho(r)\rangle\sim \frac{1}{r^2}\cos(2k_\F r).
\end{equation}
What this means
is that, for all intents and purpose, the Friedel oscillations are all
but invisible in the
Fourier transform of any conceivable STM experiment on a simple metal
in $D>1$. Some
non-trivial method of data analysis is necessary instead
\cite{brin98,spru97}.

It is worthwhile considering the effects of interactions on this
picture.  For weak enough
interactions, the effects  can be treated in a Hartree-Fock
approximation.  Thus, if we absorb any
interaction-induced changes in the band parameters into a
renormalized band structure, the only
change in the above analysis is that the external perturbation,
$V_{\bf k}$ in Eqs.  (\ref{eq:nQ}) and  (\ref{linear})
must be replaced by an effective potential, $V_{\bf k}\to V_{\bf k}
+ U_{\bf k}\rho_{\bf k}$,
where $U$ (which can be weakly ${\bf k}$ dependent) is the strength of
the electron-electron
repulsion.  This leads to the usual RPA expression for the
susceptibilities of the interacting
system, and to
\begin{eqnarray}
\chi_{\rm ch}({\bf k})=&&[1-U_{{\bf k}}\chi_0({\bf k})]^{-1}\chi_0({\bf k})
\nonumber \\
\chi_{\rm DOS}({\bf k},E)=&&[1-U_{{\bf k}}\chi_0({\bf k})]^{-1}
\chi_0({\bf k},E).
\label{RPA}
\end{eqnarray}
(It is the fact that $E$ is a probe energy, not a frequency, which is
responsible for the fact that it is simply $\chi_0({\bf k})$, rather
than a
frequency dependent factor, which appears in the expression for
$\chi_{\rm DOS}({\bf k},E)$.) Not surprisingly, since
$\chi_0(\bf k)$ is finite for all ${\bf k}$, for small
$U$ there is little qualitative difference between $\chi_0({\bf k})$
and $\chi({\bf k})$.  However, if we imagine, as is
often done (although we are not aware of any reason it is justified) that
this RPA expression applies
qualitatively for larger magnitudes of $U$, then as a function of
increasing magnitude of $U$ we
would eventually satisfy a Stoner criterion for a CDW
$U_{\bf k}\chi_0({\bf k})=1$.  Here, due to the relatively weak ${\bf
k}$ dependence
of $\chi_0({\bf k})$, the CDW ordering vector ${\bf Q}_{\rm ch}$ is determined
as much by
the ${\bf k}$ dependence of $U_{\bf k}$
as by effects intrinsic to the 2DEG.

\begin{figure}[bht]
\begin{center}
\leavevmode
\vspace{.2cm}
\noindent
\includegraphics[width=0.45\textwidth]{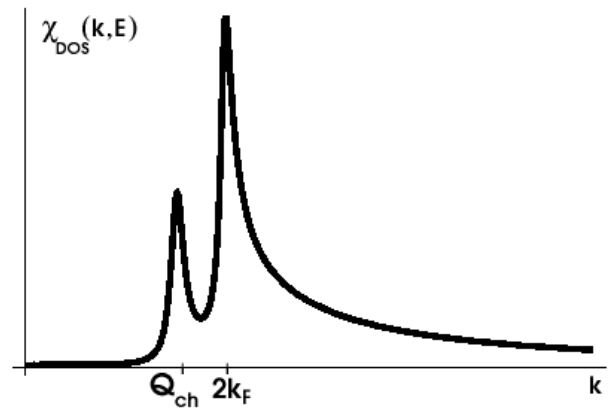}
\vspace{.2cm}\end{center}
\caption
{The RPA expression for $\chi_{\rm DOS}({\bf k},E)$ is plotted as a
function of
momentum at a fixed, low energy $E-\mu=.2 \mu$.
A small decay rate, $\Gamma=.025 \mu$,
is included to round the singularity and produce a finite $\chi_0(k,E)$ at
all k. We have taken the 
phenomenological form for $\chi_{ch}({\bf k})$ discussed in the text. 
The figure was plotted with  $\xi=5{k_\F}^{-1}$ and  $Q_{ch}=1.8 k_\F$.
}
\label{2DEGfig}
\end{figure}

Following this line of analysis, let us
consider the behavior of
$\chi_{\rm DOS}({\bf k},E)$ in the quantum disordered phase close to such
a  putative Stoner instability, where $\chi_{\rm ch}({\bf k})$ is 
highly peaked at ${\bf k}={\bf
Q}_{\rm ch}$.   For
${\bf k}$ far from the ordering vector, both the ${\bf k}$ and $E$
dependence of $\chi_{\rm DOS}$ are determined largely by $\chi_0({\bf k},E)$.
However,
at fixed voltage, $\chi_{\rm DOS}({\bf k},E)$ will also exhibit a peak
at a  ${\bf k}\approx
{\bf Q}_{\rm ch}$, with a  voltage dependent intensity proportional
to
$\chi_0({\bf Q}_{\rm ch},E)$.
Note, particularly, that as long as there is a finite correlation length
associated with the incipient order, {\it i.e.\/}, so long as the peak in
$\chi_{\rm ch}({\bf k})$ has a finite width, the corresponding peak in 
$\chi_{\rm DOS}({\bf k},E)$ will not generally occur precisely at
${\bf Q}_{\rm ch}$.  To illustrate this point, here and in Fig. 
\ref{2DEGfig} we adopt as a simple
phenomenological model,
$\chi_{\rm ch}({\bf k}) = A (\xi k_\F)^{2} 
\exp[-\frac{\xi^2}{2} (|{\bf k}|-Q_{ch})^2]$, 
where $\xi$ is the stripe correlation
length. From Eq.\ (\ref{RPA}), it follows that in addition to the singularity
inherited from $\chi_0({\bf k},E)$, $\chi_{\rm DOS}({\bf K},E)$ has a peak at a
momentum ${\bf k}$  which satisfies
\ba
{\bf k}={\bf Q}_{\rm 
ch}-{\xi^{-2}}\left[\frac{\epsilon_{\bf k}-2E}{\epsilon_{\bf k}-4E}\right]
{\bf\nabla}_{\bf k}\log[\epsilon_{{\bf k}}].
\label{Kstar}
\ea
In short, the peak associated with incipient order is weakly 
dispersing (especially at energies far from
any quasiparticle resonance condition) but so long
as $\xi$ is finite, the peak is never ``non-dispersing."  (In 
contrast, the 1D example discussed
above is quantum critical, so  $\xi$ is infinite and the feature 
associated with fluctuating order is
strictly non-dispersive.)


Although the calculations are somewhat more involved (and therefore must
be implemented numerically), the same sort of weak coupling analysis can
be carried out in the superconducting state.  Oscillations induced by
a  mean-field with period 4, representing stripes with various
internal structures in a d-wave superconductor, were carried out by
\textcite{podo02}.  As expected, the stripes
induce oscillations in the LDOS with the period 4, but with
energy dependences which reflect both the quasiparticle dispersion,
and the specific stripe structure assumed.  Impurity induced
oscillations in the LDOS in a d-wave superconductor were computed by
\textcite{byer93} and \textcite{wang02}, and the
effects of proximity to a stripe-ordered state, at RPA level, were
investigated by \textcite{polk02}.  In all cases,
interference and
stripe-related effects interact in fairly complex
ways that require detailed analysis to disentangle.

\noindent{\bf Lesson \#5:}
Depending on
what regions of ${\bf k}$ space are
probed, the STM spectrum can either be dominated largely by
band structure effects, or by the
incipient CDW order.  The most singular enhancement of the
STM signal is expected to occur
at energies such that a dispersing feature reflecting the underlying
band structure passes through
the ordering wave-vector.

\section{Regarding experiments in the cuprates}
\label{detection}

Even where  broken symmetry associated with stripe order
has ultimately been proven to exist, establishing this fact has often
turned  out to be difficult for a number of practical reasons.  In
addition, since quenched disorder is always a relevant perturbation (in the
renormalization group sense), macroscopic manifestations of broken
spatial symmetries are sharply defined only in the zero disorder limit.
Nevertheless,  the existence of some form of order which coexists with
superconductivity has implications for  the phase diagram  which can, in
principle and sometimes in practice, be tested by macroscopic
measurements.  A particularly revealing set of phenomena occur when the
strength of the superconducting order is modulated by the application of
an external magnetic field.\footnote{For experimental studies, see
\protect\cite{ono00,tyle98,ando97,ando02c,kata00,lake02,lake01,khay02,hoff02,
liu03,haw03};
for theoretical analysis, see
\cite{zhan97,arov97,deml01,zhan01,polk02,kive02,chak01,ivan00}.} Moreover,
as discussed in Sec.~\ref{nematic}, from measurements of macroscopic
transport anisotropies, electronic nematic order ({\it e.g.\/}, point-group
symmetry breaking) has been identified  beyond all reasonable doubt in
quantum Hall systems \cite{lill99,du99}, and very compelling evidence for
its  existence has been reported in the last year in underdoped {\LSCO}
and {\YBCO} \cite{ando02,ando02c}.

However, most searches for stripe order rely on more microscopic
measurements, especially elastic neutron scattering. One aspect of this
that has caused considerable confusion is that, unless an external
perturbation (such as weak crystalline orthorhombicity) aligns the
stripes in one direction, one generally finds equal numbers of $y$
directed and $x$ directed domains, leading to quartets of apparently
equivalent Bragg peaks, rather than the expected pairs. Fortunately,
where sufficiently long-ranged order exists, it is possible, by carefully
analyzing the scattering data, to distinguish this situation from a
situation in which the peaks arise from a more symmetric ``checkerboard"
pattern of translation-symmetry breaking.   From elastic neutron
scattering ({\it i.e.\/}, measurements of both the magnetic and the
nuclear $S({\bf k},\omega=0)$), and to a lesser extent from X-ray
scattering, it has been possible to establish the existence of
stripe-ordered phases in a wide range of members of the lanthanum cuprate
family of high-temperature superconductor:  {\LNSCO} over the whole
range of doped hole concentrations~
\footnote{\protect{\textcite{tran97a,ichi00,waki01}}}
 from $0.05<x\le 0.2$, {\LBCO}
\cite{fuji02}, {\LSCO} for 
~\footnote{\protect{\textcite{suzu98,kimu99,nied98,waki99,mats00,waki01}}} 
$0.02<x<0.13$, and
{\LCOplus} \cite{lee99} (including optimally doped material
with a doped hole concentration $\approx0.15$
and a superconducting $T_c$ of 42~K).   Exciting preliminary evidence of
charge-stripe order has been recently reported, as well, from elastic
neutron scattering studies on underdoped {\YBCO} (with $y=0.35$ and
$T_c=39$~K, \cite{mook02}) and optimally doped {\YBCO}
(with $y=0.93$ and $T_c=93$~K, \cite{mook02b}).  And, it is
worth mentioning, stripe order has also been similarly detected in a
number of non-superconducting doped antiferromagnets, including
~\footnote{\protect{\textcite{tran98b,lee97,yosh00,kaji02}}}
 {\LSNiO} and the colossal
magneto-resistance manganites
~\footnote{\protect{\textcite{rada99,wang00,mori98a,mori98b}}}.

Conversely, it is important to note that so far no evidence of stripe order, 
or incipient stripe order, 
has been found in any of the electron-doped cuprate superconductors. 
Indeed, all the low-energy magnetism that has been reported to date 
\cite{yama03} 
is peaked at the commensurate ordering wavevector, ${\bf Q}_{AF}$, 
rather than at an incommensurate wavevector. This is rather strong evidence 
that at least 
spin-stripe order is absent in these materials. This does not rule out a 
possible role for charge 
inhomogeneity; perhaps the electron-doped materials are more prone to form 
``bubble phases" {\it i.e.\/}, crystalline phases with more than one doped 
electron per unit cell~\cite{seul95,fogl96}.

We now turn to the core problem:  Given a system which in the absence of
quenched disorder or explicit symmetry breaking terms is in an isotropic
fluid state, how is the existence of substantial local stripe order
identified?

\subsection{Diffraction from stripes}
\label{diffraction}

Peaks in $S(\bf k,\omega)$ at the characteristic stripe ordering vectors
indicate a degree of local stripe order.  The $k$ width of these peaks can be
interpreted as an indication of the spatial extent of local stripe order, and
the low frequency cutoff as an indication of the typical stripe fluctuation
frequency.  So long as there is no spontaneous symmetry breaking,
$S(\bf k,\omega)$  necessarily respects all the point-group symmetries of the
crystal, and thus will necessarily always show peaks at quartets of $\bf k$
values, never the pairs of $\bf k$ values of a single-domain stripe ordered
state.

\begin{figure}[t]
\centerline{\includegraphics[width=3.0in]{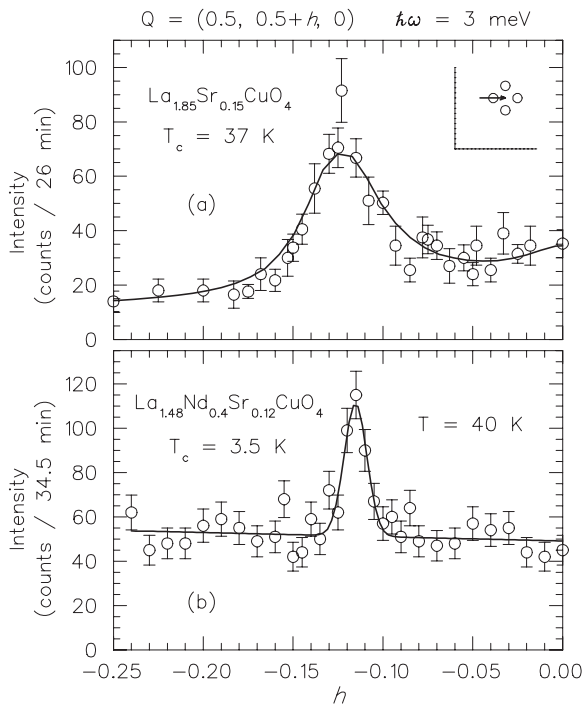}}
\caption{Comparison of constant-energy scans at $\hbar\omega = 3$~meV through
an incommensurate magnetic peak (along path shown in inset) for (a)
La$_{1.85}$Sr$_{0.15}$CuO$_4$ and (b) La$_{1.48}$Nd$_{0.4}$Sr$_{0.12}$CuO$_4$.
Both scans are at $T=40\ \mbox{K}>T_c$.  Measurement conditions are described
in \protect\cite{tran99a}.}
\label{fg:lnsco}
\end{figure}

Low frequency spin fluctuations with  relatively sharp peaks at
incommensurate wave vectors were detected many years ago in inelastic
neutron scattering studies \cite{thur89,cheo91,thur92} of
``optimally-doped'' ($x\approx0.15$, $T_c\sim 38$~K) {\LSCO}.  However,
not until the discovery \cite{tran95a} of ``honest'' stripe-ordered phases
in the closely related compound  {\LNSCO} was the
interpretation of these peaks as being due to stripe fluctuations made
unambiguously clear.  For instance, as shown in Fig.~\ref{fg:lnsco},
the magnetic structure factor at low temperature and small but finite
frequency, $\hbar\omega=3$~meV, looks very similar (in absolute magnitude
and width)  in both \LNSCO\ with $x=0.12$, where elastic scattering
indicating statistically ordered stripes has been detected \cite{tran95a}, and
in {\LSCO} with $x=0.15$, where no such static order is
discernible \cite{yama95a}.   As discussed in the previous section, this
is precisely the expected$^{\arabic{savefootb}}$ behavior near a quantum
critical point, where presumably the partial substitution of La by Nd has
moved the system from slightly on the quantum disordered to slightly on
the ordered side of a stripe ordering quantum critical
point.\footnote{This interpretation is further supported by the finding
that the magnetic $S({\bf k},\omega)$ in near optimally doped {\LSCO} has
scaling properties consistent with its being dominated by a nearby stripe
ordering quantum critical point; see \cite{aepp97}.}
We can now confidently  characterize {\LSCO} over an extremely broad
range of doping as either being in a stripe-ordered, or a nearly ordered
stripe-liquid phase.

\begin{figure}[t]
\centerline{\includegraphics[width=3.4in]{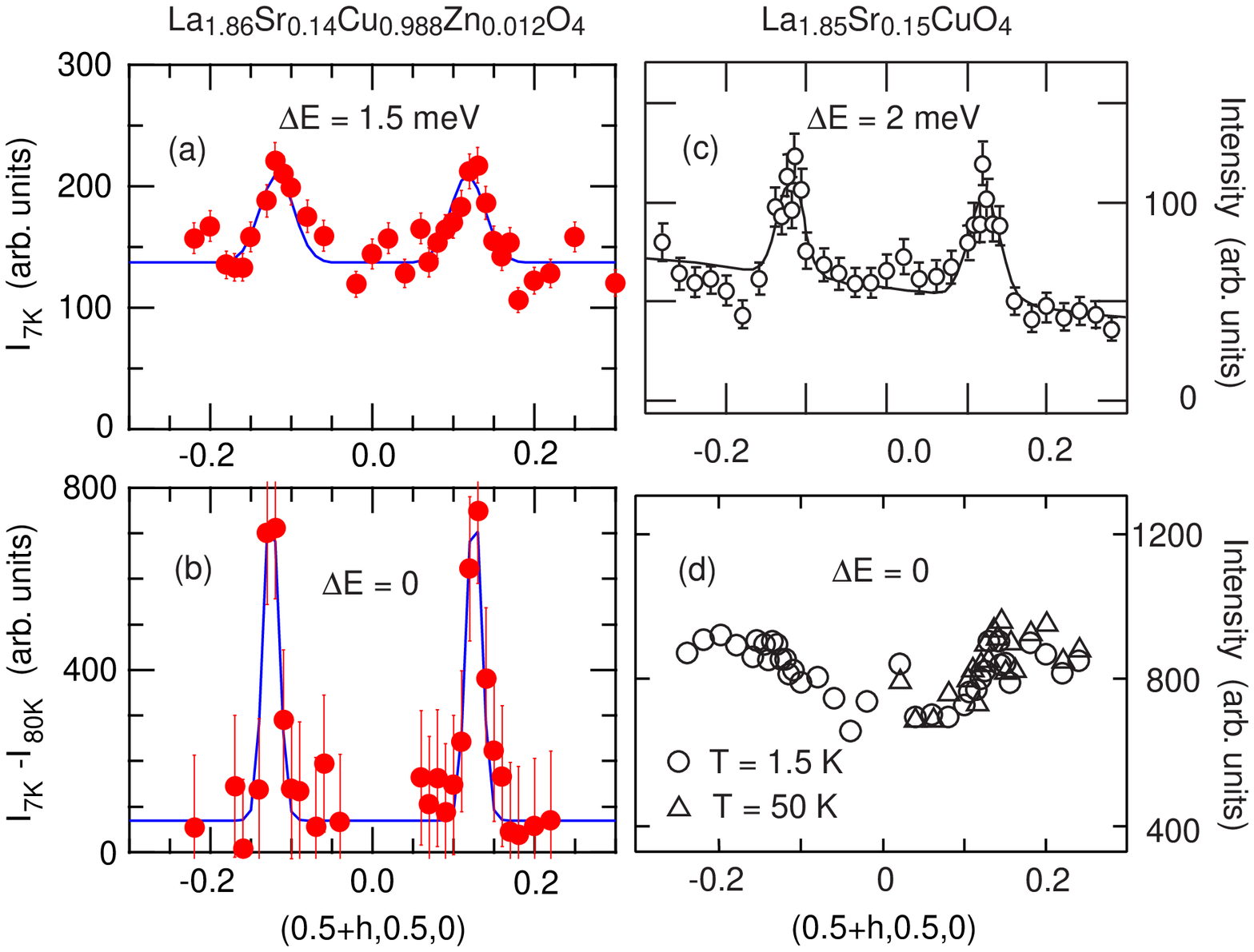}}
\caption{Comparison of magnetic scattering measurements in \LSCO\ with and
without Zn; all scans are along $Q=(\frac12+h,\frac12,0)$, measured in
reciprocal lattice units.  (a) Scan at $ E=1.5$~meV and $T=7$~K, and (b)
difference between elastic scans measure at 7~K and 80~K, both for
La$_{1.86}$Sr$_{0.14}$Cu$_{0.988}$Zn$_{0.012}$O$_4$
($T_c=19$~K) \protect\cite{hiro00}.  (c) Scan at $ E=2$~meV and $T=38$~K
\cite{yama98a}, and (d) elastic scans at $T=1.5$~K
(circles) and 50~K (triangles) \protect\cite{kimu99}, for
La$_{1.85}$Sr$_{0.15}$CuO$_4$ ($T_c=38$~K).  Measurement conditions were
different for each panel; see references for details.}
\label{fg:lsczo}
\end{figure}

An important test of this idea comes from studies of the changes in
$S(\bf k,\omega)$ produced by weak disorder.  Specifically, in
Fig.~\ref{fg:lsczo}, we compare the low frequency \cite{mats93} and
elastic \cite{hiro98} pieces of the magnetic structure factor of {\LSCO}
in the presence and absence of a small concentration (1.2\%) of Zn
impurities.  (Zn substitutes for Cu.)  As one might have expected, the Zn
slightly broadens the
$\bf k$ space structure, although not enormously \cite{kimu99,tran99b}.
Most dramatically, the Zn ``pins'' the stripe fluctuations, in the sense
that what appears only as finite frequency fluctuation effects in the 
Zn-free material, 
is pushed to lower frequencies and even to $\omega=0$ by
the quenched disorder.\footnote{It is certain that one effect of the Zn
impurities is to pin the charge stripe fluctuations.  It may also be that
the missing Cu spin plays an important role in slowing the
spin-fluctuations in the neighborhood of the Zn; the potential
importance of this form of coupling,  based on the behavior of a
Kondo impurity in a system close to a magnetic quantum critical point, has
been stressed in \cite{sach99c}.}

The issue still remains actively debated whether or not various fluctuation
effects seen in neutron scattering studies of the other widely
studied families of high
temperature superconductors, especially {\YBCO} and {\BSCCO}, can be
associated with stripe
fluctuations.  We will not review this debate here, but will touch on
it again in Secs.~\ref{STM} and \ref{nematic}, below.

One question arises \cite{chen02} concerning how we can
distinguish a fluctuating stripe phase from  a fluctuating checkerboard
phase (which breaks translation symmetry but preserves the symmetry under
exchange of $x$ and $y$). Of course, the strongest indication that
stripes, rather than checkerboards are responsible for the observed
fluctuations comes from the presence of ``nearby" stripe-ordered phases,
and the absence of any clear evidence of actual ordered checkerboard
phases in any doped antiferromagnet, to date.  However, in theoretical
studies\cite{low94,chen02,fogl96,moes96} both types of order
appear to be close  to each
other in energy, so this argument should not be given undue weight.

Where both spin and charge peaks are observed, it turns out that there
is a straightforward way to distinguish.  From Landau theory, it follows
\cite{zach98} that there is a preferred relation between the spin and
charge-ordering wave-vectors,
${\bf Q}_{\rm s}+{\bf Q}_{\rm s}^{\prime}={\bf Q}_{\rm ch}$.  For stripe
order, this means that the spin and charge wave-vectors are parallel to
each other, and related (up to a reciprocal lattice vector) by the
relation
$2{\bf Q}_{\rm s}={\bf Q}_{\rm ch}$. However, in the case of checkerboard
order, the dominant spin-ordering wave-vector is {\em not} parallel to
the charge-ordering wave vectors;  if
${\bf Q}_{\rm ch}= (2\pi/a)(\pm\delta_{\rm ch},0)$ and
${\bf Q}_{\rm ch}^{\prime}= (2\pi/a)(0,\pm\delta_{\rm ch})$, the
corresponding spin-ordering vectors,
${\bf Q}_{\rm s}= {\bf Q}_{\rm AF} \pm(2\pi/a)(\frac12\delta_{\rm ch},
\frac12\delta_{\rm ch})$ and
${\bf Q}_{\rm s}^{\prime}={\bf Q}_{\rm AF}
\pm(2\pi/a)(\frac12\delta_{\rm ch}, -\frac12\delta_{\rm ch})$
satisfy the requisite identities. Thus, the relative orientation of the
spin and charge peaks can be used to distinguish fluctuating stripe order
from fluctuating checkerboard order \cite{tran99b}.

\subsection{STM measurements and stripes}
\label{STM}

STM is a static probe and thus cannot detect any structure associated
with fluctuating order unless something pins it
\footnote{\protect\cite{yazd99,huds99,pan00,howa02b,pan01,lang02,pan00b,renn98b,
hoff02,spru97}}.
Density or ``Friedel oscillations" \cite{frie58} in simple metals
produced by the presence of a defect are directly related to
Fermi-surface-derived  non-analyticities in the susceptibility,
$\chi(\bf k)$. However, ``generalized Friedel oscillations'' can occur
in more diverse systems in which the relevant structure in $\chi(\bf k)$
is not directly related to  any feature of a Fermi surface.
For instance, a bosonic superfluid on a lattice close to a second
order transition to an insulating, bosonic crystalline phase would exhibit
generalized Friedel  oscillations with the characteristic wave length of
the roton-minimum - these oscillations, in a very direct sense, would
image the fluctuating crystalline order present in the fluid state.

There are a few important features of generalized Friedel
oscillations which follow from general principles.  If the liquid
state is proximate to a highly anisotropic state, such as a stripe
state, the values of
$\bf k=\bf Q$ at which $\chi$ has maxima will reflect the pattern of spatial
symmetry breaking of the ordered state, but $\chi(\bf k)$ will respect the
full point-group symmetry of
the crystal unless the liquid state spontaneously breaks this symmetry,
{\it e.g.\/}, is a nematic.  So, the generalized Friedel oscillations
around a point impurity in a
stripe-liquid phase will inevitably form a checkerboard pattern,
unless some form of
external symmetry breaking field is applied \cite{polk02}.

There is another form of spatial modulation of the
density of states, one with a period which disperses as a function of
the probe energy,
which is sometimes (incorrectly, we believe) referred to in the STM
literature as Friedel
oscillations.  This latter effect, which was first demonstrated by
\textcite{crom93},
is produced by the elastic scattering of quasiparticles
of given energy off an impurity.  The resulting interference between
scattered waves
leads to variations of the local density of
states at wave vectors $\bf{Q}=\bf k-\bf k^{\prime}$, where
$\bf k$ and $\bf k^{\prime}$ are the wave-vectors of states with
energy $E=\epsilon(\bf
k)=\epsilon(\bf k^{\prime})$, as determined by the band structure,
$\epsilon(\bf k)$.  Generalized
versions of these oscillations can occur even when there are no well defined
quasiparticles, so long as there are some elementary excitations of
the system with a well-defined
dispersion relation, as is shown in Sec. \ref{1DEG}.

Thus, in STM studies of cuprates we would expect stripe correlations to
make an appearance as generalized Friedel oscillations, while
quasiparticle-like interference is a distinct phenomenon that could also
be present.  The observation of a checkerboard pattern with a $4a$ period
about vortex cores in \BSCCO\ by \textcite{hoff02} is provocative
evidence for pinned charge stripes.  These results motivated
\textcite{howa02,howa02b} to search for similar evidence of stripes in
\BSCCO\ with no applied field.  The discussion below is focused on
approaches for distinguishing and enhancing modulations due to generalized
Friedel oscillations associated with such incipient order. Before
continuing, though, we should note that compelling evidence for a
``quasiparticle'' interference response has been reported by
\textcite{hoff02b} and by
\textcite{mcelroy03}.  Those experimental results are not in direct
conflict with those of \textcite{howa02,howa02b}; however, there has been
some controversy over interpretation, regarding whether the observed
spatial modulations in the tunneling conductance are explained entirely
by the interference mechanism, or whether interference modulations
coexist with generalized Friedel oscillations.  We will return to that
controversy in a subsection below.

Modulations which reflect the spectrum of elementary excitations are
distinguishable \cite{polk02b} from those related  to incipient order in
a variety of ways: whereas incipient order produces effects peaked
near
isolated ordering vectors, $\bf Q$,
the single
particle effects are peaked along extremal curves in
$\bf k$ space which disperse as a function of $E$.
Peaks in the Fourier transformed LDOS, $N({\bf k},E)$, produced by
incipient order tend to be phase coherent while
other features either have a random phase, or a phase that is strongly
energy dependent.
Indeed, the phase information may be the best way to distinguish the
consequences of incipient order  from interference effects. Naturally, in
the presence of true long range charge order $N({\bf k},E)$ should
exhibit sharp (resolution-limited) peaks which reflect the charge density
modulation in real space. This effect has been seen in STM measurements
of the quasi two-dimensional incommensurate CDW system $1T$-TaS$_2$,
lightly doped with Nb, by \textcite{dai91}.

To demonstrate the effect of the phase, assume we have measured
$N_L(\bf r,E)$, the LDOS at a
particular bias voltage, $V=E/e$ on a sample of size $L\times L$.
Its Fourier transform, $N_L({\bf k},E)$, has an arbitrary
${\bf k}$-dependent but $E$-independent phase,
$e^{-i{\bf k}\cdot{\bf r}_0}$, which depends on the choice of origin of
coordinates, ${\bf r}_0$.  As discussed above, $N_L({\bf k},E)$ can be
expected to have contributions from incipient order (with wave vector
${\bf{Q}}$), and dispersing quasiparticles.
Integrating  the
signal over a finite energy window yields  the quantity $\tilde N({\bf
k},E)$, in which the contributions from incipient order from the entire
range of  integration add constructively, while other features tend to
interfere destructively.~\footnote{It is important to emphasize that 
both the phase
and amplitude information encoded in $N_L({\bf
k},E)$ are physically significant. The commonly used power spectrum, 
$|N_L({\bf k},E)|^2$, will not lead to destructive interference of 
propagating features 
when integrated over energies.}
 This mode of analysis is particularly useful if
an energy window can be found in which none of  the dispersing features
expected on the basis of band structure considerations have wave vectors
equal to the expected ordering wave vector, ${\bf Q}$.

\begin{figure}[bht]
\begin{center}
\leavevmode
\vspace{.2cm}
\noindent
\includegraphics[width=0.45\textwidth]{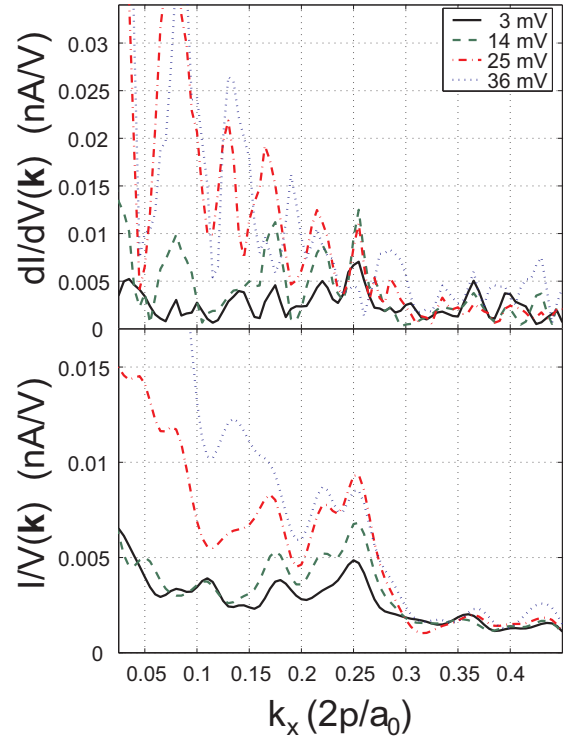}
\vspace{.2cm}\end{center}
\caption
{The real part of the Fourier transform of the local density of states 
from an STM
study of a near
optimally doped crystal of {\BSCCO}.  ${\bf k}$ is taken along the
$(1,0)$ direction, in a tetragonal convention.  a) shows $dI/dV
\propto N({\bf k}, E)$ measured at fixed voltage, $V=E/e$.  b) shows
$I/V\propto
\tilde N({\bf k},E)$, which is the integral of the quantity shown in a) from
$V=0$ to
$V=E/e$.}
\label{howald}
\end{figure}

Indeed, \textcite{howa02} have already
demonstrated the first half of this point in STM studies of a
very slightly overdoped sample of {\BSCCO} ($T_c=86$).  They
identified a peak which in $N({\bf k},E)$ at approximately
${\bf k}=\pm{\bf Q}_{\rm ch}\approx \pm (2\pi/a)(0.25,0)$ and
${\bf k}=\pm {\cal R}[{\bf Q}_{\rm ch}]=\pm (2\pi/a)(0,0.25)$ (see
Fig.~\ref{howald}), and showed that at these wave numbers
the phase of $N({\bf k}, E)$ is energy independent for $E$
between 0 and 40~meV, at which point the amplitude crosses through
zero {\it  i.e.\/}, the sign of the signal changes.  The constancy of the
phase implies that the location in real-space of the density of states
modulation is fixed at all energies, which strongly indicates that it
reflects pinned incipient order.

However, as explained above, the existence of incipient order can be
magnified if we integrate the LDOS as a function of energy. In
Fig.~\ref{howald}(a) we show data from
\textcite{howa02,howa02b} in which the values of $dI/dV$
  were obtained by a Fourier transform of
a real space image at various voltages, $V=E/e$, on a patch of
surface of size $L=160$~\AA.  The origin of coordinates was chosen that 
$N_L({\bf{k}},E)$ 
is real and positive for $\bf k-\bf Q_{\rm ch}$. In the figure ${\bf k}$ 
is taken
to lie along the $(1,0)$ direction and what is shown is the real part of 
$N_L({\bf{k}},E)$ 
(the imaginary part if generally small and noisy.)
In Fig.~\ref{howald}b, we show the same data averaged over energy
from $0$ to $E$, $I/V \propto \tilde N_L({\bf{k}},E)$. It is apparent that
integration enhances the strength of the  peak at ${\bf Q}_{\rm ch}$, and
depresses the remaining signal, especially when $E$ is smaller than  the
maximum superconducting gap value,
$\Delta_0\approx 35$~meV.  Indeed, precisely this same integration
technique was used previously by \textcite{hoff02} to enhance the
magnetic-field induced ``checkerboard" pattern in vortex cores.

\subsubsection{Differing interpretations of the STM spectra}
\label{differing}

As mentioned in the introduction, there is controversy in the
literature concerning the interpretation of the STM spectra, although
there seems to be only relatively minor
disagreements concerning the data itself.
Specifically, \textcite{hoff02b} and still more recently
\textcite{mcelroy03} have suggested that the peaks seen in
$N({\bf k},E)$ can be accounted for entirely in terms of the interference
pattern of sharply defined quasiparticles in a d-wave superconductor,
with no need to invoke incipient stripe order - or any non Fermi liquid
character of the elementary excitations, for that matter.  Let us briefly
review the line of reasoning that leads to this conclusion.

As we have seen (Lesson \#4), interference between quasiparticles  in 2D
naturally produces ridges in
$N({\bf k},E)$ along closed curves in ${\bf k}$ space, rather than
the peaks observed in
experiment.  However, as was first recognized by \textcite{wang02}, this
result is modified by the extreme eccentricity of the contours of
constant quasiparticle energy in a d-wave superconductor, where the
dispersion in one direction is $v_F$ and in the other is proportional to
$\Delta_0$.  Consequently, with finite experimental resolution, at
energies less than
$\Delta_0$, one obtains peaks in the interference patterns at
points in $\bf k$ space
which connect the tips of the contours that surround each of the four
distinct nodal points on
the Fermi surface.

\begin{figure}[bht]
\begin{center}
\leavevmode
\vspace{.2cm}
\noindent
\includegraphics[width=0.5\textwidth]{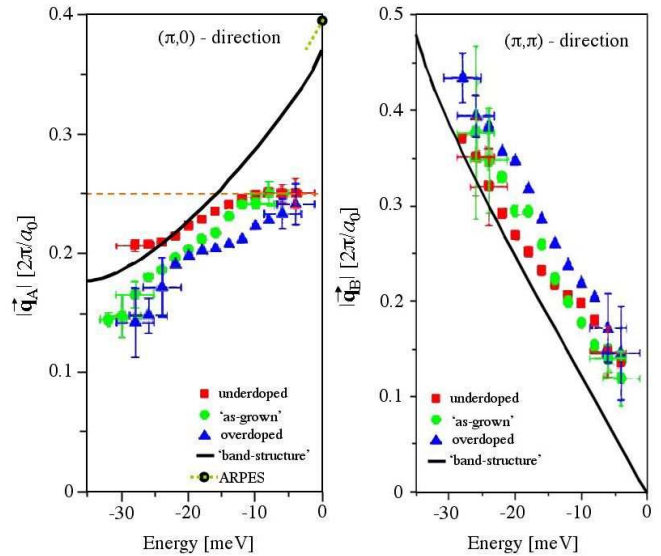}
\vspace{.2cm}
\end{center}
\caption
{Dispersion of the peaks (colored symbols) in $N({\bf k},E)$  (from
STM measurements on near
optimally doped {\BSCCO}) at wavevectors a)
${\bf k}_1$ (along the
$(1,0)$ direction) and b) ${\bf k}_2$ (along the $(1,1)$
direction) from Fig. 4 of \textcite{hoff02b}.  The
solid lines  are from a global fit to the ARPES spectrum, as
described in the text.
}
\label{davisfig}
\end{figure}

Specifically, if we assume that there are well defined quasiparticles with
energy obtained by solving the Bogoliubov-de~Gennes equations for a
d-wave superconductor, then for each energy $E$ there are eight values
${\bf q}_j$ (two in the neighborhood of each nodal point) which
simultaneously satisfy the equations $\epsilon({\bf q}_j)=0$ ({\it i.e.\/},
they lie on the normal state Fermi surface) and $\Delta_{{\bf
q}_j}=E({\bf q}_j) = E$ ({\it i.e.\/}, the quasiparticle creation energy is
$E$).  Peaks in $N({\bf k},E)$ will then occur at the various distinct
values of ${\bf k}_{ij}={\bf q}_i-{\bf q}_j$, of which  there are 7
distinct (up to symmetry) values.  For instance, the two wave vectors
with smallest magnitudes are ${\bf k}_{12}=2q_x (1,0)$ and
${\bf k}_{13}=\sqrt{2}(q_x-q_y)\left( \frac 1
{\sqrt{2}},\frac 1
{\sqrt{2}}\right)$ which come from the interference between
the state
${\bf q}_1=(q_x,q_y)$ and, respectively, the states at
${\bf q}_2=( -q_x,q_y) $ and ${\bf q}_3=(q_y,q_x) $.
The positions of the peaks in $N({\bf k}, E)$ thought to correspond to
${\bf k}_{12}$ and ${\bf k}_{13}$ from the work \textcite{hoff02b}
are reproduced in Fig.~\ref{davisfig}.
Given the quasiparticle
interpretation of the STM spectrum, then from the observed location of
any two distinct peak positions ${\bf k}_{ij}$ at a given energy, it is
possible to reconstruct the positions of all eight locations in the
Brillouin zone, ${\bf q}_j$ which give rise to all the expected peaks at
this energy.  Thus, where 7 distinct peaks are observed (as are reported
by \textcite{mcelroy03} in some ranges of energy), the quasiparticle
spectrum is highly over-constrained, and this provides a very stringent
self-consistency check on the quasiparticle interpretation.  According to
\textcite{mcelroy03}, the peaks they observe generally pass this
consistency check.  The experimental case for dispersive features in
$N({\bf k},E)$ that are consistent with the quasiparticle interference
mechanism is quite persuasive.

The controversy over interpretation revolves around the degree to which
quasiparticle interference is sufficient to explain all of the structure
in $N({\bf k},E)$.  Examining Fig.~\ref{davisfig}(a), one can see that
for energies below $\sim20$~meV, ${\bf k}_{12}$ lies close to the expected
stripe ordering vector, ${\bf Q}_{ch}$.  This is consistent with the
experimental measurements and analysis by \textcite{howa02}, who had
interpreted this behavior as evidence for stripe order; in contrast,
\cite{hoff02b} argued that the continuity with the dispersive signal
favors a single mechanism based on band structure effects.

In weighing these alternatives, it is useful to consider
a second consistency check that the quasiparticle picture should
satisfy.
While, as shown in Eq.~(\ref{convolution}),
$N({\bf k},E)$ is actually a measure of a two-particle response function,
to the extent that it is dominated by single-particle effects it should be
expressible in terms of a convolution of single-particle Green functions,
as in Eq.~(\ref{convolution}).  Thus, it should be consistent with the
behavior of the
spectral function measured in ARPES.
In evaluating this connection, significant challenges arise with respect
to the sufficiency of the quasiparticle interpretation.

In the first place, ARPES consistently reveals strong deviations from a
non-interacting line shape.\footnote{At very low energies and at low
temperatures there is good reason to believe, and compelling evidence
(although not from ARPES or STM) to confirm that the nodal quasiparticles
in the  superconducting state are well defined and long-lived; see
\textcite{may00}.  However, most of the dispersing features seen in STM
are at higher energies, typically at energies on the order of
$\Delta_0/2$, where even at low temperatures there is no  evidence of
well defined quasiparticles.}
Well below $T_c$, the measured
energy distribution curve (EDC) at  fixed ${\bf q}$ on the Fermi
surface consists of a
dispersing ``quasiparticle-like" peak with a small weight, $Z \ll
1$, and a broad, and largely
featureless multiparticle continuum which contains most of the spectral
weight.  Moreover, the quasiparticle peak is always anomalously broad
with a distinctly non-Fermi liquid
temperature and energy dependence \cite{orga01,gweo01}.  If this measured
spectral function were used to predict the structure of $N({\bf k},E)$,
then the quasiparticle interference features would be
weak \cite{capr03} (in proportion to $Z^2$), which is something of a
surprise given the apparent robustness of the experimental features.  One
might also expect the interference features to be broadened significantly
by the short quasiparticle lifetimes at the probed energies.  Perhaps the
most important consequence is that the intensities of the structures
should have a dramatic temperature dependence, going to zero as $Z$ goes
to zero at $T_c$.  This provides a crucial test to which we shall
return shortly.

In the second place, $\Delta_{{\bf q}}$ determined from the
dispersion of the quasiparticle-like
peak in ARPES differs significantly, especially at low energies, from
that obtained by \textcite{hoff02b} and \textcite{mcelroy03} in STM.
This is illustrated in Fig. \ref{davisfig}, as well.  The solid lines
in the figure were
computed (using the method of \textcite{wang02})
from a global fit to the measured ARPES
spectrum, as follows: 1)  The location of the Fermi surface,
$\epsilon({\bf q})=0$ was computed
from  the phenomenological ``band structure" that has been obtained
\cite{norm94,dama02} from
a fit to ARPES experiments.  2)  The
gap function along the Fermi surface was assumed to be of the simple
d-wave form, $\Delta_{\bf
q}={\frac{\Delta_0}{2}}[\cos(q_xa) -
\cos(q_ya)]$, a form that has been widely found to fit the measured
ARPES spectrum in optimally
and overdoped {\BSCCO}, although significant deviations from this
form are often seen in
underdoped samples \cite{meso99}.  The gap observed in ARPES experiments
corresponds to $\Delta_0 \sim$ 35-45~meV
\cite{loes97,ding95,fedo99} near optimal doping - for purposes of the
figure, we have adopted
$\Delta_0=40$~meV.
The solid black circle in the figure was obtained from the location of
the nodal point (which determines
the zero energy limit of all dispersion curves) as determined from ARPES
experiments;  this is a feature of the ARPES spectrum that has
been looked at in great detail by several groups
\cite{ding97,vall99}. There is agreement\footnote{We thank P.Johnson for
pointing this out to us.} with 1\%\ accuracy that the nodal point, where
$E({\bf q})=0$, is
${\bf q}=(2\pi/a)(0.195,0.195)$.  (Note that the phenomenological
band structure was determined from a fit to older, less precise ARPES
data, which is why the line in Fig. \ref{davisfig} does not quite approach
the correct zero energy limit;  manifestly, correcting this
discrepancy would only exacerbate the disagreement between STM and ARPES.)

It is clear from Fig. \ref{davisfig}, as originally emphasized by
\textcite{howa02}, that there are significant discrepancies between the
ARPES and STM results, especially below 20~meV.  In particular, in the
region of ${\bf k}$ space near the expected stripe-ordering wave vector,
${\bf k}={\bf Q}_{\rm ch}\sim (2\pi/a)(\frac14,0)$, the STM spectrum is
considerably less dispersive than would be expected on the basis of the
ARPES data;  it appears that the peak in $N({\bf k},E)$ largely stops
dispersing when it reaches this magic wave vector (see for instance the 
discussion surrounding Eq.(\ref{Kstar}).)\footnote{The same
discrepancy between ARPES and STM at low energies can be seen directly
from full microscopic calculations of  $N({\bf k},E)$ for non-interacting
quasiparticles. For instance,
$N({\bf k},E)$ was recently computed by
\textcite{polk02b} for non-interacting quasiparticles scattering from a
point impurity; it is apparent from their Fig.~4 that, at energies of
20~meV and below, there is a local {\it minimum} in the neighborhood of
${\bf k}={\bf Q}_{\rm ch}$, rather than a peak!  The same is apparent in
Figs.~2 and 3 of \textcite{wang02}.}

It is not clear {\em a priori} how much significance one should attach to
the discrepancies between ARPES data and the quasiparticle interpretation
of the STM.  Both are highly surface sensitive probes, so there is always
the issue of whether either is telling us anything about the bulk
properties of the materials, but this worry does not affect the
comparison between the two sets of measurements.

It seems 
significant to us that the most serious quantitative differences between the
ARPES dispersion and those inferred from a quasiparticle interpretation
of the STM data occur at low energies, and are particularly pronounced
for ${\bf k}$'s near ${\bf Q}_{ch}$, where effects of incipient stripe
order are largest. In this context, it is worth noting that in the STM
studies of \textcite{howa02b} shown in Fig.~\ref{howald},  a peak at
${\bf k}={\bf Q}_{\rm ch}$ is seen in all the curves with
$E\le 15$~meV all the way to $E=0$!  (\textcite{hoff02b} and
\textcite{mcelroy03} do not report spectra below about 6~meV.  It is
possible that, as suggested by \textcite{howa02b}, that this difference
is a consequence of the stripe signal being washed out due to the
existence of many domains in the larger field of view used in the
experiment of
\textcite{hoff02b}.)

Moreover, in their STM study of the magnetic-field-induced structure in 
vortex-core halos, \textcite{hoff02}  
observed oscillations at
precisely these same wave vectors, ${\bf k}={\bf Q}_{\rm ch}$, in the
energy-integrated LDOS  (in a 12~meV window about 0 voltage).
The amplitude of these latter oscillations is very large compared with that 
of  the disorder-induced signal, 
and the dominant contribution comes from energies
around 7~meV.

We believe that the issues of interpretation can be definitively
resolved by studies of the temperature and impurity concentration
dependence of the signal.  Since the contribution to 
$N({\bf k},E)$ from quasiparticle interference is proportional to $Z^2$, and 
since
$Z$ is observed in ARPES experiments to have a strong temperature
dependence and vanish above $T_c$, it should be very easy to quench the
interference signal by heating; what is left at temperatures
approaching $T_c$ cannot be due to quasiparticle interference.  And, of
course, above $T_c$ in overdoped samples, and above a suitable pseudogap
temperature in underdoped samples, the superconducting gap vanishes, so
only the more usual
rings can possibly arise from
quasiparticle interference. Conversely, as discussed previously, light
Zn doping is known to pin stripes in {\LSCO}, so it is reasonable to
expect the same effect in {\BSCCO}.  Thus, stripe-related signals should
be strengthened, and made less sensitive to thermal
depinning by small concentrations of Zn impurities in the Cu-O planes.

STM studies at elevated temperatures are technically
demanding, but recently \textcite{yazd03} have achieved atomic scale
resolution at temperatures well above $T_c$  on a {\BSCCO} surface.
They report that for $T \sim T_c/2$ many of the peaks in $N({\bf 
k},E)$ have, indeed,
been extinguished. However, the low energy portion of the peak
at ${\bf k}_{12}$ (in Fig. \ref{davisfig}) remains unattenuated, and 
even survives,
well above $T_c$.  This effect clearly cannot be explained by the
quasiparticle interference mechanism.
An explanation in terms of pinned
stripes seems much more plausible (although the persistence of a
stripe-induced signal at such an elevated temperature requires further
investigation).

One further feature of all the data that has been reported to date is
worth commenting on is the large (factors of 2 or 3) differences
that are observed \cite{hoff02b} in the  peak intensities at
${\bf Q}_{\rm ch}$ and
$R[{\bf Q}_{\rm ch}]$.  Systematic experiments have not, yet, been
conducted to verify whether this effect is real, or an experimental
artifact.  If it is not an artifact, then  this observation is among the
first microscopic pieces of evidence of a strong local tendency to
stripe orientational ({\it i. e.\/}, nematic) order.  In addition, such
large anisotropies are something that cannot be accounted for in any
simple way by quasiparticle interference (nor local checkerboard order).
It is important to bear in mind, however, that even if the observed
anisotropy reflects nematic order, it is expected to decrease in
magnitude \cite{howa02} in direct proportion to $A^{-1/2}$, where $A$ is
the area of the field of view, due to the unavoidable domain structure
produced by quenched disorder.

\subsection{Detecting nematic order}
\label{nematic}

While stripe order necessarily implies nematic order, the converse is not
true. Although nematic order involves the spontaneous breaking of a
spatial (point-group) symmetry, when there is no accompanying breaking of
translation symmetry, even the identification of the ordered state is
somewhat subtle, and this holds doubly for fluctuation effects.
Moreover, since quenched disorder is always relevant and results in
domain structure, true macroscopic measurements of spontaneous nematic
symmetry breaking are not possible.  The difficulty of detecting the
rotational-symmetry  breaking associated with ordered stripes is
illustrated by the case of {\LSNiO}, a  non-superconducting structural
analogue of {\LSCO}.  Even transmission electron microscopy, which is
capable  of measuring over a fairly small sample area, tends to yield
superlattice diffraction peaks for stripe order that reflect the
4-fold symmetry of the NiO$_2$ planes in the tetragonal crystal
structure \cite{chen93}.  Only recently have electron diffraction patterns
consistent with the  2-fold symmetry of an ordered stripe domain been
reported \cite{li02}. Thus, almost all tests of nematic order in solids
necessarily involve the observation of an unreasonably large, and
strongly temperature dependent anisotropy in the electronic response to a
small symmetry breaking field which favors the $x$ direction over the $y$
direction.

\subsubsection{Transport anisotropies}

An example of a system in which this approach has been successfully
applied is provided by the 2DEG in quantum Hall devices.
While the fractional quantum Hall effect dominates the physics at very
high magnetic fields, at intermediate magnetic fields (so that more than
one Landau level is occupied), it has recently been discovered
\cite{lill99,du99} that there are a set of anisotropic states which have
been identified \cite{frad99,frad00,coop02} as being quantum Hall
nematics, and which can be thought of as melted versions of the long
predicted \cite{fuku79,koul96,moes96} quantum Hall smectic (or stripe)
phases.  These phases are characterized by a large resistance anisotropy,
$\rho_{xx}\gg\rho_{yy}$, which onsets very strongly below a
characteristic temperature, $\Tn \sim 100$~mK. The precise origin of the
symmetry  breaking field which aligns the nematic domains in these
experiments has not been unambiguously determined \cite{coop01,zhu02}.
However, by applying \cite{lill99b,pan00c,coop02} an in-plane magnetic
field, the magnitude of the  symmetry breaking field can be varied, and
the transition can be significantly rounded (giving evidence \cite{coop02}
that local stripe order persists up to temperatures well in excess of
$\Tn$), and even the orientation of the nematic order can be switched,
resulting in a state with $\rho_{xx} \ll \rho_{yy}$.

Experiments that involve such fine control of an external symmetry-breaking
field are considerably harder to carry out in the context of the cuprates.
Some of the relevant materials, such as {\BSCCO} and {\LSCO} with
$x>0.05$, have an orthorhombic axis at 45$^\circ$ to the expected stripe
directions  ({\it i.e.\/}, a nematic phase can be defined in terms of
spontaneous breaking of the mirror plane which lies along the
orthorhombic $a$-axis.)  In these materials, if one wishes to align the
nematic domains, one must apply a suitable external symmetry-breaking
field such as a uniaxial strain or in-plane magnetic field; however, such
experiments tend to be challenging.

When the principle axes of an orthorhombic phase lie parallel to the
expected stripe directions, the orthorhombicity (typically $<2$\%\ in
cuprates) plays the role of a small, external symmetry-breaking field.
Examples where this occurs are superconducting {\YBCO} with $0.35\le
y\le1$ and non-superconducting {\LSCO} with $0.02\le x<0.05$.  In both of
these cases, resistivity anisotropies as large as a factor of 2 have been
observed \cite{ando02} in detwinned single crystals.  Moreover, as in  the
quantum Hall case, this anisotropy is strongly temperature dependent; in
\LSCO, $|\rho_{aa}/\rho_{bb} - 1| < 10$\%\ for temperatures in excess of
$\Tn \sim 150$~K.  Polarization dependent measurements of the infra-red
conductivity on a detwinned $x=0.03$ crystal of {\LSCO} reveal that the
frequency dependence of the conductivity anisotropy has a scale
comparable to $k\Tn$ \cite{baso02}.  Large anisotropies in the
frequency-dependent conductivity have also been
observed \cite{baso95} in \YBCO\ and in YBa$_2$Cu$_4$O$_8$,
although in those cases, some part of the conductivity anisotropy must be
due directly to the Cu-O chains.  Taken together, these various
observations are circumstantial evidence of a fair degree of local stripe
order, and of the presence of a nematic phase; the trouble being that,
without a somewhat quantitative theory (which does not exist), it is hard
to say how large an observed resistance anisotropy must be to be accepted
as being an ``unreasonably large'' response to the orthorhombicity, that
is,  one that can only be understood in terms of the alignment of nematic
domains.

More subtle investigations of orientational symmetry breaking can be
undertaken by studying the transport in the presence of a magnetic field.
A very clever approach  along these lines was introduced by
\textcite{noda99}.  They applied a  voltage  along one axis of
the CuO$_2$ planes and a magnetic field  perpendicular to the planes
in order to break the 4-fold symmetry of the crystal structure and to
obtain evidence of  one-dimensional charge transport. In a
stripe-ordered state, the geometry used should be sensitive primarily
to those domains in which the stripes are aligned parallel to the
direction of  the applied voltage.  For $x\le\frac18$, they found that,
within the stripe-ordered phase, the transverse conductivity tends to
zero at low temperature while the longitudinal conductivity remains
finite.  (This effect has been explained in terms of the electron-hole
symmetry of a $\frac14$-filled charge stripe \cite{emer00,prel01}.)

\cite{ando99} have observed a remarkable anisotropy
of the resistivity tensor induced by an in-plane magnetic field in
non-superconducting {\YBCO} with $y=0.32$ and $y=0.3$.  (Presumably,
these samples are antiferromagnetic.) The results were interpreted as
evidence for nematic stripe order.  Alternative explanations in terms of
anisotropy associated with spin-orbit coupling in the antiferromagnetic
phase  have also been proposed \cite{jano00,mosk02}.  This example
illustrates the difficulty in making a unique association between a bulk
anisotropy and microscopic stripe order.

Finally, we note that there is a very direct way to detect nematic order
using light  scattering. This approach is well known from studies of
classical nematic liquid crystals \cite{dege74}. Recently R\"ubhausen
and coworkers \cite{rubh00,yoon00} have used light scattering techniques
to study the behavior of the low-frequency dielectric tensor of the
manganite  {\BiCaMnO}  near and below the charge-ordering transition at
$T_{co} \sim 160 K$. In these experiments,  a pronounced anisotropy of the
dielectric tensor was found with a sharp temperature-dependence near
$T_{co}$. Raman scattering studies by the same group show that long range
positional stripe order sets in only at much lower temperatures. These
experiments suggest that this manganite is in a nematic state below the
transition at $T_{co}$, and long range charge-stripe order occurs at much
lower temperatures.

\subsubsection{Anisotropic diffraction patterns}

A more microscopic approach is to directly measure the nematic order
parameter, ${\cal Q}_{\bf k}$ of Eq.~(\ref{N}), by diffraction.  This was
done (more or less) by \textcite{mook00}, in neutron
scattering studies of the magnetic dynamic structure factor of  a
partially (2 to 1) detwinned sample of {\YBCO} with $y=0.6$ and
$T_c\approx60$~K.  No elastic scattering corresponding to actual stripe
order was detected; however, well developed structure was observed in the
inelastic spectrum at the two-dimensional wave vectors
${\bf Q}_{\rm s}\approx(0.5\pm0.1,0.5)$ and
${\bf Q}_{\rm s}'\approx(0.5,0.5\pm0.1)$ in units of $2\pi/a$.  Remarkably,
the intensities of the peaks at ${\bf Q}_{\rm s}$ were found to be about a
factor of 2 larger than those of the ${\bf Q}_{\rm s}'$ peaks, consistent
with the supposition that, in a single twin domain, the incommensurate
inelastic structure is entirely associated with ordering vectors
perpendicular to the chain direction, {\it i.e.\/},
${\cal Q}_{\bf Q}\approx 1$.

More recently, inelastic neutron scattering studies of
\textcite{stoc03} on a nearly single-domain sample of the Ortho-II phase
of YBa$_2$Cu$_3$O$_{6.5}$
\cite{stoc02}
have revealed substantial structure down to the
lowest energies in the magnetic structure factor.  These spectra are
highly anisotropic about the N{\'e}el ordering vector:  For a scan
perpendicular to the chain direction,  there is a broad flat-topped peak
(reminiscent of earlier results on underdoped YBCO\cite{ster94}) which
is strongly suggestive of two barely resolved incommensurate peaks at the
expected stripe ordering wavevector.  However, a scan along the chain
direction reveals a single, sharp peak at the commensurate wave vector
$\pi/a$.  Moreover, in the normal state, this   structure is observed at
all energies below the ``resonant peak'' energy, $\omega \approx
25$ meV, down to the lowest energies probed.

Although the presence of chains in this material certainly means that
there is no symmetry operation that interchanges the $a$ and $b$ axes,
the copper-oxide planes are nearly tetragonal.  Thus, it seems to us that
the extreme anisotropy of the inelastic scatterings very strong evidence
of a nematic liquid phase in this material.  As pointed out by
\textcite{mook00}, this conclusion also offers a potential explanation for
the observed \cite{baso95} large but nearly temperature independent
superfluid anisotropy  in the $a$-$b$ plane.

Interesting anisotropies have also been observed in optical phonon
branches of \YBCO\ with $y=0.6$.  The identification by \textcite{mook00}
of an anomalous broadening of a bond-bending mode at a wave vector
expected for charge-stripe order is potentially the most direct evidence
of nematic order.  However, conflicting results have been reported by
\textcite{pint02}, who have instead observed a zone-boundary softening of
the bond-stretching mode propagating along ${\bf b}$ but not for that
along
${\bf a}$.  This latter anisotropy is unlikely to be directly associated
with a stripe modulation wave vector since the anomaly occurs along the
direction parallel to the Cu-O chains.  Of course, this does not
necessarily rule out a connection with stripes, as the softening might
be associated with anisotropic screening due to charge fluctuations along
the stripes.

In an attempt to understand the effects of stripe order on phonons, a
neutron scattering study \cite{tran02} was recently performed on
La$_{1.69}$Sr$_{0.31}$NiO$_4$.  Although the charge-ordering wave
vector did not play an obvious role, the high-energy bond-stretching mode
propagating parallel (and perpendicular) to the stripe modulation
exhibited an energy splitting toward the zone boundary, while
along the Ni-O bond direction (at 45$^\circ$ to the stripes) a softening
from zone center to zone boundary was observed with a magnitude similar
to that in the cuprates.  A better understanding of the nature of the
relevant electron-phonon coupling processes is required to make
progress here.

\subsubsection{STM imaging of nematic order}

\begin{figure}[bht]
\begin{center}
\leavevmode
\vspace{.2cm}
\noindent
\includegraphics[width=0.45\textwidth]{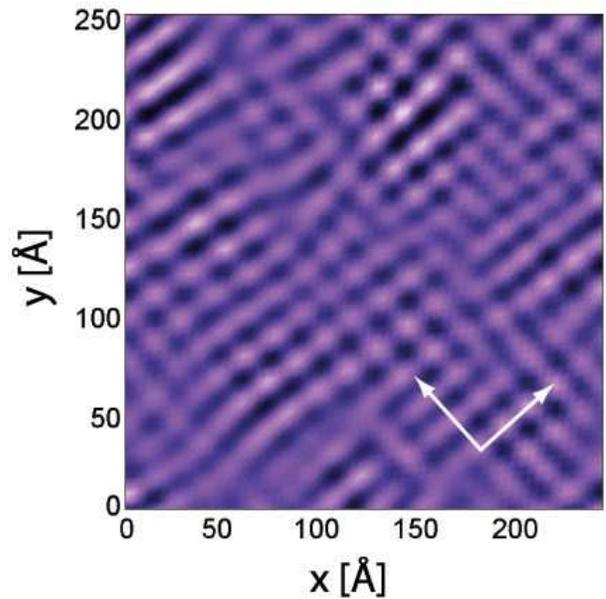}
\vspace{.2cm}\end{center}
\caption
{A filtered version of the local density of states map, $N_f({\bf r},E)$,
of the local density of states on a surface patch of a {\BSCCO} crystal.
Here, $E=15$~meV and the distances on the $x$ and $y$ axes are measured in
angstroms.  The filter is defined in Eq.~(\protect{\ref{filter}}) with
$\Lambda =(2\pi/15a)$. The arrows point along the directions of the Cu-O bonds.}
\label{fig-nematic}
\end{figure}

Because it is a local but spatially resolved probe, STM is actually
the optimal probe of nematic order.  One way it can be used, which  is
illustrated in Fig.~\ref{fig-nematic}, has been explored by
\textcite{howa02,howa02b}.  What is shown here is a filtered
version of $N({\bf r},E)$ measured on a patch of surface of a very
slightly overdoped crystal of {\BSCCO} ($T_c=86$~K).
Specifically, Howald {\it et al.\/} defined a filtered image
\begin{equation}
N_f({\bf r},E) = \int d{\bf r}^{\prime} f({\bf r}-{\bf
r}^{\prime})N({\bf r}^{\prime},E)
\label{filter}
\end{equation}
where in the present case, the filter function, $f$, has been defined so
as to accentuate the portions of the signal associated with stripe order
\begin{equation}
f({\bf r}) \propto \Lambda^2e^{-r^2\Lambda^2/2 } [\cos(\pi
x/2a)+\cos(\pi y/2a)].
\label{f}
\end{equation}
Clearly, $f({\bf r})\to \delta({\bf r})$ when $\Lambda\to\infty$, while
$N_f= N({\bf Q}_{\rm ch},E)+N({\cal R}[{\bf Q}_{\rm ch}],E) + \mbox{c.c.}$
in the limit $\Lambda\to 0$. For  intermediate values of
$\Lambda$, the filtered image shows only that portion of the signal we
have  associated with pinned stripes.  The
roughly periodic structure in the image has period $4a$. We know
independently from the analysis in Section~\ref{STM} that there is
prominent structure in the raw data with this  period, but  even were
there not, the filtering would build in such structure.  However, what is
clear  from the image is that there is a characteristic domain structure,
within which the stripes appear to  dominantly lie in one direction or
the other.  The domain size is seen to be on the order of
100~\AA~$\sim25a$, which is large compared to $\Lambda^{-1}$ and, more
importantly, roughly independent \cite{howa02b} of the precise value of
$\Lambda$.  This domain size is a characteristic correlation length
of the pinned nematic order.

This particular method of analysis builds directly on the realization
of the nematic as a melted stripe-ordered state.  Indeed,
looking at the  figure, one can clearly identify dislocations and
disclinations in what looks like a fairly  well developed locally
ordered stripe array.

More generally, STM could be used to directly measure the two independent
components of a suitably defined traceless symmetric tensorial density.
The simplest such quantities are
\ba
{\cal Q}_{\rm xx}({\bf r},E) = && [\partial_x^2-\partial_y^2] N({\bf r}, E) 
\nonumber \\
{\cal Q}_{\rm xy} ({\bf r},E) = && 2 \partial_x\partial_y N({\bf r}, E)
\ea
Of course, these quantities, like the local density of states itself,
will typically
have features which reflect the interference between elementary
excitations, and other
extraneous information.  To obtain a better view of the long
wave-length nematic
correlations, we should again integrate these densities over a suitable energy
interval, $\Omega$, and filter out the short wavelength components:
\begin{equation}
\tilde {\cal Q}_f({\bf r})= \int_0^{\Omega} \frac {dE}{\Omega} \int
d{\bf r}^{\prime}
f({\bf r}-{\bf r}^{\prime}) {\cal Q}({\bf
r}^{\prime},E)
\end{equation}
where $f$ might be a Gaussian filter, as in Eq.~(\ref{f}), but without the
cosine factors.  A map of $\tilde Q_f$ should produce a domain structure,
similar  to that shown in Fig.~\ref{fig-nematic}, but without all the
short wavelength detail.  Where the domain size, $L_N$, is large compared
to $a$, the resulting picture should look  qualitatively the same
independent of the range over which the signal is coarse-grained, so long
as $L_N \gg \Lambda^{-1} \gg a$. The above procedure should work well if
there are no other long-wavelength features in the data.  Unfortunately,
for BSCCO, the inhomogeneities in the gap structure
\cite{howa02,lang02} hamper such a procedure.

\subsection{``1/8" anomaly}
\label{18}

Many members of the lanthanum cuprate family of high-temperature
superconductors exhibit strong singularities in the doping dependence of
various interesting low-temperature properties at $x=1/8$;  together,
these phenomena are referred to as the ``1/8 anomaly.''  For instance, in
{\LBCO} and {\LNSCO}, there is \cite{mood88,craw91} a deep minimum in
$T_c(x)$; in {\LBCO} and {\LSCO} there is a pronounced
\cite{craw90,zhao97} maximum in $\alpha(x)$, the isotope exponent
$\alpha\equiv d\log(T_c)/d\log(M)$; and in {\LSCO} there is a
pronounced \cite{pana02} minimum in the superfluid density,
$n_s(x)$.  One of the central inferences drawn by \textcite{tran95a}
following the discovery of stripe order in {\LNSCO} is that this 1/8
anomaly is associated with a commensurate lock-in of the stripe
structure. At $x=1/8$ the preferred spacing between charge stripes is 4
lattice constants, so there is an additional commensuration  energy which
stabilizes stripe order at this particular hole density.

While it is possible to imagine other forms of
CDW order which would similarly be stabilized at $x=1/8$, there can be no
doubt that in this family of materials the 1/8 effect is associated with
stripe order.  This has been confirmed, for example, in {\LSCO} where
quasi-elastic magnetic scattering from spin-stripe order has been
detected \cite{suzu98,kimu99} for $x$ in the neighborhood of 1/8; in
low-energy inelastic measurements, \cite{yama98a}
have shown that the magnetic peak width is most narrow at $x=1/8$.
Correspondingly, slow (probably glassy) spin fluctuations have been
detected by $\mu$SR in the same material with a somewhat arbitrarily
defined onset temperature,
$T_g(x)$, which has \cite{pana02} a pronounced peak at $x=1/8$.  The fact
that quasi-elastic magnetic order as detected by neutron
scattering onsets at a considerably higher temperature than that
detected by $\mu$Sr is clearly a consequence of the
inevitable glassiness of a density-wave transition in the presence of
quenched disorder; it reflects the differences in the time scales of the
two probes, not the presence of two distinct ordering phenomena.  More
recently, new experiments on {\LSCO} as a function of $x$ show pronounced
singularities in the $x$ dependence of the $c$-axis Josephson plasma
edge \cite{baso02} and in the low-temperature thermal
conductivity \cite{take02}; these effects can be interpreted
straightforwardly in terms of a peak in the stability of the charge-stripe
order at $x=1/8$.  Furthermore, charge-stripe order has now been detected
directly by neutron diffraction in
La$_{1.875}$Ba$_{0.125-x}$Sr$_x$CuO$_4$ \cite{fuji02}.

The large drop in $T_c$ at $x=1/8$ found in {\LBCO} is not observed in
{\LSCO}; however, such a dip in the doping dependence of $T_c$ can be
induced in {\LSCO} (centered at $x=0.115$) by substitution of 1\%\ Zn for
Cu, as shown some time ago by \textcite{koik92}  Zn
substitution enhances local magnetic order at low temperature near the
dip minimum, as detected by $\mu$SR \cite{wata02,pana02}.   Given the
clear association between the 1/8 anomaly and stripe order in {\LSCO}, an
indirect method to look for the presence of local stripe order in other
families of cuprate superconductors is to test whether a 1/8
anomaly can be induced by Zn substitution.  In the case of
{\YBCO}, there already exists in the literature strong evidence from the
work of Tallon and collaborators \cite{tall95} that the ``60-K plateau''
in the $y$ dependence of $T_c$ is not, primarily, a reflection
of some form of oxygen ordering in the chain layers, as is commonly
assumed, but is rather a barely resolved 1/8 anomaly; this conclusion is
also supported by a Ca-substitution study \cite{akos98} and by a recent
study of transport properties \cite{ando02b}.  To further test this idea,
Koike and collaborators \cite{akos00,akos98,wata00} studied the doping
dependence of $T_c$, and of $T_g$ measured by $\mu$SR, in lightly Zn doped
{\YBCO} and {\BSCCO}.  They have found
that there is some tendency for light Zn doping to produce a dip in
$T_c(x)$  and a more impressive peak in $T_g(x)$ at
$x\approx1/8$ in  both {\YBCO} and {\BSCCO}.  Related results have been
reported \cite{yang00,ando02p} for the single-layer system
Bi$_2$Sr$_{2-x}$La$_x$CuO$_{6+\delta}$.  While it is clear that more work
is needed to test the connection between the ``1/8 anomaly'' and stripe
pinning in these various systems, we consider this a promising approach to
the problem, as it permits evidence of local order to be obtained using a
variety of probes which can be applied in materials for which large
crystals, and/or easily cleaved surfaces are not easily obtained.

\subsection{Other probes}
\label{other}

It is clear that the most direct evidence for stripes comes from
techniques that can provide images of charge, spin, and/or lattice
modulations in real space (STM) or in reciprocal space (diffraction
techniques).  We have already discussed neutron and X-ray diffraction,
but explicit mention should also be made of transmission electron
microscopy (TEM).  The charge stripes in {\LSNiO} were first detected by
TEM \cite{chen93}, and a recent study has provided high-resolution TEM
images of local stripe order in
La$_{1.725}$Sr$_{0.275}$NiO$_4$ \cite{li02}.  So far, TEM studies have not
provided positive evidence for stripes in any cuprates, but, though
challenging, it should be possible to do so.

Less direct but extremely valuable information comes from techniques that
are sensitive to local order.  We have already mentioned evidence for
local, static hyperfine fields and slowly fluctuating hyperfine fields
obtained by $\mu$SR.  Besides providing a practical measure of local
magnetic order as a function of doping \cite{nied98,klau00}, $\mu$SR can
detect the distribution of local hyperfine fields in a sample with
relatively uniform order \cite{nach98}, as well as being able to detect
inhomogeneous magnetic or superconducting order \cite{savi02}.  Related
information can be obtained by NMR and
NQR techniques, where the latter also
provides information on the distribution of local electric-field
gradients.  Some of the first evidence for spatial inhomogeneity of
magnetism in lightly-doped {\LSCO}, consistent with stripe-like behavior,
was obtained in a La NQR experiment by \textcite{cho92}.  Later
experiments \cite{tou93,goto94,ohsu94,goto97} (including La and Cu NMR and
NQR) provided evidence for local magnetic order near
$x=1/8$ in {\LSCO} and {\LBCO}.

The work by Imai's group \cite{hunt99} suggesting that local stripe order
in variants of {\LSCO} could be detected through the ``wipeout effect'' of
Cu NQR has motivated a considerable number of further
studies using Cu and La NMR and
NQR \cite{sing99,juli99,suh99,suh00,curr00,teit00,juli01,hunt01,
teit01,simo02}.
Although there has been some controversy over details of the
interpretation (and in certain cases, there may be differences between
samples), it is now generally agreed that NQR and NMR are sensitive to
the onset of slow spin fluctuations whose appearance tends to correlate
with the onset of local charge stripe pinning as determined by
diffraction.  In particular, the glassy nature of the ordering has been
investigated.  In the case of Eu-doped {\LSCO}, related information has
been obtained by electron-spin resonance detected from a very low density
of Gd impurities \cite{kata97,kata98}. A recent Cu NQR and Y
NMR study \cite{sing02b} has shown that similar signatures are observed in
Ca-doped YBa$_2$Cu$_3$O$_6$, consistent with the $\mu$SR study of
\cite{nied98}, and suggestive of the presence of
pinned stripes in that system at low temperatures.  Direct evidence for
local spatial inhomogeneities in {\LSCO} has been obtained from studies of
NMR line broadening \cite{haas00} and frequency-dependent NQR relaxation
rates \cite{sing02a}.

Very recently, a fascinating study \cite{haas03} of the NMR/NQR
spectra in {\YBCO} has found evidence of two distinct planar O
environments in the unit cell, but only one Cu environment.  This finding
is not consistent with any form of translation-symmetry breaking.  It is,
however, suggestive of substantial nematic order, in which the O sites
midway between two copper sites in one direction (for instance, the
chain direction) has substantially larger hole-density than the O's on the
bonds in the perpendicular direction.  Of course, because the crystal is
orthorhombic, this does not truly imply any symmetry breaking, but it should 
be possible, in principle, to establish whether
the magnitude of the effect is out of proportion with the small
distortions produced by the orthorhombicity.

Several recent NMR studies \cite{curr00b,mitr01,mitr02,kaku02} have
exploited the magnetic-field dependence of the technique to probe the
spatial variation of nuclear spin-lattice relaxation rates in the vortex
lattice state of cuprate superconductors. Vortex
cores are regions  of  suppressed superconductivity, so if there is a
competing spin-stripe ordered phase, it will be enhanced and
pinned \cite{deml01,zhan01,kive02} in the neighborhood of the vortex core.
Indeed, $^{17}$O NMR measurements on {\YBCO} and YBa$_2$Cu$_4$O$_8$
indicate antiferromagnetic-like spin fluctuations and a reduced density
of states associated with the vortex cores \cite{mitr02,kaku02}.  (Similar
results, along with evidence of static spin ordering in the vortex cores
at low temperatures, have been reported \cite{kaku02b} in
Tl$_2$Ba$_2$CuO$_{6+\delta}$.) A $\mu$SR study of the details of the
magnetic-field distribution in YBa$_2$Cu$_3$O$_{6.50}$ suggests the
presence of a small local hyperfine field near vortex cores \cite{mill02}.
For completeness, we note that while one neutron scattering study has
indicated very weak, field-induced, elastic antiferromagnetic scattering
in superconducting {\YBCO} \cite{vakn00}, two other studies, focusing on
the effect of a magnetic field on inelastic scattering, found no low
frequency enhancement \cite{bour97b,dai00}.

It has been demonstrated that stripe order has an impact on phonon heat
transport \cite{hess99}.  The thermal conductivity increases
slightly on cooling through the charge-ordering transition, exhibiting a
normal peak at $\sim25$~K.  The suppression of the latter peak in
superconducting {\LSCO} has been attributed to the scattering of phonons
by fluctuating stripes \cite{babe98}.  The doping dependence of
the thermal conductivity measured in {\YBCO} and in mercury-cuprates has
been interpreted as evidence for a 1/8 anomaly in those
materials \cite{cohn99}.  An enhancement of the thermal conductivity in
La$_{1.88}$Ba$_{0.12}$CuO$_4$ was originally noted by \textcite{sera90}

Indirect evidence of some form of local charge order can also be gleaned
from the response of a system to electro-magnetic
radiation.\footnote{For a discussion of the electromagnetic signatures of
CDW order, see \protect\cite{grun94}.}  For instance,  Raman
scattering \cite{blum02} was recently used to reveal charge ordering, and
more importantly (for present purposes) to detect the effect of
collective CDW motion at higher temperatures in
Sr$_{12}$Cu$_{24}$O$_{41}$, a ladder system with a local electronic
structure very similar to that of the cuprate superconductors.  Careful
measurements of the optical response of optimally doped {\YBCO}  have
revealed \cite{home00,bern01} infrared active phonons  with  oscillator
strength comparable to those in the undoped insulating compound.  This
has been plausibly interpreted as giving evidence of ``fluctuating charge
inhomogeneities'' in the copper-oxide planes - the point being that for
the  phonons to be unscreened at that frequency, their local environment
on some scale must be insulating.

ARPES experiments can also be interpreted as giving indirect evidence of
local stripe order.  At an empirical level, this can be done by
looking \cite{zhou99,zhou01} at the evolution of the ARPES spectra in
a sequence of materials in the {\LCO} family in which various types of
stripe order have been detected;  this provides a basis for identifying
similar features in the ARPES spectrum of materials in which stripe order
has not been established.  For instance, stripe long-range order produces
a tendency for the Fermi surface in the antinodal region of the Brillouin
zone to be flat (one-dimensional), and to suppress spectral weight in the
nodal region.  Some of these same features, especially the differing
impact on the nodal and antinodal regions, are induced by Zn doping of
{\BSCCO}, as well, plausibly reflecting \cite{whit99} the tendency of Zn
to pin local stripe order.  More broadly, many of the most striking
features of the observed
\cite{fedo99,vall99,camp03,dama02,ino99} ARPES spectra can be
naturally interpreted in terms of an underlying quasi one-dimensional
electronic structure \cite{carl03}. Examples of this are the disappearance
of the the coherent quasiparticle peak \cite{carl01} upon heating above
T$_c$, the dichotomy between the widths of the peaks as a function of
momentum and energy in the normal state spectrum \cite{orga01,gweo01}, and
the general structure of the spectra and  the evolution of the Fermi
surface with
doping \cite{salk96,gran02,gran01,zach00b,zach01,ichi99b,erol01}.

Finally, several probes have been used to search for the
inhomogeneous distribution
of bond-lengths expected from the lattice response to local charge
stripe order.
Pair-distribution-function (PDF) analysis of scattering data and extended
X-ray absorption fine structure (EXAFS) spectroscopy can
detect instantaneous distributions of nearest-neighbor bond lengths.
Both techniques have the potential to provide indirect evidence of
dynamical, as well as static, stripes; however, whether they have
sufficient sensitivity to detect the very small bond-length variations
associated with charge modulation in the cuprates remains a topic of some
contention.  \cite{bozi00} reported a
doping-dependent broadening of the in-plane Cu-O bond-length distribution
in {\LSCO} detected by PDF analysis of neutron powder diffraction data.
While the reported doping dependence is intriguing, the maximum
enhancement of the bond-length spread is much too large to be compatible
with the Debye-Waller factors measured in a single-crystal neutron
diffraction experiment on La$_{1.85}$Sr$_{0.15}$CuO$_4$ by
\textcite{brad01}  A much more extreme discrepancy occurs with the EXAFS
studies of {\LSCO} (and other cuprates) by Bianconi's group
\cite{lanz96,bian96,sain97}.  They report a low-temperature splitting of
the in-plane Cu-O bond lengths of 0.08~\AA, which is not only
incompatible with a large number of diffraction studies, but is also
inconsistent with EXAFS analysis performed by other groups
\cite{niem98,hask00}. Large atomic mean-square displacements detected in
{\YBCO} by ion channeling \cite{shar00} have been attributed to
transitions associated with dynamic stripes, but the magnitude of the
displacements and the temperature dependence are difficult to reconcile
with other experiments.

In short, these various local measures of the distribution of lattice
displacements are, in principle,  a very good way to look for local
charge-ordering tendencies, but it is  important to reconcile the results
obtained with different techniques.  Where apparent inconsistencies exist,
they must be resolved before unambiguous inferences can be made.

\section{Weak and strong coupling perspectives}
\label{weak}

Order in simple metals is typically thought of as arising from a Fermi
surface instability caused by the weak residual interactions between the
lowest energy quasiparticles.  Specifically, density wave order occurs
when the band structure is such that there are nearly nested segments of
the Fermi surface.  Thus, even at temperatures above the ordering
temperature, or if the interactions are slightly weaker than the
critical strength needed for ordering, there will often be
structure in the appropriate dynamical structure factors due to
this near nesting which in some sense reflect the fact that the metal is
close to an ordered state.  In this section, we will analyze the
connection between this weak-coupling Fermi-liquid theory based
perspective and the description in terms of the collective modes of the
ordered or nearly ordered state which we have taken until now.

It is worth mentioning at the outset that one difficulty arises from
the relative paucity of established theoretical results in the
strong coupling limit.  More
than half a century of concerted work has produced a rather complete
understanding of the effects
of weak residual interactions on the properties of a well formed
Fermi liquid in a simple metal.
However, only in special circumstances is a comparably sound and
complete understanding available
in a strongly interacting system.  For instance, in the weak coupling
limit, considerable useful
information about the quasiparticle spectrum can be adduced directly
from transport data, because
Boltzmann transport theory relates the two properties in a precise
manner.  The theory of transport
in a strongly interacting system is, in our opinion, not established.
The  upshot of this is that
it is much easier to point to experiments that show that simple, weak
coupling ideas {\it cannot}
be safely applied in many highly correlated materials, than to point
to experiments that show that
strong coupling notions {\it can be}.

\subsection{Distinctions in principle}

\subsubsection{Thermodynamic distinctions}

As mentioned in the introduction, stripe-ordered states are
ultimately defined by a specific set of broken symmetries.  From this
viewpoint, there is no distinction between the strong coupling limit, in
which spin stripes can be viewed as
micro phase
separation of charges into an array of rivers with strips of nearly
insulating antiferromagnet between them, and the weak-coupling Hartree-Fock
description of a SDW which opens small gaps on various nearly nested
segments of the Fermi surface.  Indeed, in many circumstances, the two
limits are adiabatically connected as a function of the interaction
strength.  However, this is not guaranteed.
For instance, a stripe-ordered state can be either metallic or insulating.  At
$T=0$, metallic and insulating stripes are thermodynamically distinct
states of matter.

There is a more subtle possible distinction possible between metallic
spin-stripe states in the weak and strong coupling limits.  Certainly, in
the weak-coupling limit, the ungapped portions of the Fermi surface
support well-defined quasiparticles, so the system remains a Fermi
liquid unless, at still lower temperatures, it suffers an additional
ordering transition, say to a superconducting state.  On the other hand,
in the strong coupling limit, the metallic behavior in the stripe ordered
phase may or may not be well characterized by Fermi liquid power laws.  It
is now established that non Fermi liquid states can, in principle, exist
in more than one dimension.\footnote{   In particular
the existence of a non Fermi liquid ``sliding" phase in a stripe ordered
system has been shown\cite{emer00,vish01} for a class of interacting Luttinger
liquids.}

The character of charge-stripe order tends to more easily distinguish
between strong and weak coupling limits.  For repulsive interactions,
weak coupling Hartree-Fock theories always produce charge order which is
parasitic on the fundamental spin order.  In Landau theory, there is a
cubic coupling between CDW and SDW order which, whenever the primary
order parameter is the SDW order, leads to CDW order which onsets at the
same $T_c$, but with a strength proportional to the square of the SDW
order:  $\langle\rho_{2\bf Q}\rangle \propto |{\bf S}_{\bf Q}|^2 \sim
|T_c - T|^{2\beta}$.  By contrast, in a strong coupling picture of
micro-phase separation, if the spin-ordering is  only triggered when the
holes agglomerate into stripes, one would typically expect the stripe
ordering transition to be first order, or for the charge ordering to
precede spin ordering.  Thus, the measured sequence of transitions
provide sharp criteria that can be
used to discriminate
between  weak and strong coupling.

In the thermal or quantum disordered phase proximate to a stripe ordered
state, sharp thermodynamic distinctions between the strong and weak
coupling limits are harder to draw.  The only possible
thermodynamic distinction is, again,  that in the weak-coupling limit the
disordered state is necessarily a Fermi liquid, while in strong coupling
it may either be a Fermi liquid or a non Fermi liquid.

\subsubsection{Quantitative distinctions}

Even where no sharp thermodynamic distinctions exist, there are very
many clear physical differences between the weak and strong coupling
limits.  Of these, the most obvious is the relative importance of
electron quasiparticles with a sharply defined Fermi
surface.  In the weak coupling limit, the quasiparticle is the
essential building block in terms of which all other properties are
derived.  In the strong coupling limit, by contrast, even where it might
be the case that well-defined quasiparticles (with very small weight,
$Z$) are recovered in the strict
$T\to 0$ limit,  such
coherence may  require such low temperatures as to be only of
academic interest.  Thus, on a
practical level, ``effective non-Fermi liquids,'' in which the
majority of the electronic
excitations in the experimentally relevant ranges of temperature,
energy, and wave-number are not
well defined quasiparticles, cannot be sensibly treated in terms of a
weak coupling
approach.  Conversely, where well defined quasiparticles are
everywhere manifest, there is an a priori reason to prefer a weak coupling
viewpoint.

Another set of important of quantitative distinctions concern the
${\bf k}$ and $\omega$
dependences of the magnetic dynamical structure factor, $S({\bf
k},\omega)$.   In a one band-model
of non-interacting electrons, the integral of $S$ over all ${\bf k}$
and $\omega$ is
$(1/8)[1-x^2]$ while in the large interaction (local moment) limit it is
$(1/4)(1-x)$, which is not dramatically different.  However, in
weak coupling, the integrated
intensity comes from a very broad range of energies, of order $E_F$,
and momenta, of order $2k_\F$,
while for an antiferromagnet the integral is dominated by energies
less than the exchange energy,
$J$, and is highly peaked near the magnetic ordering vector, whether
or not the system is
actually ordered.  Put another way, in weak coupling the structure
factor is dominated by the
particle hole continuum, and there are signatures of well defined
collective modes (spin waves)
only
at energies $\omega \lesssim T_c$,
while in the strong coupling
limit, deep in the ordered phase, spin-waves dominate the structure
factor even at
energies of order $J$.  Also, in the ordered state at weak coupling,
the ordered moment $m$ at
$T=0$ is small, $m \sim
\mu_B (T_c/E_F)$, while in the strong coupling limit, not too close
to a quantum critical point,
$m \sim \mu_B$ is large.

As a concrete example, let us first consider the structure factor
of a non-interacting electron gas in 2D. For momenta  $q > 2 k_\F$ it is
\begin{equation}
S(q,\omega)=\frac{k_F^2}{2\pi^2 v_F q}
\sqrt{1-(\frac{q}{2k_F}-\frac{\omega}{v_F q})^2}
\end{equation}
when $v_F q (\frac{q}{2k_\F}-1)< \omega < v_F q (\frac{q}{2k_\F}+1)$ and zero
otherwise. (The expression for $q < 2k_\F$ is somewhat more complicated.)
Clearly, the spectral weight is
distributed smoothly over the entire support of this particle-hole 
continuum, {\it i.e.\/}, over a range of
momenta of order $2k_\F$ and of energies  of order the band width. From 
the weak coupling perspective
the chief qualitative effect of interactions is to
produce (quasi)bound states that manifest themselves as peaks at
or below threshold of the continuum. In the broken-symmetry phase
these bound states are Goldstone modes, which descend down to zero
energy and loose weight to  the new Bragg peaks. What is important to note here
is that even in the broken-symmetry phase, most of the spectral 
weight remains in the incoherent
background spread more or less uniformly over energies of order
band width.

In contrast, consider the zero temperature structure factor of an
insulating antiferromagnet computed to first order in a $1/S$
expansion, where $S$ is formally the magnitude of the spin, but
more physically should be viewed as the ``distance" of the system
from the nearest quantum disordered phase.  Here, on a square lattice,
\begin{eqnarray}
S_{ij}(\bf q,\omega) &&\!\!\!\! \!\! =S(S-A) \; \hat{m}_{i} \; \hat{m}_j \; 
\delta(\omega) \; \delta({\bf q} - {\bf Q}_{\rm AF}) 
\\
  +\frac{S}{8J} && \!\!\!\!  \!\!\!
 \sqrt{\frac{2-\cos q_x -\cos q_y}{2+\cos q_x+ \cos q_y}} 
\left(\delta_{ij}-\hat{m}_i \hat{m}_j\right)
\;
\delta(\omega-\omega_{\bf q})
\nonumber 
\end{eqnarray}
where $\omega_{\bf k}=2SJ \sqrt{4-(\cos k_x +\cos k_y)^2}$ is the spin wave 
dispersion (which depends
on the lattice and the details of the exchange interactions), $\hat m$ is a
unit vector parallel to the magnetization, and $A=0.3932+\ldots$.  
(Higher order terms in $1/S$ typically produce finite lifetimes for all
but the lowest energy spin waves, and also give rise to a
multi spin wave continuum.)  As promised, the
basic energy scale is set by $J$, and a large portion of the total
spectral weight is found in a small region of ${\bf k}$ space
around ${\bf k}= {\bf \pi}$.

Note that the energy scale $J$ can  be deduced from the value
of the spin wave energy near the zone boundary, $\omega_{{\bf
k}_0}=4SJ=2J$, for ${\bf k}_0=(\pi,0)$. Even in the thermally
disordered state, so long as $T \ll J$, the zone edge spin wave can
be studied to get an idea of the relevant energy scales for
magnetic excitations.

Finally, distinctions can be drawn based on the sensitivity of the
various signatures of order or
proximate order to various weak perturbations:
weak disorder, as we have discussed extensively above, can
serve to pin fluctuating
order, and hence enhance the small $\omega$ portion of $S(\bf q,\omega)$.  
However,
in weak coupling, the
principal effect of weak disorder is to scatter the quasiparticles,
and thus broaden any Fermi
surface features ({\it i.e.\/}, nested segments) which might give rise
to peaks at particular ${\bf
q}$'s in $S(\bf q,\omega)$.  Thus, in weak coupling, we expect weak quenched
disorder to  suppress, rather than
enhance low frequency signatures of stripe order or incipient stripe order.
Distinct responses to an applied magnetic field can also be expected 
to differentiate the weak and
strong coupling limits.  In particular, where the zero-field 
ground state is superconducting, induced
order in vortex cores occurs under broad circumstances in the 
strong-coupling limit where there are
competing orders;  this sort of phenomena is generally much less 
prominent in the weak coupling limit.
These issues are discussed in a number of good recent reviews - see 
footnotes 23 and 24.

\subsubsection{Intermediate coupling}

It is worth remembering that in many materials, the
interaction strength is comparable to the Fermi energy, {\it i.e.\/},
stability of matter (or some  memory of the Virial theorem) conspires to
place systems in the awkward regime of intermediate coupling.  Here,
neither the weak nor strong coupling approaches are well justified, and
one is typically forced to extrapolate results beyond their regime of
validity.  In this case, it is generally sensible to study the problem
from both the strong and weak coupling perspectives. Some features (such
as those that are most sensitive to the presence of a particle-hole
continuum) may be best viewed from the weak coupling perspective, while
others (such as the role of collective modes) may be better viewed from
strong coupling.

One traditional way to understanding this regime is to include the
effects of interactions in the context of the random phase approximation
(RPA).  For the complex spin susceptibility $\chi$, one writes
\begin{equation}
   \chi({\bf k},\omega) = 
\frac{\chi_0({\bf k},\omega)}
     {1-U_{\bf k}\chi_0({\bf k},\omega)},
\end{equation}
where $\chi_0$ is the susceptibility for noninteracting electrons, and
$U$ characterizes the interaction.  In the
case of dominantly forward-scattering interactions (such as the long-range
piece of the Coulomb interaction), the RPA summation of bubble diagrams is
justified even when the interactions are relatively strong.  For the case
of short-range interactions, no such systematic justification exists to
the best of our knowledge.  However, the results obtained from RPA are
explicit and often intuitively appealing.  It is certainly reasonable, at
least on a phenomenological level, to compare the results of an RPA
treatment with experiment to determine whether the physics is simply
connected to the weakly-interacting limit, or on the other hand whether
the interactions produce qualitatively new phenomena.

One of the first RPA calculations was performed by \textcite{bulu90} to
model the antiferromagnetic spin fluctuations detected by NMR; for the
interaction, they selected $U_{\bf k}=U$, the on-site Coulomb repulsion.
\textcite{litt93} found significant structure in the bare
$\chi_0"({\bf k},\omega)$ from nearly-nested features of the Fermi
surface.  By suitable choice of the chemical potential, and inclusion of
band narrowing by a factor of 2--4 compared to band structure
calculations, they were able to find reasonable agreement with the early
inelastic neutron scattering measurements of low-energy magnetic
excitations in \LSCO\ \cite{cheo91}.  Calculations for \YBCO\
\cite{si93,mont93} have typically found it necessary to employ the RPA
expression with
$U_{\bf k} = -J[\cos(k_xa)+\cos(k_ya)]$ (or a similar form) in
order to obtain reasonable consistency with
experiment.\footnote{\textcite{ande97} has made a strong and compelling
argument that antiferromagnetic exchange cannot be properly treated in
RPA, and that the approach is internally inconsistent.}  Several recent
calculations \cite{brin99,norm00,onuf02} have addressed the resonance peak
in \YBCO\ and the downward-dispersing excitations that appear at slightly
lower energies \cite{bour00,arai99,ito02}.

In addition, we note that the lack of dispersion of the spin gap that
develops below $T_c$ in \LSCO \cite{lake99}, together with the
enhancement of intensity above the gap \cite{maso96}, have been
interpreted in terms of the spin response of the electrons that
participate in the superconducting state \cite{lake99,maso96,morr00}.

\subsection{Experimental evidence supporting a strong-coupling perspective in
the cuprates}

{From} a purely theoretical
perspective, it may not be possible to determine {\it a priori} whether
there is an advantage to either a weak- or strong-coupling approach;
however, we believe that numerous experimental results point decisively
towards the strong-coupling picture.  Many of the relevant results involve
measurements of magnetic correlations, and we have collected selected
experimental quantities in Table~\ref{tab:mag}.  (Some explanations of
how the parameter values are taken from experiment are given in App.~B.)
Below we highlight some of the important observations, and explain their
significance.

\newlength{\blanksp}
\settowidth{\blanksp}{0}
\newfont{\mathvs}{cmex10 scaled 1000}

\begin{table*}[t]
\caption{Various quantities characterizing the
strength of antiferromagnetic correlations in doped and undoped cuprates
and nickelates.  All are determined at low temperature.  The quantity $p$
is the nominal hole concentration per planar Cu; for \YBCO\ and
YBa$_2$Cu$_4$O$_8$ this was estimated from the ``universal'' formula
relating $T_c/T_c^{\rm max}$ and $p$ from
\textcite{pres91}.  The acronyms used under ``phase'' are: AFI =
antiferromagnetic insulator, CSO = charge-stripe order, SSO = spin stripe
order, SC = superconducting order. $m$ is the magnetic moment per planar
Cu, and $m/m_0$ is the moment relative to the AFI parent.
{\protect\raisebox{2ex}{\mathvs\symbol{82}}} $S({\bf k},\omega)$ is the
dynamical structure factor, integrated over a Brillouin zone and over
energy up to 100~meV, and normalized to the AFI parent.  $J_{\rm eff}$ is
the effective super-exchange energy characterizing the maximum energy of
the magnetic excitations ($2J_{\rm eff}$) measured by neutron scattering,
or obtained from Raman scattering assuming that the two-magnon peak is at
$2.7J_{\rm eff}$.  Further discussion is given in Appendix B.}
\begin{ruledtabular}
\begin{tabular}{ldclllll}
  material & \multicolumn{1}{c}{p} & phase & \multicolumn{1}{c}{$m$} &
\multicolumn{1}{c}{$m/m_0$} &
\multicolumn{1}{c}{{\protect\raisebox{2ex}{\mathvs\symbol{82}}}
$S({\bf k},\omega)$} & \multicolumn{1}{c}{$J_{\rm eff}$} &
\multicolumn{1}{c}{$T_{\rm crit}$\quad\quad} \\
   & & & \multicolumn{1}{c}{$(\mu_{\rm B})$} & & &
\multicolumn{1}{c}{(meV)} & \\
\colrule
  \LCO & 0 & AFI & 0.60(5)\footnote{From neutron diffraction
   \cite{yama87}.} & 1 & 1 &
   152(6)\footnote{From inelastic neutron scattering by \cite{cold01}.}&
   $T_{\rm N}\approx300$~K \\
  La$_2$Cu$_{0.8}$(Zn,Mg)$_{0.2}$CuO$_4$ & 0 & AFI &  &
0.72(5)\footnote{From neutron diffraction \cite{vajk02}.} &  &  &
   $T_{\rm N}\approx100$~K \\
  La$_{1.88}$Ba$_{0.12}$CuO$_4$ & 0.12 & CSO, SSO & &
0.60(2)\footnote{Local moment relative to \LCO\ from $\mu$SR
\cite{nach98}.}\setcounter{savefoot}{\value{footnote}}  & & &
$T_{\rm sso}\approx 35$~K \\
  La$_{1.48}$Nd$_{0.4}$Sr$_{0.12}$CuO$_4$ & 0.12 & CSO, SSO &   &
0.60(2)$^d$
&  &120(25)\footnote{From two-magnon Raman scattering, assuming that the
observed peak is at $2.7J$ \cite{nach02,goza03}.} &
$T_{\rm sso}\approx50$~K \\
  La$_2$CuO$_{4.11}$ & 0.16 & SSO, SC &  & 0.62(2)\footnote{Local moment
relative to \LCO\ from $\mu$SR \cite{savi02}.} & & &
   $T_{\rm sso} = T_c = 42$~K \\
  La$_{1.86}$Sr$_{0.14}$CuO$_4$ & 0.14 & SC & 0 & 0 & 0.4\footnote{From
inelastic neutron scattering \cite{hayd96a}.} & 140(10)$^g$ &
   $T_c=35$~K \\

\null \\

  YBa$_2$Cu$_3$O$_{6.1}$ & 0 & AFI & 0.55(3)\footnote{From neutron
diffraction \cite{casa94}.} & 1 & 1 & 125(8)\footnote{From inelastic
neutron scattering \cite{hayd96b} and two-magnon Raman scattering
\cite{blum96,blum94}.} &
$T_{\rm N}\approx 400$~K
\\
  YBa$_2$Cu$_3$O$_{6.5}$ & 0.09 & SC & 0 & 0 & 0.5\footnote{Based on
inelastic neutron scattering results \cite{fong00,hayd96b}.} &   &
   $T_c\approx52$~K \\
  YBa$_2$Cu$_3$O$_{6.6}$ & 0.10 & SC & 0 & 0 & 0.6\footnote{Based on
inelastic neutron scattering results \cite{dai99,hayd96b}.}  &
110(15)$^k$ & $T_c=63$~K \\
  YBa$_2$Cu$_3$O$_{6.7}$ & 0.11 & SC & 0 & 0 & 0.3\footnote{Based on
inelastic neutron scattering results \cite{fong00,hayd96b}.} &
125(5)\footnote{From two-magnon Raman scattering \cite{blum96,blum94}.}  &
$T_c\approx70$~K \\

\null \\

  YBa$_2$Cu$_4$O$_8$ & 0.13 & SC & 0 & 0 &  & 125(5)$^m$  & $T_c=80$~K \\

\null \\

  La$_2$NiO$_4$ & 0  & AFI & 1.0(1)\footnote{From neutron
   diffraction \cite{wang92}.} & 1 &   &
\hspace{\blanksp}31(1)\footnote{From inelastic neutron scattering
\cite{yama91}.} & $T_{\rm N}\approx330$~K\\
  La$_2$NiO$_{4.133}$ & 0.27 & CO & 0.8(1)\footnote{From neutron
diffraction \cite{tran95b}.} & 0.80(5) &  &  &
$T_{\rm sso}=110$~K \\
  La$_{2-x}$Sr$_x$NiO$_4$ & \sim0.33 & CO &   &
1.11(5)\footnote{Local moment relative to La$_2$NiO$_4$ from $\mu$SR
\cite{jest99}.} &   &
  \hspace{\blanksp}20(1)\footnote{From inelastic neutron scattering
\cite{bour03,boot03}.} &
$T_{\rm sso}\approx 200$~K \\

\end{tabular}
\end{ruledtabular}
\label{tab:mag}
\end{table*}

\begin{itemize}

\item Magnetism in undoped cuprates is quantitatively consistent with
super-exchange between local magnetic moments on copper ions.

\end{itemize}

The undoped parent compounds, such as La$_2$CuO$_4$ and
YBa$_2$Cu$_3$O$_6$ are antiferromagnetic insulators with a
charge-transfer gap of $\sim2$~eV \cite{kast98}.  The ordered magnetic
moments are consistent with one unpaired spin per planar Cu ion, after
zero-point spin fluctuations, given accurately by spin-wave theory
\cite{igar92}, are taken into account.  The effective super-exchange
energy, $J_{\rm eff}\sim0.1$~eV, determined from measurements of the
spin-wave dispersion, has been calculated
from {\it ab
initio} cluster models \cite{oost96,muno00}, consistent with the local
super-exchange mechanism.

If the antiferromagnetic order corresponded to a spin-density wave due to
Fermi-surface nesting, one would expect the magnetic correlations and the
optical gap to disappear at the N\'eel temperature.  Neither of these
things happens; these materials remain correlated insulators in the
disordered state, and the spin correlations are well described by a
nonlinear sigma model \cite{chak89} with parameters derived from the
ordered state
\cite{kast98}.\footnote{The recent observation by
\protect\textcite{cold01} that the spin-wave dispersion at large energies
is better fit if a four-spin exchange interaction is included in the
microscopic Heisenberg model is still consistent with a system of 
localized spins.}

\begin{itemize}

\item With light doping, local magnetic moments change little, even
though long-range N\'eel order is destroyed.

\end{itemize}

In a $\mu$SR study of lightly-doped \LSCO\ and
Y$_{1-x}$Ca$_x$Ba$_2$Cu$_3$O$_6$, \textcite{nied98} showed that there is
essentially no change in the low-temperature ordered moment per planar Cu
as the long-range antiferromagnet order is destroyed by increasing the
hole concentration.  This behavior has also been detected in NMR
studies \cite{chou93a}.  In the ``spin-glass'' regime that occurs in the
doping range between the antiferromagnetic and superconducting phases,
the ordered moments decrease gradually \cite{nied98}.  We now know that
for \LSCO\ with $0.02\lesssim x<0.06$ the low-temperature phase involves
ordering of magnetic moments in a diagonal spin-stripe structure
\cite{waki99,fuji02}.  That identical behavior occurs in
Y$_{1-x}$Ca$_x$Ba$_2$Cu$_3$O$_6$ is supported by an NQR/NMR study
\cite{sing02b}.

\begin{itemize}

\item The energy scale of magnetic excitations and the strength of the
dynamic structure factor at frequencies less than $J$ change only 
modestly with doping.

\end{itemize}

One can see from Table~\ref{tab:mag} that $J_{\rm eff}$ decreases a
relatively small amount with doping. There is also evidence in some
underdoped \YBCO\ samples for effects reminiscent of spin-wave-like
dispersion at energies above 50~meV \cite{bour97,fong00}.  The
partially-integrated dynamical structure factor decreases with doping,
but remains substantial.

If the magnetic excitations corresponded to
electron-hole excitations across the Fermi surface, one would expect that
the magnetic energy scale should increase to a value comparable to the
Fermi energy.  For that scale to be the same as $J_{\rm eff}$ of the
undoped phase would seem to be an incredible coincidence.  With a
substantial increase in energy scale, one would expect a corresponding
decrease in the integrated dynamical structure factor.

Similar trends in terms of a modest reduction of  $J_{\rm eff}$ with 
doping and the survival
of spin-wave-like excitations are found in stripe-ordered
La$_{2-x}$Sr$_x$NiO$_4$ \cite{bour03,boot03}.  This system is definitely
in the strong-coupling limit, as it remains semiconducting above the
charge-ordering temperature
\cite{kats96}.

\begin{itemize}

\item The {\bf k} dependence of the magnetic neutron scattering measured
from superconducting samples of \YBCO\ is consistent with spin density on
planar Cu ions.

\end{itemize}

The magnetic scattering cross section is proportional to the square of
the magnetic form factor (Fourier transform of the magnetization
density); this is true for both elastic and inelastic scattering.  An
early analysis of inelastic magnetic scattering at different {\bf k}
points in superconducting \YBCO\ samples revealed an anisotropy
\cite{ross92} that was later shown to be consistent with the anisotropy
due to the magnetic form factor for a single spin in a $3d_{x2-y2}$
orbital \cite{sham93}.

The coherence of spin excitations within the CuO$_2$ bilayers of \YBCO\
also results in a bilayer structure factor.  The structure factor is a
sinusoidal function of momentum transfer perpendicular to the bilayers,
with a period that is proportional to the spacing between the layers of
the centers of spin density.  The observed period is quantitatively
consistent with the intra bilayer spacing of the Cu atoms, but is
incompatible with the spacing of the oxygen atoms, which is
significantly smaller \cite{tran92}.

For electronic states at the Fermi level, there is a substantial amount
of weight from planar O $2p$ states \cite{taka88}.   As a result, one
would expect that magnetic scattering due to electron-hole excitations
should, in real space, have significant weight associated with the oxygen
sites.  The measured modulations \cite{tran92,fong96,bour96} are not
compatible with a significant component corresponding to the oxygen
spacing.

\begin{itemize}

\item When stripe order is observed in more heavily doped samples, 
the magnetic moments are large.

\end{itemize}

As indicated in the table, the local magnetic moments detected in
stripe-ordered phases by $\mu$SR are comparable to the moment found in
undoped La$_2$CuO$_4$.  The large moments imply a strong modulation of
hole density.  The magnetic form factor determined by neutron diffraction
is consistent with that expected for spin moments \cite{tran96b}.

\begin{itemize}

\item When stripe order is observed, charge orders before the spins, and
the ordering wave vector grows with doping.

\end{itemize}

When charge and spin-stripe order are observed, as in \LNSCO, the charge
order appears at a higher temperature than the stripe order
\cite{tran96b,ichi00}.  (In La$_{1.875}$(Ba,Sr)$_{0.125}$CuO$_4$, the
ordering temperatures for charge and spin are very close \cite{fuji02};
nevertheless, the charge order parameter grows more rapidly than the spin
order parameter.)  It follows \cite{zach98} that the charge order is not
driven by the spin ordering.

This result is natural for a strong-coupling picture of stripe
correlations, but it poses a considerable challenge for the weak-coupling
approach.  In the latter, if one attributes the spin-stripe order to a
nesting instability of the Fermi surface, then one must find a distinct
nesting feature, with a spanning wave vector corresponding to
${\bf Q}_{\rm ch}$, to explain the charge-stripe order.  Of course it is
also necessary to explain the doping dependence of the ordering wave
vectors.  According to ARPES studies, the spanning wave vector near
${\bf k}=(2\pi/a)(\frac12,0)$ is close to ${\bf Q}_{\rm ch}$.
However, its variation with doping is opposite to that of
${\bf Q}_{\rm ch}$ \cite{ding97,ino02};  with increasing $x$, the 
hole-like Fermi-surface approaches
closer to the $(\pi,0)$ point, so the spanning wave 
vector decreases, while ${\bf Q}_{\rm
ch}$ increases.  We are not aware of a plausible explanation for the 
combined charge and spin modulations
from the weak-coupling perspective.

\begin{itemize}

\item Substituting Zn impurities into the CuO$_2$ planes
pins stripe
order in \LSCO.

\end{itemize}

The substitution of Zn into the CuO$_2$ planes causes a significant
increase in normal-state resistivity without modifying the carrier
density \cite{fuku96}.  It also wipes out the dispersive feature near the
nodal point observed by ARPES in the normal state \cite{whit99}.  We have
already discussed the fact that Zn-doping can induce static stripe order
in \LSCO\ (see Fig.~\ref{fg:lsczo}).  These results are incompatible with
a mechanism for stripe order based on Fermi-surface nesting.  The Zn
impurities break translational symmetry and cause considerable scattering
of the charge carriers.  Any sharp features at the Fermi surface are
smeared out in the presence of the Zn.  Thus the Zn should destroy, not
induce, a nesting instability.

\begin{itemize}

\item There is no evidence of well-defined quasiparticles in the normal
state of under- and optimally-doped cuprates.

\end{itemize}

There are many features of the doped system which suggest that, for the
most part, there are no well defined quasiparticles.  ARPES spectra in
\BSCCO\ and \LSCO\ in some cases do exhibit features with a well defined
dispersion, and by looking at the spectral intensity integrated over a
small energy window about the Fermi energy, a Fermi surface of sorts is
observed \cite{dess93,ding96b,ino02}.  However, in no case has a peak been
observed with a width small compared to its mean, which we 
take to be the
definition of a well defined quasiparticle.  In some ranges of
temperature and ${\bf k}$, there is no well defined quasiparticle peak,
at all.  Near the nodal points, there are marginally defined
quasiparticle, in the sense that there is certainly a clear peak in the
spectral function, but its width is approximately twice its mean energy
\cite{vall99}.  There is corroborating evidence that any quasiparticles
are at best ``marginal" which comes from the
$T$-linear dependence of the normal state resistivity, and various other
indirect (but bulk) measurements \cite{ande87,ande92,varm89,palee99}.

In addition, there is considerable indirect evidence that the familiar
Fermi liquid power laws are strikingly absent.  NMR experiments reveal
significant temperature dependence to the nuclear
$1/T_1T$ and dramatic violations of the Korringa relation, not only in
the pseudogap regime, but even in the high temperature regime at optimal
doping \cite{taki91}.  The optical conductivity has a clearly non-Drude
form \cite{coll89}, above and beyond the peculiar temperature dependence
of the DC conductivity.  The Hall number, as well, is anomalously
temperature dependent \cite{wang87}.

All together, these results strongly imply that a picture of weakly
interacting quasiparticles has limited validity in the cuprates.  The
one exception to this concerns the low-energy behavior ($E\ll\Delta_0$)
of the cleanest superconducting materials, deep in the superconducting
state.  Here, considerable indirect evidence exists that there are
remarkably long lived nodal quasiparticles which dominate the physics
\cite{suth03}.  However, direct evidence of these sharply defined
excitations has yet to emerge in any single-particle experiment such as
ARPES or STM.

\section{Conclusions}
\label{conclusion}

There are many  reasons why identifying and studying fluctuations
associated with the presence of ``nearby'' ordered  states has become one
of the main thrusts in the study of cuprate  superconductors and related
materials. To some extent, these states are interesting just because
they occur somewhere in the (multidimensional) phase diagram of these
intriguing materials.  Quantum critical points associated with some of
these orders have been proposed to be the  explanation of the notoriously
peculiar high-temperature (``normal state")  behavior observed in many
experiments.  The  possibility that certain local orders are inextricably
linked with the phenomenon of high-temperature superconductivity is the
most enticing reason of all.  However, before we can determine whether a
particular form of local order, such as local stripe order, could
possibly be central to the problem of high temperature superconductivity,
we need to determine whether or not it is ubiquitous in the families of
materials which exhibit high-temperature superconductivity.
Regardless of one's motivation, it is clear that the ability to identify
the proximity to an  ordered phase  by the signatures of
``fluctuating order" is  very useful. This paper has focused on
practical  considerations  pertinent to this task.

A combination of rather general scaling considerations concerning quantum
critical phenomena, and specific insights gleaned from the solvable
models studied, has lead us to articulate a number of ``lessons''
concerning the optimal way of obtaining information about nearby ordered
states from experiment.  {\bf 1) }  The information is best obtained from
the low-frequency part of the dynamic structure factor, preferably
integrated over a small, but non-zero range of frequencies, with the
scale of frequencies set by the characteristic frequency of quantum
fluctuations, $E_{\rm G}/\hbar$.  {\bf 2) } Weak disorder can make it
easier, especially for static probes, to image the local order, as it can
pin the fluctuations without greatly disrupting the intrinsic
correlations.   {\bf 3)}  Experiments which reveal strongly 
dispersing features generally give
information about the elementary excitations of the system;  however, 
distinguishing dispersing
features that arise from well defined quasiparticles
from the multiparticle continua
characteristic of quantum critical points can be exceedingly subtle.
Specifically, even where one set of experiments can be sensibly
interpreted in terms of band structure effects, care must be taken in
interpreting this as evidence of well defined electron-like
quasiparticles. {\bf  4)}  Several aspects of the $E$ and
${\bf k}$ dependence of the LDOS (measurable by STM) in the presence 
of weak disorder allow
one, in principle,  to distinguish interference effects due to the 
scattering of the elementary
excitations from impurities, from the effects of pinned incipient 
order.  Interference effects in 2D
produce peaks along {\it curves} in
${\bf k}$ space which disperse as a function of energy in a manner which
is directly related to the quasiparticle dispersion relations,
such as could be measured in
ARPES, and they may (or may not) have a strongly energy dependent phase.
Pinning of incipient order produces peaks at well defined {\it points} 
in ${\bf
k}$ space which depend only weakly on energy, and generally
have an energy independent phase.   {\bf 5)}  It will often be
true that interference effects and collective pinning will jointly produce
complicated ${\bf k}$ and $E$ dependent properties in the local density of 
states 
that arise from a
combination of both effects, especially in energy ranges in which
the interference and pinning features lie at nearby values of
${\bf k}$.

In Section IV of this paper, we applied these ideas to an analysis of the
evidence of local stripe order in a number of neutron scattering and STM
measurements on various cuprate superconductors.  The evidence of both
spin- and charge-stripe order is unambiguous in the {\LCO} family of
high-temperature superconductors.  However, there is increasingly strong
evidence of substantial local charge-stripe order, and probably nematic
order as well, in {\BSCCO} and {\YBCO}.Conversely, no evidence of incipient 
stripe order has been reported to date in the electron-doped cuprates, 
{\NCCO} and {\PCCO}.

\bigskip

{\textrm Note Added:}  After this work was completed, we received an
advanced copy of a review article by \cite{sach02b} which
has overlapping material with the present paper, and which reaches similar
conclusions concerning the effects of incipient stripe order on the 
STM spectra.

\section*{Acknowledgments}

We would like to dedicate this paper to our friend and collaborator,
Victor J. Emery, whose untimely death we mourn.
Many of the ideas discussed in this paper have been strongly
influenced by Vic's seminal contributions to this field.
We thank  K. B. Cooper, S. L. Cooper, J.C.Davis, J. P. Eisenstein, J.Hoffman,
P. Johnson, D-H.Lee,
P.B.Littlewood, A. Polkovnikov, S.Sachdev, D.J.Scalapino,  Z-X. Shen, A.
Yazdani, and X. Zhou for useful and stimulating  discussions.
This work was supported in part by the National Science Foundation
through the grants No. DMR 01-10329 (SAK at  UCLA), DMR 01-32990
(EF, at the University of  Illinois),  DMR 99-78074 (VO, at Princeton
University), by the David and Lucille
Packard Foundation (VO, at Princeton University),  and (AK and CH at
Stanford), and by the Department of Energy's Office of Science under
contracts No. DE-AC02-98CH10886  (JMT, at Brookhaven National
Laboratory),  DE-FG03-01ER45929-A000 (AK and CH at
Stanford University), and DE-FG03-00ER45798 (IPB at UCLA).

{
\appendix
\section{Luttinger liquids as one-dimensional quantum critical charge
ordered states} 
\label{sec:luttinger}

In Section \ref{1DEG} we discussed how a single impurity induces charge
order in the  simplest and best understood quantum critical system, the
Tomonaga-Luttinger Model of a one-dimensional electron gas. In this
Appendix we give a summary of the physics and of essential technical
aspects of the discussion of Section \ref{1DEG}.
In this Appendix we will only present the aspects of
the theory relevant to the derivation of the expressions used in
Section \ref{1DEG}. There are a number of excellent reviews which cover
the theory of the 1DEG at great depth and we refer the interested reader
to that literature \cite{emer79,frad91,stone94,gogo98}.

Consider a one-dimensional system of interacting spin one-half fermions
(electrons). We will denote by $\Psi_\sigma(x)$ the Fermi field for an
electron with spin $\sigma=\uparrow,\downarrow$, and by
$\psi_{\pm,\sigma}(x)$ its right and left moving components respectively:
\begin{equation}
\Psi_\sigma(x)=e^{\displaystyle{ik_\F x}}\;
\psi_{+,\sigma}(x)+e^{\displaystyle{-ik_\F x}}\; \psi_{-,\sigma}(x)
\label{eq:decomposition}
\end{equation}
where $k_\F$ is the Fermi wave vector. In what follows we will assume that
the electron density is incommensurate and ignore umklapp scattering
effects. Thus we will be working in a regime in which the dynamics of the
right and left moving components of the electron are slowly varying and
hence are well described by an effective continuum Hamiltonian density
${\cal H}={\cal H}_0+{\cal H}_{\rm int}$ where
  \begin{equation}
{\cal H}_0=-i v_\F \sum_{\sigma=\uparrow,\downarrow}
\left(\psi_{+,\sigma}^\dagger \partial_x 
\psi_{+,\sigma}-\psi_{-,\sigma}^\dagger \partial_x 
\psi_{-,\sigma}\right)
\end{equation}
is the Hamiltonian density for non-interacting electrons in the low-energy
regime where the dispersion is linearized; here $v_\F$ is the Fermi
velocity. As usual, irrelevant operators which account for corrections to
the linear dispersion are not included (see however below for caveats).
The effects of interactions are included in ${\cal H}_{\rm int}$.

The best way to describe the physics of the 1DEG is by means of
bosonization methods. Bosonization
is the statement that the low-energy spectrum of the 1DEG is exhausted by
a long lived bosonic excitation described by a field $\phi_\sigma(x)$
which represents particle-hole fluctuations of spin $\sigma$ near the
Fermi points $\pm k_\F$. It turns out that the field $\phi_\sigma$ also
describes the phase fluctuations of a $2k_\F$ CDW.
Consequently the electron density operator $\rho_\sigma(x)$ is decomposed
into a long wavelength piece $j_0^\sigma$  and a $2k_\F$ piece related to
the density wave  order parameter. The long wavelength electron density for
spin projection $\sigma$, $j_0^\sigma(x)$, is:
\begin{equation}
j_0^\sigma(x)=\psi_{+,\sigma}^\dagger
\psi_{+,\sigma}+\psi_{-,\sigma}^\dagger \psi_{-,\sigma}
\end{equation}
and the long wavelength current density with spin projection 
$\sigma$, $j_1^\sigma(x)$, is:
\begin{equation}
j_1^\sigma(x)=\psi_{+,\sigma}^\dagger
\psi_{+,\sigma}-\psi_{-,\sigma}^\dagger \psi_{-,\sigma}
\end{equation}
Here $j_0^\sigma$ and $j_1^\nu$  are operators  normal-ordered with
respect to the filled Fermi sea.

The  density and current density operators obey the equal-time 
bosonic commutation relations
\begin{equation}
\left[j_0^\sigma(x),j_1^{\nu}(y)\right]=\frac{i}{\pi}
\partial_x \delta(x-y) \delta_{\sigma,\nu}
\end{equation}
where $\sigma,\nu=\uparrow,\downarrow$. One can identify $j_0^\sigma(x)$ 
and $j_1^\sigma(x)$ with
\begin{eqnarray}
j_0^\sigma(x)&=&\frac{1}{\sqrt{\pi}} \; \partial_x \phi_\sigma(x)
\nonumber \\
j_1^\sigma(x)&=&\frac{1}{\sqrt{\pi}} \; \Pi_\sigma(x)
\label{bosonization-ccr}
\end{eqnarray}
where $\phi_\sigma(x)$ is a bose field and $\Pi_\sigma(x)$ is its 
canonically conjugate momentum; they obey the canonical equal-time 
commutation relations
\begin{equation}
\left[ \phi_\sigma(x), \Pi_\nu (y) \right]=i \delta_{\sigma,\nu} \; 
\delta(x-y).
\label{ccr}
\end{equation}
The (slowly varying) Fermi fields $\psi_{\pm,\sigma}(x)$ can themselves
be reconstructed from the bose field $\phi_\sigma(x)$:
\begin{equation}
\psi_{\pm,\sigma}(x)=\frac{1}{\sqrt{2\pi \alpha}} \;
{\cal N}_{\sigma}\; e^{\displaystyle{i \sqrt{\pi} (\pm 
\phi_\sigma(x)-\theta_\sigma(x) )}}
\label{mandelstam}
\end{equation}
(here $\alpha\sim  v_\F/D$ is a short-distance cutoff and $D$ is 
the fermion band width) where we have introduced the dual field 
$\theta_\sigma(x)$, defined by the identity
\begin{equation}
\partial_x \theta_\sigma(x)=\Pi_\sigma(x)
\label{theta}
\end{equation}
and ${\cal N}_\sigma$, the Klein factor, is an operator which 
guarantees that Fermi fields with different spin labels anti-commute 
with each other.

It is convenient to introduce the charge and spin bose fields 
$\phi_c$ and $\phi_s$,
\begin{eqnarray}
\phi_c&=&\frac{1}{\sqrt{2}}(\phi_\uparrow+\phi_\downarrow)
\nonumber \\
\phi_s&=&\frac{1}{\sqrt{2}}(\phi_\uparrow-\phi_\downarrow).
\label{c-s}
\end{eqnarray}
The Hamiltonian density of the interacting system can be written a 
sum of operators which
are marginal (with scaling dimension equal to 2), relevant (with 
scaling dimension smaller than 2) and irrelevant (with scaling
dimension larger than 2). For an incommensurate 1DEG the effective 
low-energy Hamiltonian, which contains only marginal operators,
has the spin-charge separated form
\begin{equation}
{\cal H}={\cal H}_c+{\cal H}_s
\end{equation}
where the charge Hamiltonian density ${\cal H}_c$ in bosonized form 
has the universal (Tomonaga-Luttinger) form
\begin{equation}
{\cal H}_c=\frac{v_c}{2} K_c \; \Pi_c^2+ \frac{v_c}{2K_c} 
\left(\partial_x \phi_c\right)^2
\label{Hc}
\end{equation}
where $v_c$ is the charge velocity and $K_c$ is the charge Luttinger 
parameter. T
For repulsive interactions, the spin Hamiltonian density ${\cal H}_s$ 
has the same form
\begin{equation}
{\cal H}_s=\frac{v_s}{2} K_s \; \Pi_s^2+ \frac{v_s}{2K_s} 
\left(\partial_x \phi_s\right)^2
\label{Hs}
\end{equation}
The non-universal charge and spin velocities and the Luttinger 
parameters encode the dependence of this low
energy theory on the microscopic parameters of the system. Quite 
generally,  for a system with repulsive interactions the charge
Luttinger parameter obeys the inequality $K_c<1$, and for spin
rotationally invariant  interactions the spin Luttinger parameter 
satisfies the equality $K_s=1$. Typically the charge and spin
velocities  satisfy the inequality $v_c>v_s$. In the weak coupling 
limit, and neglecting all irrelevant operators, simple expressions
for $K_c$, $v_c$ and $v_s$ in terms of the backscattering and forward 
scattering amplitudes can be written down. \footnote{At
intermediate and strong coupling the effective low-energy theory 
still has the same form but with significant renormalization of
these parameters.}

Physical observables of the 1DEG have simple expressions in the 
bosonized theory. Since the Tomonaga-Luttinger model is strictly 
quadratic in bose fields, it allows for a straightforward computation 
of the correlation functions of all observables of interest, both at 
zero and finite temperature, as well as for different types of 
boundary conditions. For the purposes of this review it will be 
sufficient to note that the electron density has the decomposition
\begin{eqnarray}
\rho(x)&=&\frac{2k_\F}{\pi}+j_0(x)
\nonumber \\
&+&e^{\displaystyle{i2k_\F x}} {\cal O}_{\rm CDW}(x)+ 
e^{\displaystyle{-i2k_\F x}}
{\cal O}_{\rm CDW}^\dagger (x) \\
&+&e^{\displaystyle{i4k_\F x}} {\cal O}_{4k_\F}(x)+ 
e^{\displaystyle{-i4k_\F x}} {\cal O}_{4k_\F}^\dagger (x)
\nonumber
\label{rho-bose}
\end{eqnarray}
where
\begin{equation}
j_0=j_0^\uparrow+j_0^\downarrow=\sqrt{\frac{2}{\pi}}\; \partial_x \phi_c
\label{j0}
\end{equation}
is the long wavelength charge density, and
\begin{eqnarray}
{\cal O}_{\rm CDW}&=&\sum_\sigma \psi_{+,\sigma}^\dagger \psi_{-,\sigma}
\\
&=& \frac{1}{\pi \alpha} \cos(\sqrt{2\pi} \phi_s) \; 
e^{\displaystyle{-i\sqrt{2\pi} \phi_c}}
\nonumber
\label{cdw}
\end{eqnarray}
is the CDW order parameter ({\it i.e.\/ }, the $2k_\F$ amplitude of 
the charge density), and ${\cal O}_{4k_\F}$ is the $4k_\F$ CDW order 
parameter (which we will not discuss here).

In particular this implies that the charge dynamical structure factor 
for wave vectors close to $2k_\F$, discussed in Section \ref{1DEG}, 
is  the (retarded) correlation function of the CDW order parameter:
\begin{equation}
S_{CDW}(x,t)=\langle {\cal O}_{\rm CDW}^\dagger (x,t) {\cal O}_{\rm 
CDW}(0,0)\rangle
\label{SCDW}
\end{equation}
whereas for small wave vectors it is given instead by the (retarded) 
density correlator
\begin{equation}
S_{0}(x,t)=\langle j_0(x,t) j_0(0,0)\rangle
\label{S0}
\end{equation}
In particular, the spectral function  for $S_{CDW}(x,t)$, which we 
will denote by ${\tilde S}_{CDW}(k,\omega)$, has the scaling form
\begin{equation}
{\tilde S}_{CDW}(k,\omega)=\frac{v_c}{16\pi^4} \left(\frac{\pi T 
\alpha}{v_c}\right)^{K_c-1} \Phi_{CDW}\left(\frac{v_s k}{\pi 
T},\frac{\omega}{\pi T}\right)
\label{CDW-spectral}
\end{equation}
where $\Phi_{CDW}(x,y)$ is the (dimensionless) scaling function~\cite{orga01}
\begin{eqnarray}
\Phi_{CDW}(x,y)\equiv&& \!\!\!\!
   \int_{-\infty}^\infty du \int_{-\infty}^\infty dv \;\;
\displaystyle{  h_{\frac{K_c}{2}} \left(\frac{u+v}{2}\right)}
\nonumber \\
  \times \;\;\;
\displaystyle{h_{\frac{K_c}{2}}^*\left(\frac{u-v}{2}\right)}
&&  \!\!\!\!
\displaystyle{h_{\frac{1}{2}}\left(-\frac{r u +v}{2}\right) \; 
h_{\frac{1}{2}}^*\left(\frac{r u -v}{2}\right)
}
\nonumber \\
&&
\label{PhiCDW}
\end{eqnarray}
Here $r=v_s/v_c<1$ and $h_\alpha(z)$ is given by ~\cite{orga01}:
\begin{equation}
h_\nu(z)=
{\rm Re}
\left[
(2i)^\nu \;
{\displaystyle{B \left(\frac{\nu-i z}{2},1-\nu \right) }}
\right]
\label{h-alpha}
\end{equation}
where $B(x,y)$ is the Euler Beta function:
\begin{equation}
B(x,y)={\displaystyle{\frac{\Gamma(x) \Gamma(y)}{\Gamma(x+y)}}}
\label{beta}
\end{equation}
and $\Gamma(z)$ is the Gamma function.

We now turn to the calculation of the tunneling density of states 
discussed in Section \ref{1DEG}.
There we presented the behavior of the $2k_\F$ component of the 
tunneling density of states $N(2k_\F+q,E)$, the  Fourier Transform in 
$x$ and $t$, of the $2k_\F$ component of the electron spectral 
function. At zero temperature and for a semi-infinite system with 
$0\leq x <L$ (with $L \to \infty$), the $2k_\F$ component of the 
fermion Green's function, $g_{2k_\F}(x,t)$, is given by 
\cite{egge00,egge96,matt97}
\begin{eqnarray}
g_{2k_\F}(x,t)&\equiv& \langle \psi_{+,\sigma}^\dagger(x,t) 
\psi_{-,\sigma}(x,0)  \rangle
\\
&\sim&
\displaystyle{ \frac{1}{2\pi} 
\frac{\alpha^{a+b-1/2}\left({2x}/{v_{\rm c}\tau}\right)^{c}}
{\left(v_{\rm s}\tau-2x\right)^{1/2}
\left(v_{\rm c}\tau-2x\right)^{a}\left(v_{\rm c}\tau+2x\right)^{b}
}}
\nonumber 
\label{eq:GF-semi}
\end{eqnarray}
where $\tau=t+i0^+$. Notice that in Eq. (\ref{eq:GF-semi}) we have 
kept the dependence on the short distance cutoff $\alpha$; consequently, 
$g_{2k_\F}(x,t)$ naively has units of $\alpha^{-1}$. However, in a 
Luttinger liquid the fermion operator has an anomalous scaling dimension. 
In for this semi-infinite geometry the scaling dimension is governed by
the exponents $a$, $b$ and
$c$, which are given by
\begin{equation}
a=\displaystyle{\frac{\left(K_c+1\right)^2}{8K_c}}\;, \quad 
b=\displaystyle{\frac{\left(K_c-1\right)^2}{8K_c}} \; ,\quad 
c=\displaystyle{\frac{1}{4}\left(\frac{1}{K_c}-K_c\right)}
\label{exponents}
\end{equation}
At finite temperature $T>0$, $g_{2k_\F}(x,t;T)$ becomes
\begin{widetext}
\begin{eqnarray}
g_{2k_\F}(x,t;T) \sim&&
{\displaystyle{
\frac{1}{2\pi \alpha}
\left(\frac{\pi T\alpha}{ v_s}\right)^{\frac{1}{2}}
\left(\frac{\pi T\alpha}{ v_c}\right)^{\frac{1}{4}(K_c+K_c^{-1})}
}}
\\
&&\times
\displaystyle{
\left(
\frac{-i}{\sinh
\left(
\frac{\pi T}{ v_s}(v_s \tau-2x)
\right)}
\right)^{\frac{1}{2}}
\left(
\frac{-i}{\sinh
\left(
\frac{\pi T}{ v_c}(v_c \tau-2x)
\right)}
\right)^{a}
\left(
\frac{-i}{\sinh
\left(
\frac{\pi T}{ v_c}(v_c \tau+2x)
\right)}
\right)^{b}
\left(
\frac
{\sinh
\left(
\frac{2\pi Tx}{ v_c}
\right)}
{\sinh
\left(
\frac{\pi T \tau}{v_c}
\right)}
\right)^{c}
}
\nonumber 
\label{g2kF-T}
\end{eqnarray}
\end{widetext}
As with the structure factor,
$N(q+2k_\F,E)$ can also be expressed in terms of
a scaling function $\Phi$ as
\begin{eqnarray}
N(q+2k_\F,E) \equiv &&\!\!\!\!
\int_0^\infty dx \int_{-\infty}^\infty dt \; 
e^{\displaystyle{-i(q x-E t)}} \; g_{2k_\F}(x,t)
 \nonumber \\
= 
\frac{ B}{2E}
\left(\frac{\alpha E}{ v_c}\right)^{2b}&&\!\! 
\Phi\left(\frac{2E}{ v_{\rm
c}q},\frac {E}{k_BT}\right) 
\end{eqnarray}
which depends also on the charge Luttinger parameter $K_c$ and on the 
ratio of the charge and spin velocities $v_c/v_s$. $B$ is the dimensionless 
quantity
\begin{equation}
B=
\displaystyle{
\frac{e^{-i\pi (c+1)/2}}{\Gamma(a+b+c-\frac{1}{2})}
}.
\label{eq:B}
\end{equation}

The behavior of $N(2k_\F+q,E;T)$ for $T>0$ is shown  in Fig.\ 
\ref{Nofk} and Fig.\ \ref{lowT} in Section \ref{1DEG}. At $T=0$ the 
scaling function $\Phi(2E/ v_c q)$ is given by
\begin{widetext}
\begin{equation}
\Phi(u)=
\displaystyle{ \sqrt{\frac{v_c}{v_s}} \;
u^{c+1} \;
\int_0^1dt \; t^c \left(1-t \right)^{a+b-\frac{3}{2}} \; \left(1-u t 
\right)^{-a} \; \left(1+u^* t \right)^{-b} \;
\left(1- \frac{v_c}{v_s} u t \right)^{-\frac{1}{2}}
}
\label{eq:R2kF-A}
\end{equation}
\end{widetext}
Here $u=2 |E|/q v_c+i0^+$. 
We have obtained analytic
expressions for the singular pieces of $N(k,E)$ at
$T=0$, for a general value of the charge Luttinger parameter in the range
of interest
$0<K_{\rm c}<1$, and general $v_{\rm c}/v_{\rm s}$.
$N(k,E)$ has power law
singularities as  $2E/ v_{\rm c} q\to \pm 1$, $2E/ v_{\rm
c} q\to 0$, and
$2E/ v_{\rm c} q\to \infty$ and $2E/ v_{\rm c} q\to v_{\rm
s}/v_{\rm c}$.

Physically the non-analyticities at $2E/ v_{\rm c} q=\pm 1$ and
$2E/ v_{\rm s}q =1$  represent the threshold of the continuum of propagating 
excitations of appropriate type
moving to the right and to the left respectively. The singularity at
$2E/ v_{\rm c} q\to \infty$ corresponds to the
$2k_\F$ CDW stabilized (pinned) by the boundary.

Close to the
right-dispersing charge-related singularity at
$q\to 2E/ v_{\rm c}$
we find
\begin{equation}
N(2k_\F+q,E) \sim   \frac{1}{2E}
\left(\frac{\alpha E}{ v_c}\right)^{2b}\!\!\!
    A_+ \left[1-\frac{2E}{ v_{\rm c}q}\right]^{b-\frac{1}{2}}
\label{eq:u+1}
\end{equation}
where $A_+$ is a finite complex coefficient determined by
the strength of the singularity. An important feature of this result
is that as $E$ goes through
this singularity at
fixed $q$, the phase of $N(2k_\F+q,E)$ jumps by $\pi (b-1/2)$.
(See Fig. \ref{lowT}.)
Close to the
right-dispersing spin-related singularity at
$q\to 2E/ v_{\rm s}$
we find
\begin{eqnarray}
N(2k_\F+q,E) \sim  &&  \\
  \frac{1}{2E}
\left(\frac{\alpha E}{ v_s}\right)^{2b} &&\!\!\!\!\!\!\!\!\! 
\left(\frac{v_s}{v_c}\right)^{a+b+c-1}
  \!\!\!\!\!   A^{\prime}_+ \left[1-\frac{2E}{ v_{\rm 
s}q}\right]^{2b-\frac{1}{2}}
\nonumber 
\label{eq:u+vs}
\end{eqnarray}
where $A'$ is another complex coefficient.
In the non-interacting limit $K_{\rm c} =1$ and $v_{\rm s}=v_{\rm c}$,
these two singularities coalesce into a simple
pole for a particle
moving to the right. However, for
$K_{\rm c}<1$, the TLL does not have quasiparticles but massless soliton
states instead. As usual, the power laws reflect the
multi-soliton continuum.  (More generally, the limit  $v_{\rm s}/v_{\rm
c}\to 1$ is somewhat subtle \cite{bind02}.)  Similarly, for
$q\to -2E/ v_{\rm c}$
\begin{equation}
N(2k_\F+q,E)  \sim \frac{1}{2E}
\left(\frac{\alpha E}{ v_c}\right)^{2b}\!\!\!
A_-\left[1+\frac{2E}{ v_{\rm c}q}\right]^{b}
\label{eq:u-1}
\end{equation}
where, once again, $A_-$ is a finite complex
coefficient specific to this singularity.
However, unlike the singularity at $E=+ v_{\rm c}q$, the
behavior near $E=- v_{\rm c}q$ although
non-analytic is not divergent since $b>0$ for repulsive interactions.

In addition to propagating excitations, we also find that
$N(2k_\F+q,E)$ has  a  singularity
associated  with the non-propagating
CDW as $q \to 0$;
\begin{equation}
N(2k_\F+q,E)  \propto   \frac{1}{2E}
\left(\frac{\alpha E}{ v_c}\right)^{2b} \; 
\left(\frac{v_s}{v_c}\right)^{1/2} \; 
\left[\frac{2E}{ v_{\rm c}q}\right]^{(1-K_{\rm c})/2}\!\!\!
\label{eq:q0}
\end{equation}
which diverges as $q \to 0$.
This singularity also exhibits a phase jump as $q \to 0^{\pm}$, equal
to $\pi (1-K_{\rm c})/2$.

At low voltages
$E\to 0$ and at fixed $q$,
we find
\begin{equation}
N(2k_\F+q,E)   \propto    \frac{ 1}{2E}
   \left(\frac{\alpha E}{ v_c}\right)^{2b} \; 
\left(\frac{v_s}{v_c}\right)^{1/2} \; \left[\frac{2E}{ v_{\rm 
c}q}\right]^{c+1}.
     \label{eq:E0}
\end{equation}

Finally, we summarize here the calculation of the induced density of 
states in the weak-impurity limit, $E>T_K$. In Section 
\ref{sec:response} we showed that the density of states at wavevector 
$k$ and voltage $E$ induced by an impurity potential with amplitude 
$V(k)$ at wave vector $k$  defines the susceptibility $\chi_{\rm 
DOS}(k,E)$ given in  Eq. (\ref{eq:chi-dos}). Here we will sketch the 
calculation of $\chi_{\rm DOS}(2k_\F+q,E)$ (to leading order in 
$V(k)$)  for a Luttinger liquid. We must first find the Green's 
function
\begin{eqnarray}
&&\!\!\!\!\!\!\!\!\!\! \!\!\!\!\!\!\!\!\!\!
G(z,_1,z_2,z_3)=
 \\
&&\displaystyle{\langle
\psi_{+,\uparrow}^\dagger(z_1) \psi_{-,\uparrow}(z_2) 
\left(\psi_{-,\uparrow}^\dagger(z_3) \psi_{+,\uparrow}(z_3)+\uparrow 
\leftrightarrow \downarrow \right)
\rangle
}
\nonumber 
\label{four}
\end{eqnarray}
which can be computed readily using bosonization methods. At $T=0$ 
and for an infinite system
the time-ordered Green's function $G(z_1,z_2,z_3)$, in imaginary time, 
is given by
\begin{eqnarray}
G(z_1,z_2,z_3)&=&
\displaystyle{
\frac{1}{2\pi^2 \alpha^2}
\left|z_1-z_2\right|^{-\frac{1-K_c^2}{4K_c}}}
\nonumber \\
&&\!\!\!\!\!\!\!\!\!\!\!\!\!\!\!\!\!\!\!\!\!\!\!\!\!\!\!\!\!\!\!\!
\times \;
\displaystyle{
\left(z_1-z_3\right)^{-\frac{K_c-1}{4}}
\left(z_2-z_3 \right)^{-\frac{K_c+1}{4}}
\left(w_1-w_3\right)^{-\frac{1}{4}}
}
\nonumber \\
&&\!\!\!\!\!\!\!\!\!\!\!\!\!\!\!\!\!\!\!\!\!\!\!\!\!\!\!\!\!\!\!\!
  \times \;
\displaystyle{
\left(\bar z_1- \bar z_3\right)^{-\frac{K_c+1}{4}}
\left( \bar z_2 - \bar z_3\right)^{-\frac{K_c-1}{4}}
\left(\bar w_2- \bar w_3\right)^{-\frac{1}{4}}
}
\nonumber \\
&& 
\label{three-FT}
\end{eqnarray}
where we have used the charge and spin complex coordinates $z=x+i v_c \tau$ 
and $w=x+ i v_s \tau$.
The susceptibility $\chi_{\rm DOS}(2k_F+q,E)$ is obtained, upon analytic 
continuation to real frequencies, from 
\begin{eqnarray}
\chi_{\rm DOS}({2k_\F+q},E) = && \!\!\!\!
{\textrm Im} \int_{-\infty}^\infty dx 
\int_{-\infty}^\infty d\tau e^{\displaystyle{-i (kx-E\tau)}} 
\nonumber \\
&& \!\!\!\!\!\!\!\!\!\!\!\! 
\times \; \int d\tau^{\prime} \; G(x,\tau; x, 0; 0, \tau')
\label{eq:chi-FT}
\end{eqnarray}
The result for the integrated density of states $\tilde N(2k_\F+q,D)$ 
presented in 
Section \ref{1DEG}, Eq.\  (\ref{eq:int-high}) follows from integrating Eq.\ 
(\ref{eq:chi-FT}) over  energies large compared with the low-energy 
cutoff $T_K$.

\section{Experimentally determined scales of magnetism in various
materials}

Here we explain some of the assumptions behind the parameter values
specified in Table~\ref{tab:mag}.  For antiferromagnetic phases, the
meaning of the parameters is relatively straightforward, but extracting
parameters from measurements on doped systems can require extrapolations
from simple models.  Note that we have not addressed important issues 
related to
possible novel forms of orbital magnetism that have been proposed
\cite{chak01b,varm97}.

\subsection{Magnitude of the ordered moments}
\subsubsection{Absolute moments from neutron scattering}

In the case of long-range antiferromagnetic order, the absolute magnitude
of the average ordered moment per planar Cu, $m$, is obtained
from an elastic neutron diffraction measurement by comparing the
intensities of the magnetic Bragg superlattice peaks with those of Bragg
peaks from the chemical lattice.  It has been observed that $m$ is
correlated with the N\'eel temperature, $T_{\rm N}$, in both
\LCO\ \cite{yama87} and \YBCO\ \cite{tran88}.  In the earliest
measurements of \LCO, samples typically contained excess oxygen,
resulting in values of $m$ and $T_{\rm N}$ significantly lower than those
of the stoichiometric material.  Another complication is that
determination of $m$ requires knowledge of the magnetic form factor,
which turns out to be considerably more anisotropic \cite{sham93} than
was appreciated initially.

For comparison, it is worth noting that for a spin $\frac12$ Heisenberg
model on a square lattice with only nearest neighbor interactions, the
magnitude of the average ordered spin per Cu, to second order in a $1/S$
expansion \cite{igar92}, is $\langle S\rangle=0.307$.  For a moment due to
spin only, $g=2$ and
$m=g\langle S\rangle\ \mu_{\rm B}=0.61\ \mu_{\rm B}$.  Although it has not
been possible to measure the $g$ factor for Cu in the layered
cuprates,\footnote{Despite numerous attempts, it has not been possible to
detect an electron-spin resonance signal from Cu$^{2+}$ in planar
cuprates \cite{simo93}.}  a typical value for a Cu$^{2+}$ ion in a
distorted octahedral environment is 2.2 \cite{abra70}, which implies a
typical moment of 0.67~$\mu_{\rm B}$.  The fact that the observed maximum
moments in \LCO\ and \YBCO\ are 10 to 20\%\ smaller than this could be a
result of hybridization of the Cu
$3d_{x2-y2}$ orbital with O $2p_\sigma$ orbitals.  (By symmetry, the
net spin density on oxygen sites is zero in the N\'eel structure.)

\subsubsection{Relative moments from $\mu$SR}

In $\mu$SR, one measures a precession frequency, which is directly
proportional to the local hyperfine field at the $\mu^+$ site.
Determination of the absolute moment requires knowledge of the $\mu^+$
location (obtained by calculation rather than experiment) and a
calculation of the magnetic field at that site due to the ordered Cu
moments.  Alternatively, under the assumption that the $\mu^+$ site is
unchanged by doping, one can determine the local moment in a doped sample
relative to that in the parent insulator.  An advantage of the $\mu$SR
technique is that its sensitivity does not depend on the existence of
long-range order.  A beautiful example of this is the study of
\textcite{nied98} in which it was shown that the local ordered moment (at
very low temperature) changes little as long-range antiferromagnetic
order is destroyed by doping in \LSCO\ and
Y$_{1-x}$Ca$_x$Ba$_2$Cu$_3$O$_6$.  In a complementary study,
\textcite{klau00} have shown that substantial moments associated with
local antiferromagnetic order (presumably stripe order pinned by the
tetragonal structure) are present in La$_{1.8-x}$Eu$_{0.2}$Sr$_x$CuO$_4$
for $0.08\lesssim x\lesssim 0.18$.

In a stripe-ordered sample, the muons sample a distribution of local
hyperfine fields, resulting in a fairly rapid damping of the precession
signal \cite{nach98,klau00,savi02}.  The relative moment quoted in the
table corresponds to the maximum of that distribution.

\subsection{Integrated low-energy spectral weight}

Neutron scattering measures the dynamical magnetic structure factor
$S({\bf Q},\omega)\equiv S_{\rm zz}(\bf Q,\omega)$. Actually, the spin 
structure factor is 
the tensorial quantity $S_{ \alpha \beta}(\bf Q,\omega)$, which obeys 
the sum rule
\begin{equation}
  \int_0^\infty d\omega \int_{\rm BZ} d{\bf Q}\, 
  \sum_{\alpha}S_{\alpha \alpha}({\bf Q},\omega) =
    S(S+1),
\end{equation}
where $S$ is the spin per magnetic ion.  For cuprates in general, and
superconducting samples in particular, it is challenging to measure the
magnetic scattering to sufficiently high energies (with sufficient
signal-to-noise ratio) that the sum rule can be evaluated; however, there
do exist a few studies that allow one to evaluate the energy integral up
to 100 meV.  While there are significant uncertainties in evaluating the
integral from published data, these should be smaller than the
uncertainty in the calibration of the sample volume contributing to the
signal, which for superconducting samples can be on the order 30\%\
\cite{fong00}.

Since we are interested in the change of this integrated quantity with
doping, we have normalized it to the value measured for the relevant
parent antiferromagnet.  Note that for the latter case, the integral
includes the signal in the magnetic Bragg peaks as well as that from the
spin waves.

Another quantity that one can evaluate is the relative weight under the
resonance peak, which appears below $T_c$ in the superconducting state.
Values, normalized to the full sum rule weight, have been tabulated by
\textcite{kee02}.  For \YBCO\ samples with $T_c$ varying from 52 to 93~K,
they found the normalized weight to be about 1.5\%; for a sample of
\BSCCO\ with $T_c=91$~K, the value is $\sim6$\%.  It should be noted that
integrating $S({\bf Q},\omega)$ up to 100~meV for the experimental
measurements on antiferromagnetic YBa$_2$Cu$_3$O$_{6.15}$ \cite{hayd96b}
gives just 20\%\ of the full sum rule weight; hence, using the same
normalization as for $\int S({\bf Q},\omega)$ in Table~\ref{tab:mag}, the
weight of the resonance peak in \YBCO\ is about 0.08.

\subsection{Energy scale of spin fluctuations}

\subsubsection{From neutron scattering}

In the spin-wave theory of a Heisenberg model on a square lattice with
only nearest-neighbor coupling, the zone-edge magnon with
${\bf q}_0 =(\pi,0)$ has energy
$\epsilon_{{\bf q}_0}=4SJ_{\rm eff}=2J_{\rm eff}$, where the final
equality applies for $S=\frac12$.  Longer-range couplings have observable
effects in the spin-wave dispersion curves measured for \LCO\
\cite{cold01}; nevertheless, it is useful to characterize
$\epsilon_{{\bf q}_0}$, which is essentially the local energy
to flip a spin, in terms of $J_{\rm eff}$.  Assuming that a local
super-exchange coupling is still active between local moments in doped
samples, we have used the same formula to define $J_{\rm eff}$ in terms
of the maximum spin-excitation energy observed in superconducting
cuprates.

\subsubsection{From two-magnon Raman scattering}

While Raman scattering is not sensitive to individual spin waves, it does
provide a valuable probe of two-magnon correlations in 2D square-lattice
antiferromagnets \cite{lyon88}.  The dominant part of the response has
$B_{1g}$ symmetry and appears at low temperature as a strong peak in
intensity, with the peak occurring at an energy of $2.8J_{\rm eff}$ for
$S=\frac12$ and $6.8J_{\rm eff}$ for $S=1$ \cite{cana92}.  The scattering
mechanism is believed to involve relatively short-range excitations, so
that the response is not strongly sensitive to the existence of
long-range order.  Although there is no formal justification for it, we
have used the formula for peak energy in the antiferromagnetic state in
order to extract a characteristic value for $J_{\rm eff}$ from the
2-magnon signal measured from superconducting samples.  Note that there
are also results available on \BSCCO\ \cite{blum97,suga00} which we have
not included in Table~\ref{tab:mag}.

}


\end{document}